\documentclass{jfm}

\pdfoutput=1 

\usepackage{graphicx}
\usepackage{natbib}
\usepackage{hyperref}
\usepackage{amsmath}
\usepackage{ar}
\usepackage{tikz}
\usepackage{pgfplots}
\usepackage{tikz-3dplot}
\usepackage{bm}
\usepackage{multirow}
\usepackage{subfig}
\usepackage{overpic}
\usepackage[normalem]{ulem}

\newcommand{\tcb}[1]{\textcolor{blue}{#1}}
\newcommand{\tcr}[1]{\textcolor{red}{#1}}

\title{On the symmetry-breaking instability of the flow past axisymmetric bluff bodies}

\author[A. Chiarini, R. Gauthier and E. Boujo]
{Alessandro Chiarini\aff{1}\aff{2}\corresp{alessandro.chiarini@polimi.it},  Romain Gauthier\aff{3} and Edouard Boujo\aff{3}\corresp{edouard.boujo@epfl.ch}
}
\affiliation{
}

\affiliation{
\aff{1} Complex Fluids and Flows Unit, Okinawa Institute of Science and Technology Graduate University, 1919-1 Tancha, Onna-son, Okinawa 904-0495, Japan
\aff{2} Dipartimento di Scienze e Tecnologie Aerospaziali, Politecnico di Milano, via La Masa 34, 20156 Milano, Italy
\aff{3}Laboratory of Fluid Mechanics and Instabilities, \'Ecole Polytechnique F\'ed\'erale de Lausanne, CH-1015 Lausanne, Switzerland
}

\begin{document}
\maketitle

%%%%%%%%%%%%%%%%%%%%%%%%%%%%%%%%%%%%%%%%%%%%%%%%%%
\begin{abstract}
The primary bifurcation of the flow past three-dimensional axisymmetric bodies is investigated. We show that the azimuthal vorticity generated at the body surface is at the root of the instability, and that the mechanism proposed by \cite{magnaudet-mougin-2007} in the context of spheroidal bubbles extends to axisymmetric bodies with a no-slip surface. The instability arises in a thin region of the flow in the 
near wake and is associated with the occurrence of strong vorticity gradients. We propose a simple yet effective scaling law for the prediction of the instability, based on a measure of the near-wake vorticity and of the radial extent of the separation bubble. At criticality, the resulting Reynolds number collapses approximately to a constant value for bodies with different geometries and aspect ratios, with a relative variation that is one order of magnitude smaller than that of the standard Reynolds number based on the free-stream velocity and body diameter. The new scaling can be useful to assess whether the steady flow past axisymmetric bodies is globally unstable, without the need for an additional stability analysis.
\end{abstract}


\section{Introduction}
\label{sec:introduction}

\subsection{Axisymmetric 3D bluff bodies}

The flow past a stationary isolated axisymmetric body may be considered as a simplified case of a more general family of immersed three-dimensional (3D) bluff bodies, which are ubiquitous in human life and engineering applications. In spite of the symmetries of the body, instabilities are known to generate  asymmetric and possibly unsteady flows; different regimes are indeed possible depending on the Reynolds number $Re = U_\infty H / \nu$, 
based on the body cross-stream dimension $H$, the free-stream velocity $U_\infty$ and the fluid kinematic viscosity $\nu$. In this work, we focus on the primary symmetry-breaking instability.
  
The flow past 3D axisymmetric bodies becomes asymmetric before transitioning to an unsteady regime \citep{magarvey-etal-1961,magarvey-etal-1961b,magarvey-etal-1965}. The steady axisymmetric flow past a sphere, for example, is known to transition to a steady asymmetric state at %a Reynolds number of 
$Re \approx 211$ \citep{johnson-patel-1999} through the regular bifurcation of an eigenmode of azimuthal wavenumber $m=1$ \citep{tomboulides-orszag-2000}. This bifurcation gives origin to a pair of counter-rotating steady streamwise vortices in the wake, that are not aligned with the flow but possess a reflectional symmetry about a longitudinal plane of arbitrary azimuthal orientation, and exhibit an eccentricity that increases with the distance from the body. 
The same bifurcation has been observed for the flow past other axisymmetric bodies, for example disks and bullet-shaped bodies (slender cylindrical bodies with a smooth leading edge and a blunt trailing edge). 
The flow past a disk placed perpendicular to the incoming flow exhibits the primary regular bifurcation towards an asymmetric state at 
a Reynolds number  $Re \approx 115$ \citep{natarajan-acrivos-1993,fabre-etal-2008,meliga-etal-2009}.  
The flow past a bullet-shaped body undergoes the same bifurcation, but at a larger Reynolds number \citep{bruckner-2001,bohorquez-etal-2011} %found that the flow undergoes the first regular bifurcation at a Reynolds number 
that increases with the length-to-diameter aspect ratio $\AR=L/H$ of the body, 
for example $Re \approx 216$ for $\AR=1$ and $Re \approx 435$ for $\AR=6$. 
A similar dependence on $\AR$ has been observed by \cite{chrust-etal-2010} for the flow past flat cylinders whose 
axis is parallel to the free stream. 
They observed that the flow bifurcates towards an asymmetric state at $Re \approx 115-120$ for $\AR \rightarrow 0$ and at $Re \approx 270$ for $\AR=1$.

The instability mechanism of the primary bifurcation of the flow past axisymmetric bodies has been extensively investigated over the years.
\cite{monkewitz-1988} investigated the linear stability of an analytic two-parameter family of 
axisymmetric, 
parallel and incompressible wake profiles. He observed that the first helical mode with $m=1$ displays the largest growth rate for all cases, and that it is the only mode to become absolutely unstable for velocity profiles approximating those found in the near wake.
\cite{pier-2008} studied the local absolute instability features of the flow past a sphere, with the aim of linking the local flow properties with the fundamental mechanism driving the global flow bifurcation. 
He neglected the strong non-parallelism of the near-wake region, and highlighted the local properties by freezing the flow at different streamwise coordinates and studying the equivalent axially parallel shear flows. 
In doing so, he demonstrated the existence of absolutely unstable regions in the near wake, and found that the strength and spatial extent of these regions increase with the Reynolds number. 
The results of \cite{monkewitz-1988} and \cite{pier-2008} clearly hint at the existence  of a wavemaker in the  wake (an absolutely unstable region where the fluctuations self-sustain), that feeds the convectively unstable region downstream. However, these results can not fully explain the symmetry breaking corresponding to the first bifurcation of the flow. 

\subsection{MM's model for free-slip bodies}

In the context of free-slip oblate spheroidal bodies, \cite{magnaudet-mougin-2007} (hereafter referred to as MM) proposed an instability mechanism that accounts for the strongly non-parallel nature of the flow. 
The theoretical arguments they put forward are based on the idea that the regular bifurcation of the flow is driven by the azimuthal vorticity $\omega_\theta = \partial u_r/\partial z - \partial u_z/\partial r$  generated at the body surface and then transported into the wake (where $(u_r,u_z)$ are the velocity components in the  radial and axial directions $r$ and $z$). 
For free-slip ellipsoids, in fact, they found that the wider range of $Re$ for which the flow becomes unstable as $\AR$ increases (see figure 7 of their paper) well correlates with the increase in 
maximum surface vorticity $\omega_{\theta,\max}$ and 
normal diffusive vorticity flux at the body surface (see section 3 of their paper). 
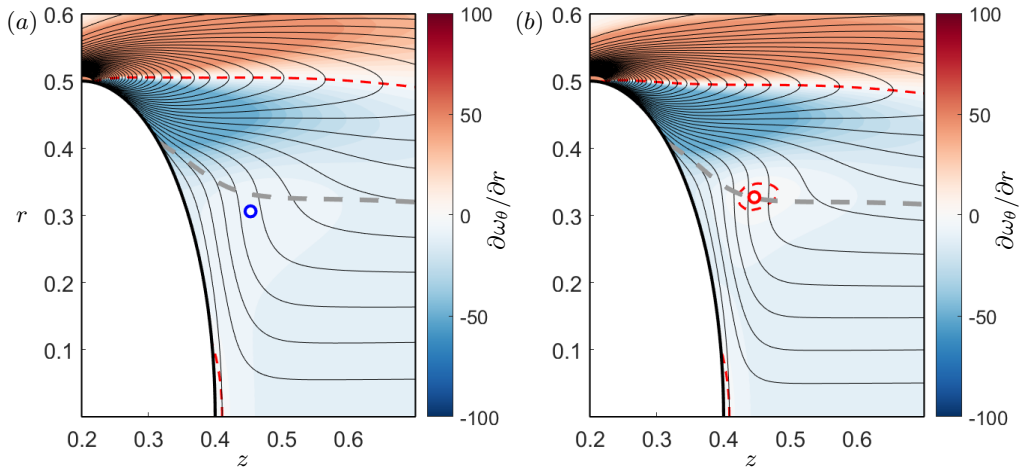
\begin{figure}
\centering
\centerline{  
    \begin{overpic}[width=0.49\textwidth, trim=15mm 0mm 20mm 5mm, clip=true]{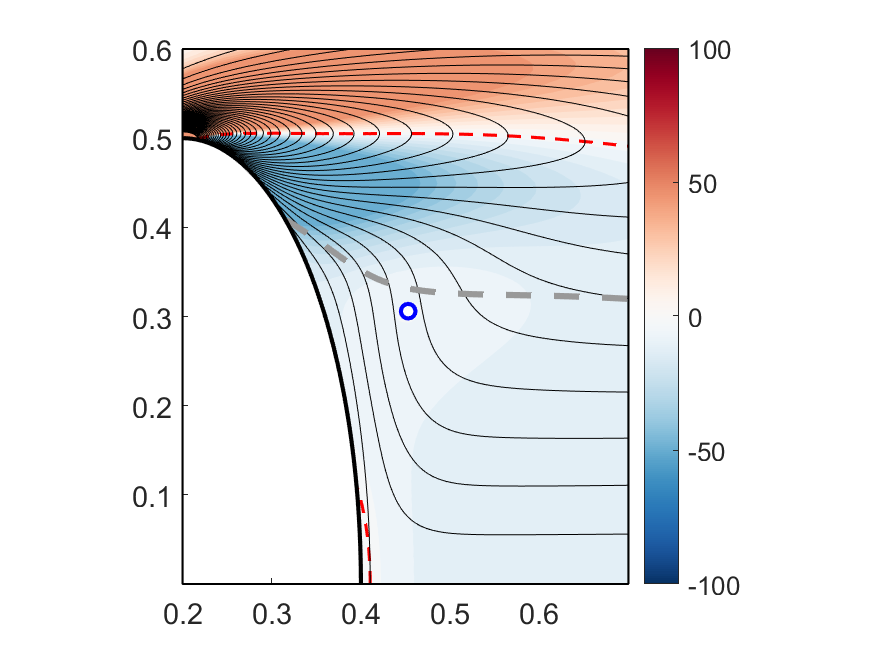}
        \put(-1,88){$(a)$}    
        \put(1,49.5){$r$} 
        \put(45,1){$z$}
        \put(95,44){\rotatebox{90}{$\partial \omega_\theta/\partial r$}}
 	\end{overpic} 
    \begin{overpic}[width=0.49\textwidth, trim=15mm 0mm 20mm 5mm, clip=true]{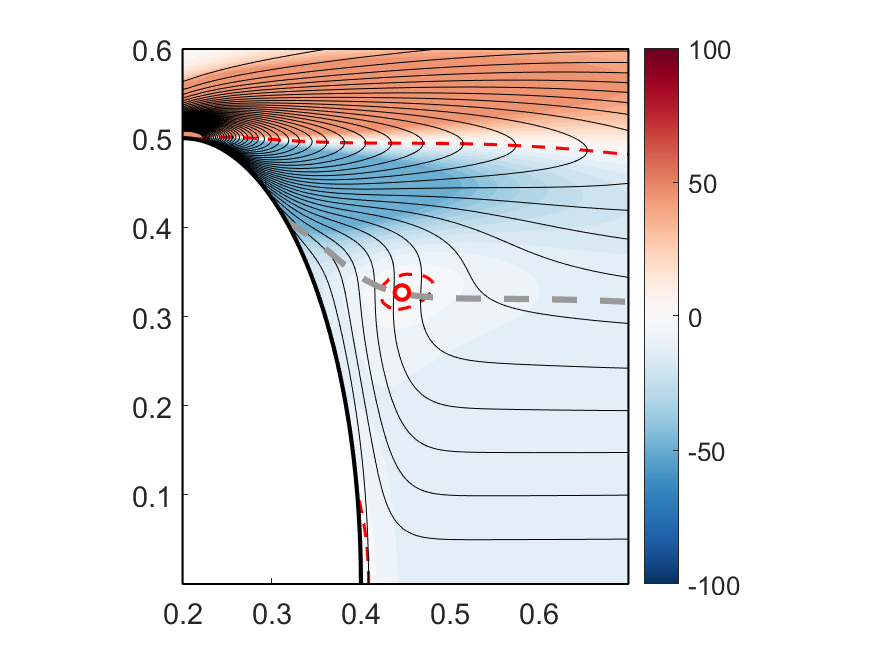}
        \put(-1,88){$(b)$}   
        \put(45,1){$z$}
        \put(95,44){\rotatebox{90}{$\partial \omega_\theta/\partial r$}}
 	\end{overpic}
}
\caption{
Near-wake distribution of the azimuthal vorticity around an ellipsoid with free-slip surface and $\AR=0.4$ for $(a)$~$Re=200$ and $(b)$~$Re=270$. 
Black lines are isocontours of the azimuthal vorticity $\omega_\theta$, and coloured contours show $\partial \omega_\theta/\partial r$.
The grey dashed line shows $u_{z} = 0$, and the red dashed line  $\partial \omega_\theta/\partial r = 0$.  
Blue/red circles indicate the negative/positive maximum of $\partial \omega_\theta/\partial r$ in the near-wake region.}
\label{fig:MM}
\end{figure}
They suggested that the instability originates in a thin region of the flow downstream the body. By inspecting  the  vorticity field, they observed that two distinct regions may be identified  (see figure 21 of their paper and figure \ref{fig:MM}):
(i)~very close to the body, isocontours or $\omega_\theta$  are almost parallel to the free-slip surface; %, which behaves substantially as a vorticity-free boundary. 
(ii)~farther downstream in the wake recirculation region, instead, isocontours of $\omega_\theta$  are almost parallel to the symmetry axis, because of the tendency of $\omega_\theta/r$ to approach a constant value as $Re$ increases.
At large Reynolds numbers, indeed, the spatial distribution of the vorticity in the recirculation region of the steady viscous flow past an axisymmetric body resembles that of a Hill's spherical vortex \citep{hill-1894,batchelor-1967}; see for example figure 10 in \cite{fornberg-1988} and the related discussion. 
MM thus conjectured that the instability arises in the transition region where isocontours of $\omega_\theta$  have to  turn %sharply 
to satisfy both the conditions at the body surface and in the wake recirculation region. As $Re$ increases this transition region shrinks, and isocontours of $\omega_\theta$ must turn more and more sharply, resulting into stronger and stronger vorticity gradients. 

In an attempt to provide a rational instability criterion, MM related the flow bifurcation with the appearance of points where $\partial \omega_\theta/ \partial r = 0$ in the near wake. 
This comes directly from the balance equation for the steady axisymmetric base-flow vorticity, that reads in cylindrical coordinates 
\begin{equation}
\underbrace{ u_r \dfrac{\partial \omega_\theta}{\partial r } }_{\mathrm{I}} 
\underbrace{ - u_r \dfrac{\omega_\theta}{r} }_{\mathrm{II}} 
+
\underbrace{ u_z \dfrac{ \partial \omega_\theta }{\partial z} }_{\mathrm{III}}
=
\underbrace{ \dfrac{1}{Re} \dfrac{\partial}{\partial r} \left( \frac{1}{r} \dfrac{ \partial ( r \omega_\theta ) }{\partial  r} \right)}_{\mathrm{IV}} 
+
\underbrace{ \dfrac{1}{Re} \dfrac{\partial^2 \omega_\theta}{\partial z^2} }_{\mathrm{V}}.
\label{eq:vort}
\end{equation}
We focus on the near-wake region where $\omega_\theta<0$, $u_z<0$ and $u_r>0$ (see figure \ref{fig:MM}). 
Here, terms $\mathrm{II}$ and $\mathrm{III}$ at the l.h.s. 
are positive everywhere. 
As long as $\omega_\theta$ becomes more negative with $r$ (i.e. $\partial \omega_\theta/\partial r <0$), term $\mathrm{I}$ at the l.h.s. is negative and may, at least partially, balance terms $\mathrm{II}$ and $\mathrm{III}$. When instead an isocontour of $\omega_\theta$ becomes perpendicular to the symmetry axis (i.e. $\partial \omega_\theta/\partial r=0$), term $\mathrm{I}$ vanishes and the positive l.h.s. can only be balanced by the positive viscous terms $\mathrm{IV}$ and $\mathrm{V}$. 
Note that, where $\partial \omega_\theta/\partial r = 0$, term $\mathrm{IV}$ reduces to $- \omega_\theta/r^2/Re \sim Re^{-1}$. 
Therefore, as the Reynolds number increases, term $\mathrm{V}$ is observed to dominate over term $\mathrm{IV}$,
and the balance equation reduces to:
\begin{equation}
- u_r \frac{ \omega_\theta}{r} + u_z \frac{ \partial \omega_\theta }{\partial z} \sim \frac{1}{Re} \frac{ \partial^2 \omega_\theta }{\partial z^2}.
\end{equation}
This implies that, to balance the l.h.s., the streamwise gradient of $\omega_\theta$ in term $\mathrm{V}$ at the r.h.s. has to vary more and more sharply (over a shorter distance along $z$) as $Re$ increases. 
 MM's idea is that this tendency of the vorticity to become discontinuous in the points where $\partial \omega_\theta/\partial r = 0$ underlies the instability mechanism. 

Notably, the numerical simulations of \cite{magnaudet-mougin-2007} corroborate their arguments, showing a good correlation between the first appearance of points with $\partial \omega_\theta/\partial r = 0$ and the onset of the bifurcation in the context of free-slip axisymmetric bluff bodies. However, despite some hints provided by the same authors (see the concluding discussion in their paper), it is not clear whether axisymmetric bodies with free-slip and no-slip surfaces share this same instability mechanism driven by the vorticity generated at the surface. It is well known, indeed, that the mechanism of vorticity generation on a surface changes with the boundary condition; see the seminal works of \cite{truesdell-1954,lighthill-1963, morton-1983, leal-1989, wu-wu-1993, wu-1995, lundgren-koumoutsakos-1998} and the more recent works of \cite{brons-etal-2014, terrington-etal-2020}. 
For no-slip surfaces, the normal diffusion vorticity flux depends on the tangential shear stress and on the pressure gradient, and it is non null also in the limit case of a flat wall \citep{wu-wu-1993}. On a free-slip surface, instead, the vorticity appears as a consequence of the continuity of the tangent stresses, and is non null only in the case of curved surfaces \citep{wu-1995}. 
Thus, the generality of MM's mechanism and the influence of the different vorticity generation mechanisms at no-slip and free-slip surfaces on the primary flow bifurcation are still open questions that deserve further investigation.

\subsection{Focus of the present work}

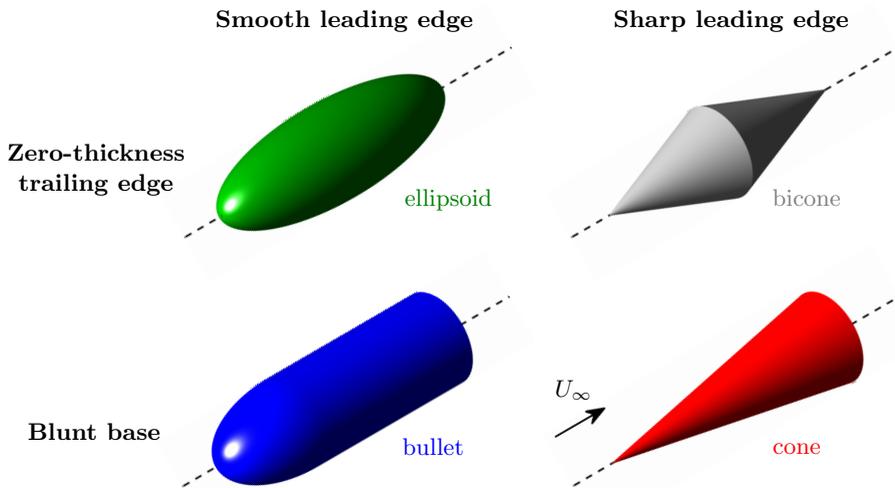
\begin{figure}
\vspace{1.cm}
\centerline{   
\hspace{2cm}
    \begin{overpic}[height=5cm, trim=0mm 0mm 0mm 0mm, clip=true, angle=-60]{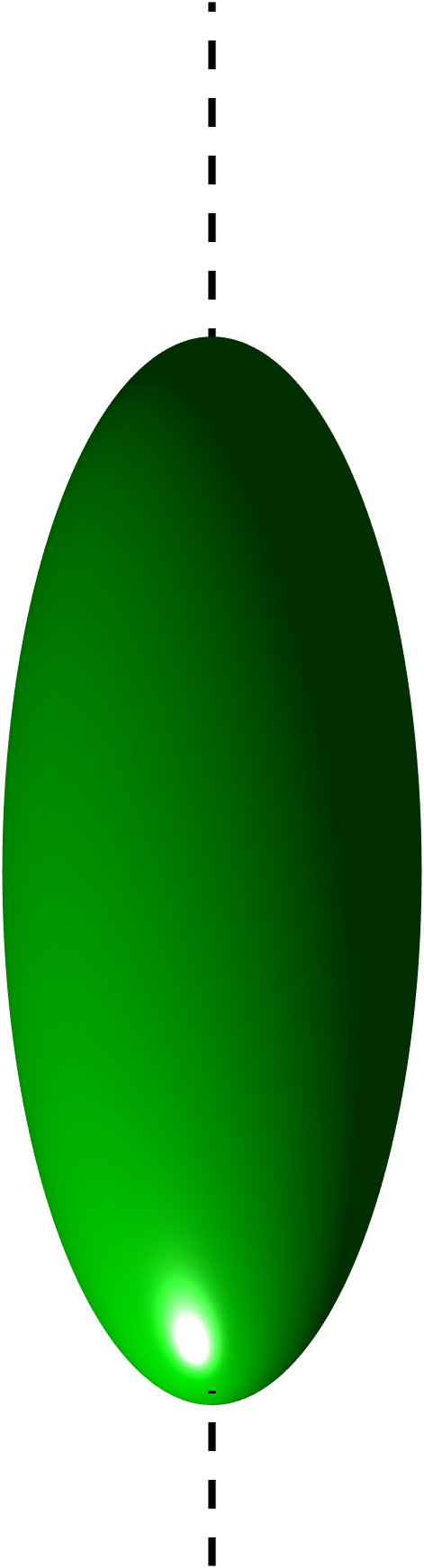}
 	      \put(15,67){\textbf{Smooth leading edge}}    
 	      \put(-40,32){\textbf{Zero-thickness}}    
 	      \put(-37,24){\textbf{trailing edge}}
        \put(65,20){\textcolor[rgb]{0,0.5,0}{ellipsoid}}       
    \end{overpic} 
    \begin{overpic}[height=5cm, trim=0mm 0mm 0mm 0mm, clip=true, angle=-60]{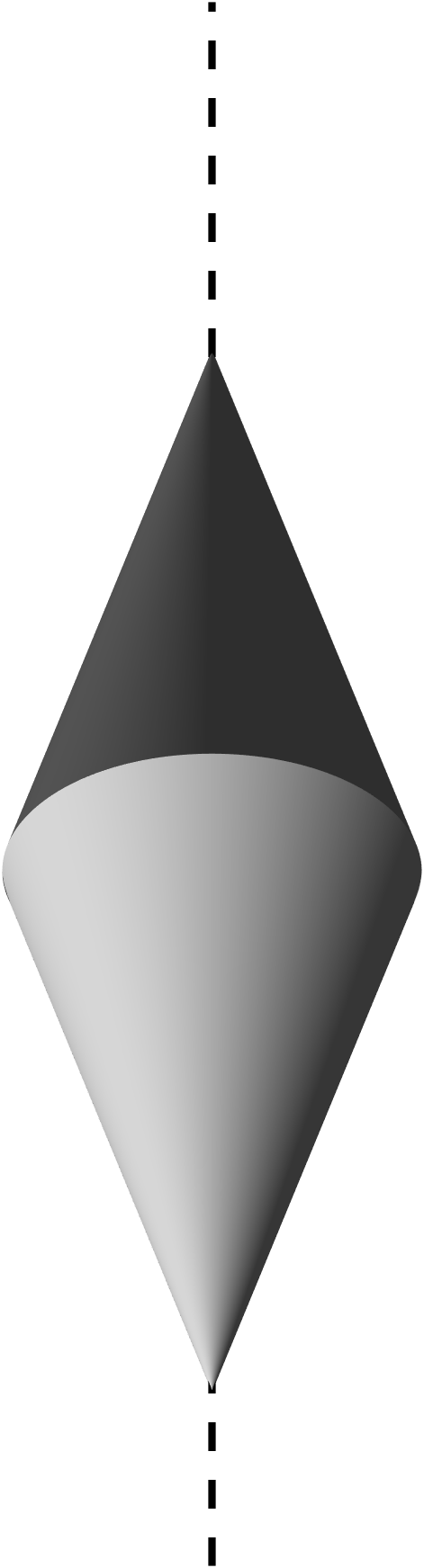}
 	      \put(18,67){\textbf{Sharp leading edge}}  
        \put(60,20){\textcolor{gray}{bicone}}   
    \end{overpic} 
}
\vspace{-0.5cm}
\centerline{   
\hspace{2cm}
    \begin{overpic}[height=5cm, trim=0mm 0mm 0mm 0mm, clip=true, angle=-60]{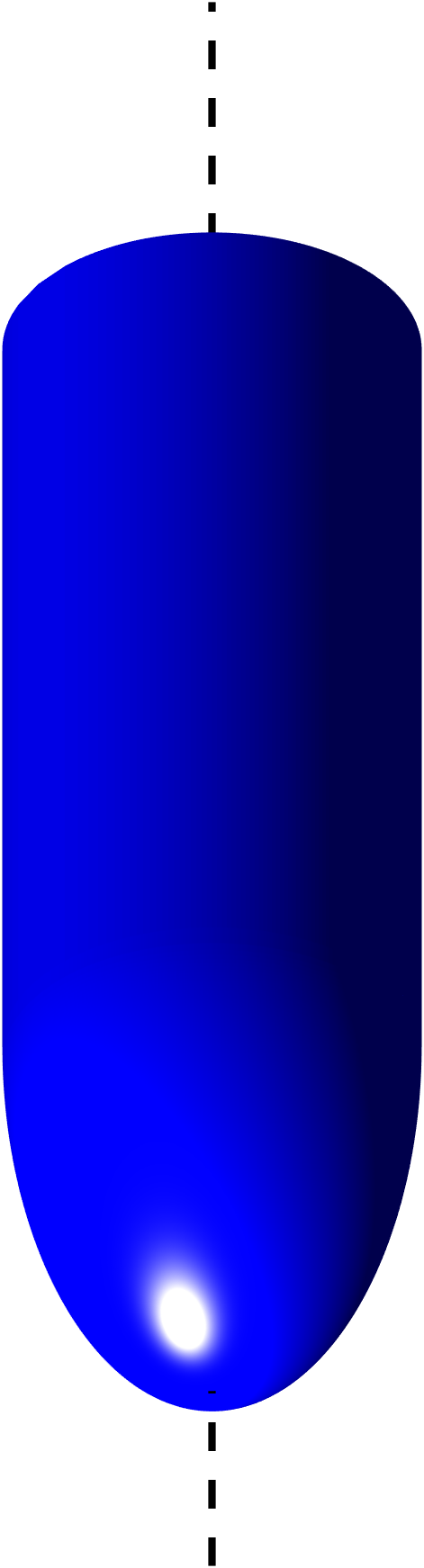}
        \put(-34,24){\textbf{Blunt base}}  
        \put(65,20){\tcb{bullet}}   
    \end{overpic} 
    \begin{overpic}[height=5cm, trim=0mm 0mm 0mm 0mm, clip=true, angle=-60]{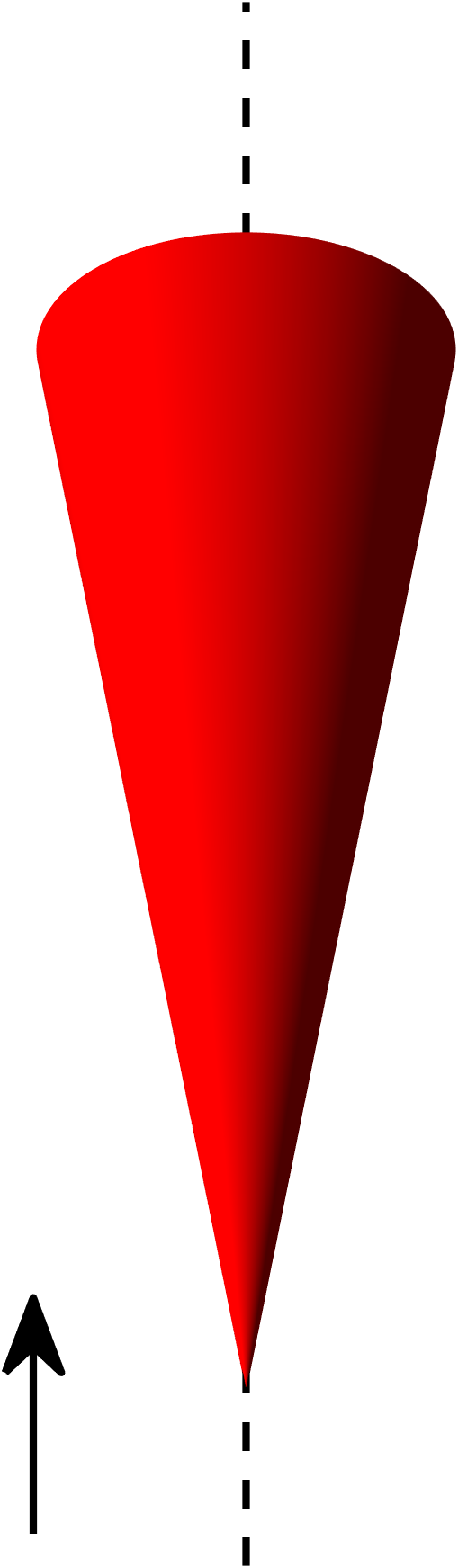}
        \put(3,35){$U_\infty$} 
        \put(60,20){\tcr{cone}} 
    \end{overpic}
}
\caption{Sketch of the considered geometries. Here $\AR=3$.}
\label{fig:geometry}
\end{figure}

To address these questions, we study the flow past no-slip axisymmetric bluff bodies with different shapes and aspect ratios, and focus on the primary symmetry-breaking bifurcation. 
We consider four geometries,  ellipsoids, bullets, cones and bicones, that feature different combinations of smooth/sharp leading edge and smooth/blunt trailing edge (see figure \ref{fig:geometry}). 
The aspect ratio of the ellipsoid and the bicone is varied between $1 \le \AR \le 5$, and that of the bullet and the cone between $1 \le \AR \le 8$.

We give evidence that  MM's instability mechanism captures rather well the primary instability of the flow past axisymmetric bodies with different shapes and surface types. Indeed, we find that the onset of the bifurcation correlates well with the first appearance of points where $\partial \omega_\theta/\partial r = 0$ in the near wake not only for free-slip surfaces but also for no-slip surfaces, despite the different vorticity generation mechanism at the
surface. 
As first hinted by \cite{magnaudet-mougin-2007}, this shows that although the vorticity is at the root of the instability, the way it is produced at the wall does not play a major role. 
We then propose a simple yet effective scaling law, that may be used to predict the onset of the instability without the need for 
an additional stability analysis. 
Indeed, the large variability of the critical Reynolds number with the geometry (for example, as mentioned above, $Re_c \approx 115$ for the disk, $Re_c \approx 211$ for the sphere,  and $ 126 \le Re_c \le 435$ for bullet-shaped bodies with $1 \le \AR \le 6$) clearly shows that the free-stream velocity $U_\infty$ and the cross-stream dimension $H$ of the body are not the appropriate velocity and length scales for describing this bifurcation. 
In the same spirit as \cite{chiarini-quadrio-auteri-2022b} in the context of two-dimensional symmetric bodies, the new scaling is based on quantities related to the physics of the problem, i.e. measures of the vorticity at the body surface and of the radial extent of the wake recirculation region. 
We show that the resulting Reynolds number, when evaluated at the onset of the bifurcation, collapses rather well to a constant value for all the considered bodies, and thus provides a simple criterion to assess the flow stability.

\section{Methods}
\label{sec:methods}

\subsection{Problem formulation}

We investigate the primary symmetry-breaking bifurcation of the incompressible flow past 3D axisymmetric bluff bodies with different geometries and aspect ratios. The bodies have length $L$ and maximum diameter $H$, and are placed in a uniform flow with velocity $U_\infty$ aligned with their symmetry axis (see figure \ref{fig:geometry}). Standard cylindrical coordinates are used (with $r$, $\theta$ and $z$  the radial, azimuthal and axial directions), and the origin is placed at the leading edge of the bodies. 
The flow is governed by the incompressible Navier--Stokes (NS) equations for the velocity and pressure fields $\{\bm{u},p\}$,
\begin{equation}
\frac{\partial \bm{u}}{\partial t} + \bm{u} \cdot \bm{\nabla} \bm{u} = - \bm{\nabla} p + \frac{1}{Re} \bm{\nabla}^2 \bm{u}, \ \bm{\nabla} \cdot \bm{u} = 0,
\label{eq:NS}
\end{equation}
with $Re = U_\infty H /\nu$.
Unless otherwise stated, all quantities are made dimensionless using $U_\infty$ and $H$.

We consider four different geometries, that yield a base flow with distinct features:
ellipsoids, bullets, cones and bicones (see figure \ref{fig:geometry}). The aspect ratio of the bodies $\AR = L/H$ is varied between $ 1 \le \AR \le 5$ for ellipsoids and bicones and between $1 \le \AR \le 8$ for bullets and cones.
Ellipsoids and bullets possess a smooth leading edge (LE), where the flow separation is driven by the adverse pressure gradient and the position of the separation point changes with the Reynolds number; conversely, cones and bicones feature a zero-thickness sharp LE. 
Bullets and cones have a blunt trailing edge (TE), where the separation point  is enforced by the geometry, and the cross-stream size of the wake recirculation region does not depend on $Re$ and $\AR$; by contrast, ellipsoids and bicones have a zero-thickness TE, where the flow separation is not set by the geometry and the size of the wake recirculation region varies with $Re$ and $\AR$. 

\subsection{Linear stability}

The onset of the primary instability is studied using linear theory \citep{theofilis-2003,theofilis-2011}. The field $\{\bm{u},p\}$ of velocity and pressure is divided into a time-independent axisymmetric base flow $\{\bm{u}_0,p_0\}$, and an unsteady perturbation $\{\bm{u}_1,p_1\}$ of small amplitude $0<\epsilon \ll 1$, 
\begin{equation}
\bm{u}(\bm{x},t) = \bm{u}_0(\bm{x}) + \epsilon \bm{u}_1(\bm{x},t) \ \text{and} \ p(\bm{x},t) = p_0(\bm{x}) + \epsilon p_1(\bm{x},t).
\end{equation}
Introducing this decomposition in the NS equations, one obtains at order $\epsilon^0$ the steady nonlinear  NS equations for the base flow $\{\bm{u}_0,p_0\}$. 
At order $\epsilon^1$, one obtains the linearised Navier--Stokes equations (LNSE) for the perturbation field $\{\bm{u}_1,p_1\}$. 
By using a normal mode ansatz and a Fourier transform in the azimuthal direction, for each mode the perturbation field takes the form
\begin{equation}
  \{{\bm{u}}_{1},{p}_{1}\} (\bm{x},t)=  \{ \hat{\bm{u}}_{1}, \hat{p}_{1} \}(r,z) e^{\lambda t + i m \theta} + c.c.,
\end{equation}
where $m$ is the azimuthal wavenumber and $c.c$ designates the complex conjugate terms.
Introducing this expansion in the LNSE yields, for each $m$, an eigenvalue problem for the complex eigenvalue $\lambda  = \lambda_r+ i \lambda_i$ and the complex eigenvector $\{\hat{\bm{u}}_1,\hat{p}_1\}$, 
\begin{equation}
\lambda \hat{\bm{u}}_1 + \mathcal{L}_m\{\bm{u}_0,Re\} \hat{\bm{u}}_1 + \bm{\nabla}_m \hat{p}_1 = 0, \ 
\bm{\nabla}_m \cdot \hat{\bm{u}}_1=0;
\label{eq:LNSE}
\end{equation}
$\bm{\nabla}_m$ is the gradient operator relative to the azimuthal wavenumber $m$, and $\mathcal{L}_m$ stands for the Fourier-transformed linearised Navier--Stokes operator
\begin{equation}
  \mathcal{L}_m\{\bm{u}_0,Re\}\hat{\bm{u}}_1 = \mathcal{C}_m ( \hat{\bm{u}}_1, \bm{u}_0 ) - \frac{1}{Re} \bm{\nabla}_m^2 \hat{\bm{u}}_1
\end{equation}
where
\begin{equation*}
  \mathcal{C}_m( \bm{u}_A, \bm{u}_B ) = \bm{u}_A \cdot \bm{\nabla}_m \bm{u}_B + \bm{u}_B \cdot \bm{\nabla}_m \bm{u}_A.
\end{equation*}
The flow stability is ascertained by the solution of the generalised eigenvalue problem (\ref{eq:LNSE}) for the complex frequency $\lambda$. When $\lambda_r<0$ the flow is linearly stable, while when $\lambda_r>0$ the associated global mode is linearly unstable and grows exponentially in time. When $\lambda_i \neq 0$ the unstable mode is time dependent. When $m \neq 0$ the unstable mode is modulated in the azimuthal direction. Since the focus of the present work is on the primary symmetry-breaking regular bifurcation of the flow past axisymmetric bodies, we restrict our analysis to $m=1$.

\subsection{Numerical method}

The  analysis is based on finite elements, and the simulations have been carried out using the COMSOL Multiphysics software \citep{comsol-1998}. 

Given the axisymmetric nature of the bodies, we consider only the $(r,z)$ plane, and build the mesh in the numerical domain $\{r,z \ | \ 0 \leq r \leq 50; -50 \leq z \leq 150\}$, with nodes strongly clustered near the body. We employ a quadrilateral mesh with $8$ layers and a growth factor of $1.2$ in the vicinity of the body, and a triangular mesh in the remaining part of the domain. Several sub-domains are used to have control of the mesh size. The element size decreases from $1$ in the farfield to $1/120$ on the body surface; at the corners the element size is approximately $10^{-5}$. The number of elements changes with the geometry and the aspect ratio. For the shortest and longest bodies %with $\AR=1$ and $\AR=8$, 
the number of elements is $79 \times 10^3$ and $126 \times 10^3$ for ellipsoids, $83 \times 10^3$ and $146 \times 10^3$ for bicones, $74 \times 10^3$ and $122 \times 10^3$ for bullets and $77 \times 10^3$ and $143 \times 10^3$ for cones.
See Appendix~\ref{sec:appendix_convergence} for details about the mesh convergence.

The low-$Re$ steady axisymmetric base flow $\{\bm{u}_0,p_0\}$ is obtained by solving the axisymmetric steady version of the NS equations (\ref{eq:NS}) using the Newton's iteration method.
The NS equations are completed with the following boundary conditions:  uniform velocity field $\bm{u}_0 = U_\infty \bm{e}_z$ at the inlet, stress-free $p_0\bm{n} - Re^{-1} \bm{\nabla} \bm{u}_0 \cdot \bm{n} = \bm{0}$ at the outlet and farfield ($\bm{n}$ is the unit normal vector), no-slip and no-penetration $\bm{u}_0=\bm{0}$ on the body surface, and axisymmetry conditions $u_{0r} = \partial_r u_{0z} = \partial_r p_0 = 0$ on the axis $r=0$.
The generalised eigenvalue problem \eqref{eq:LNSE} for the onset of the primary instability is then solved using the Arnoldi algorithm, with standard boundary conditions: $\bm{u}_1=\bm{0}$ at the inlet and on the body surface, stress-free $p_1\bm{n} - Re^{-1} \bm{\nabla} \bm{u}_1 \cdot \bm{n} = \bm{0}$ at the outlet and farfield, and $m=1$ conditions $\partial_r u_{1r} = \partial_r u_{1\theta} =  u_{1z} =  p_1 = 0$ on the axis.
To avoid singularity on the axis, the equations %(\ref{eq:NS}) 
are multiplied by $r^2$ before taking the variational form. For the computation of the base flow, the finite elements formulation employs the higher-order Lagrange P3 and P2 elements for velocity and pressure, respectively. For the linear stability analysis, instead, P2 and P1 elements are used. 
\section{The symmetry-breaking bifurcation}
\label{sec:bifurcation}

\subsection{Base flow}

%------ AR=1 ------
\begin{figure}
\vspace{0.5cm}
\centerline{ 
    \begin{overpic}[width=0.49\textwidth, trim=0mm 0mm 0mm 0mm, clip=false]{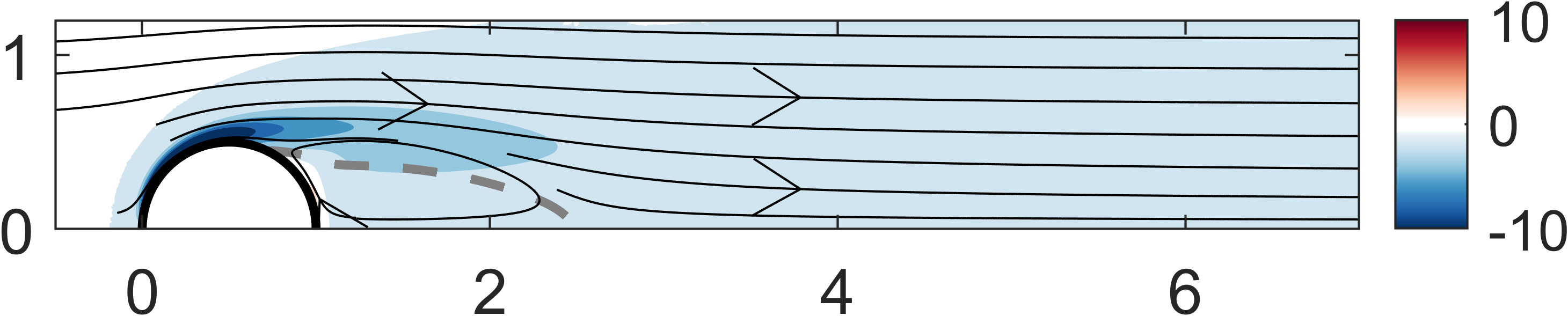} 
       \put(-8,19){$(a)$}
       \put(-5,10){$r$} 
 	\end{overpic}
    \begin{overpic}[width=0.49\textwidth, trim=0mm 0mm 0mm 0mm, clip=false]{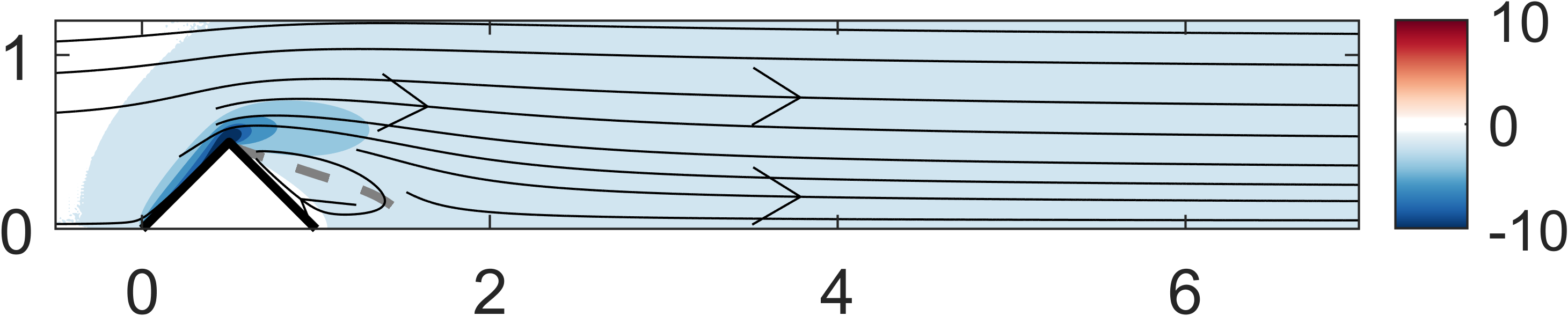}  
 	\end{overpic}
}
\centerline{ 
    \begin{overpic}[width=0.49\textwidth, trim=0mm 0mm 0mm 0mm, clip=false]{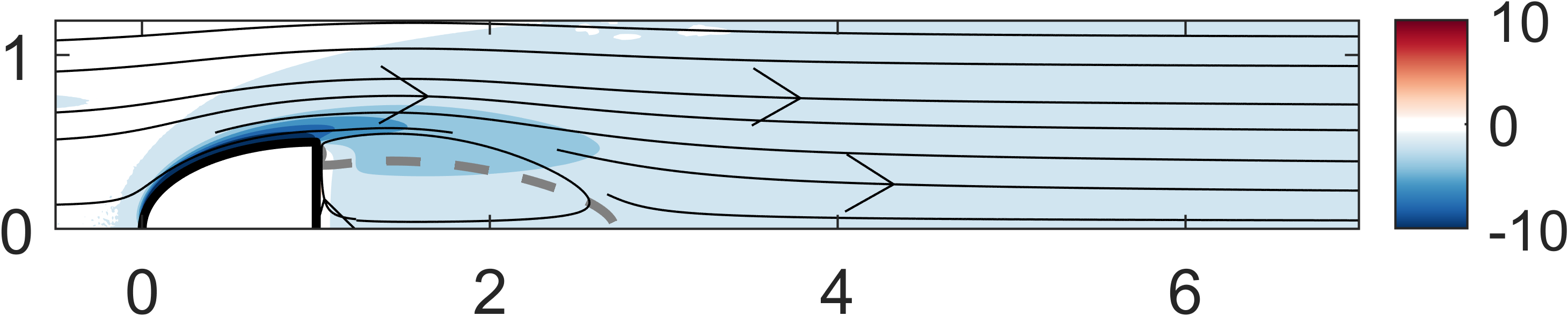}  
       \put(-5,10){$r$}
       \put(47,-4){$z$}
 	\end{overpic}
    \begin{overpic}[width=0.49\textwidth, trim=0mm 0mm 0mm 0mm, clip=false]{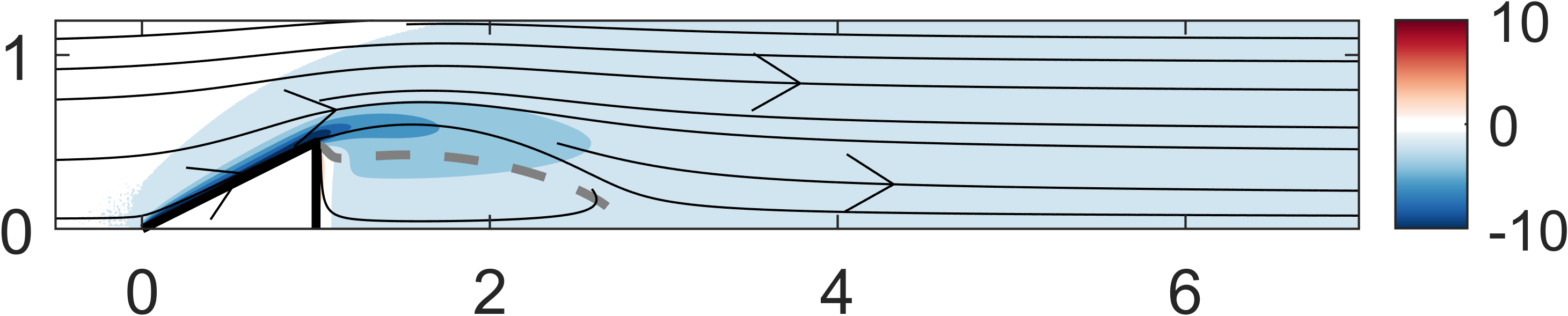}  
       \put(47,-4){$z$}
 	\end{overpic}
}
%----------------------------------
\vspace{0.7cm}
%----------------------------------
\centerline{ 
    \begin{overpic}[width=0.49\textwidth, trim=0mm 0mm 0mm 0mm, clip=false]{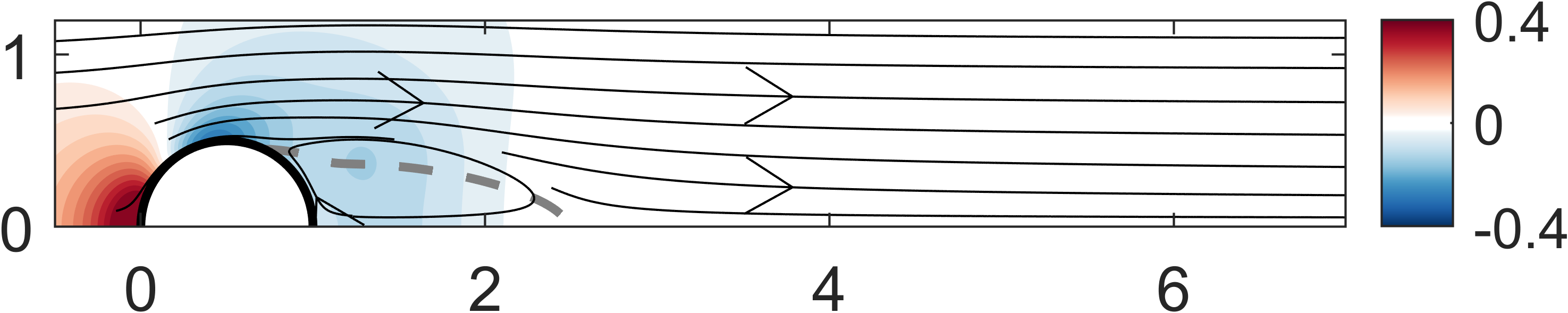}  
       \put(-8,19){$(b)$}
       \put(-5,10){$r$}
 	\end{overpic}
    \begin{overpic}[width=0.49\textwidth, trim=0mm 0mm 0mm 0mm, clip=false]{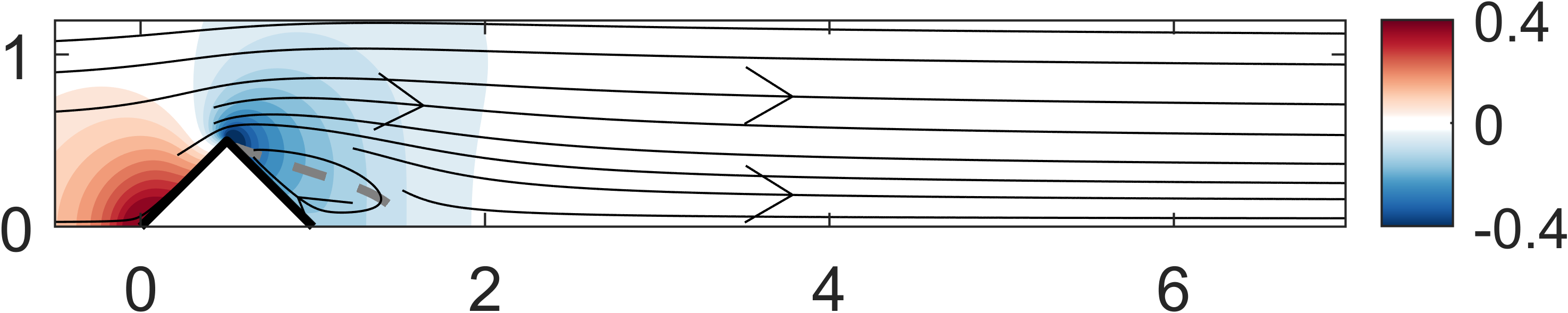} 
 	\end{overpic}
}
\centerline{ 
    \begin{overpic}[width=0.49\textwidth, trim=0mm 0mm 0mm 0mm, clip=false]{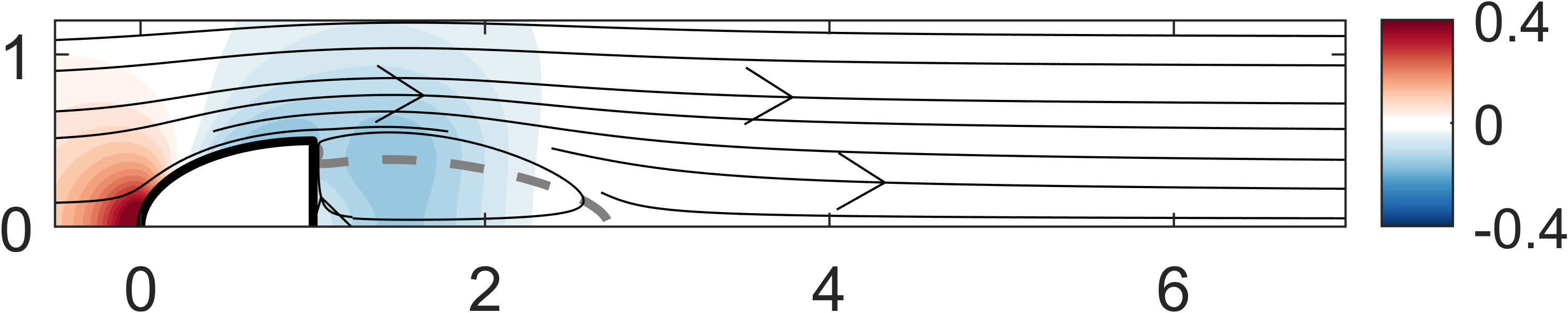}  
       \put(-5,10){$r$}
       \put(47,-4){$z$}
 	\end{overpic}
    \begin{overpic}[width=0.49\textwidth, trim=0mm 0mm 0mm 0mm, clip=false]{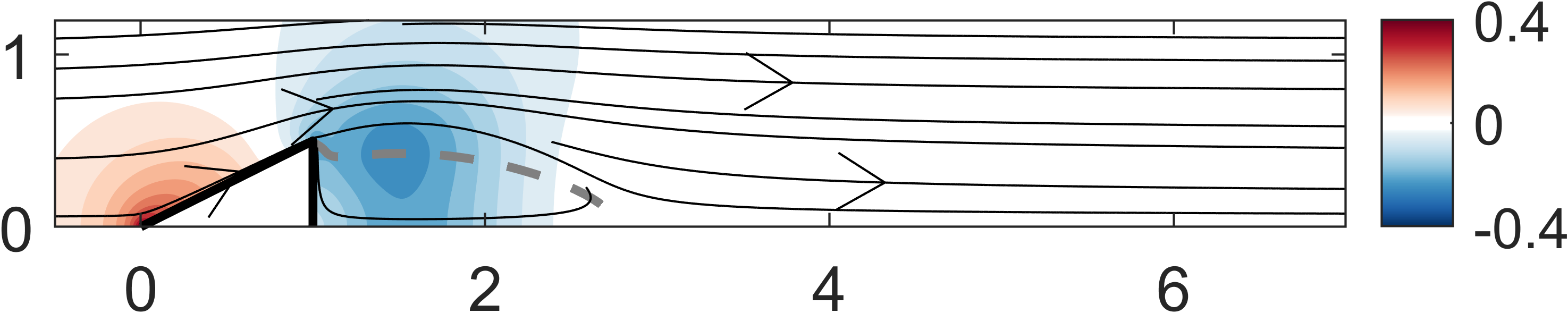}  
       \put(47,-4){$z$}
 	\end{overpic}
}
\vspace{0.2cm}
\caption{
Base flow for $\AR=1$ at $Re \simeq Re_c$: streamlines and contours of
$(a)$~azimuthal vorticity,
$(b)$~pressure.
Ellipsoid, $Re=210$;
bicone, $Re=135$;
bullet, $Re=220$;
cone, $Re=160$. 
Dashed line: $u_{0z}=0$.
}
\label{fig:BF-vort-press-AR1}
\end{figure}

%------ AR=4 ------
\def\field{vort}
\begin{figure}
\centerline{ 
    \begin{overpic}[width=0.49\textwidth, trim=0mm 0mm 0mm 0mm, clip=false]{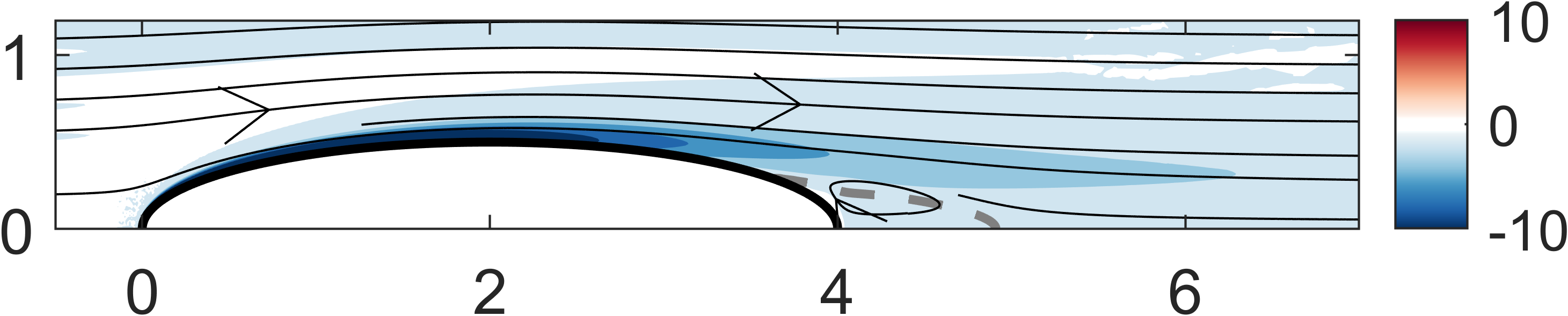}  
       \put(-8,19){$(a)$}
       \put(-5,10){$r$}
 	\end{overpic}
    \begin{overpic}[width=0.49\textwidth, trim=0mm 0mm 0mm 0mm, clip=false]{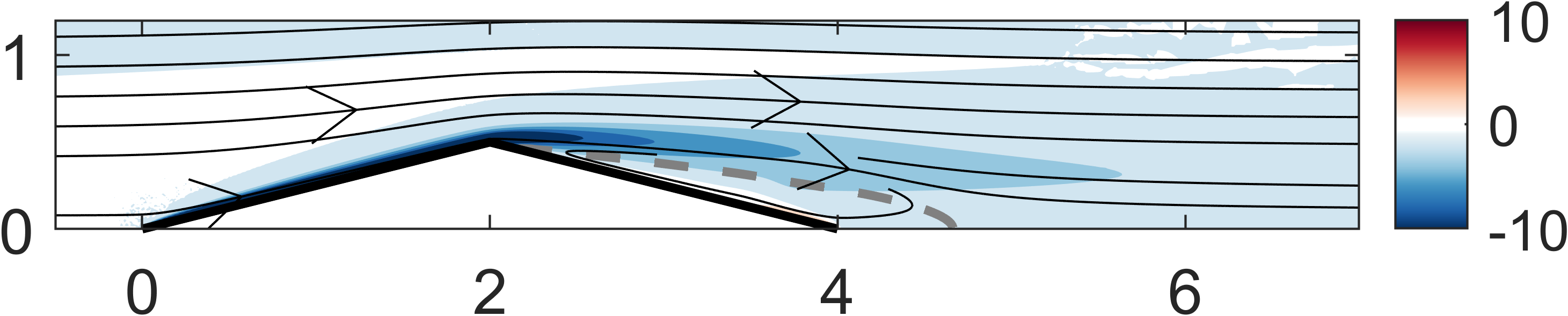}  
 	\end{overpic}
}
\centerline{ 
    \begin{overpic}[width=0.49\textwidth, trim=0mm 0mm 0mm 0mm, clip=false]{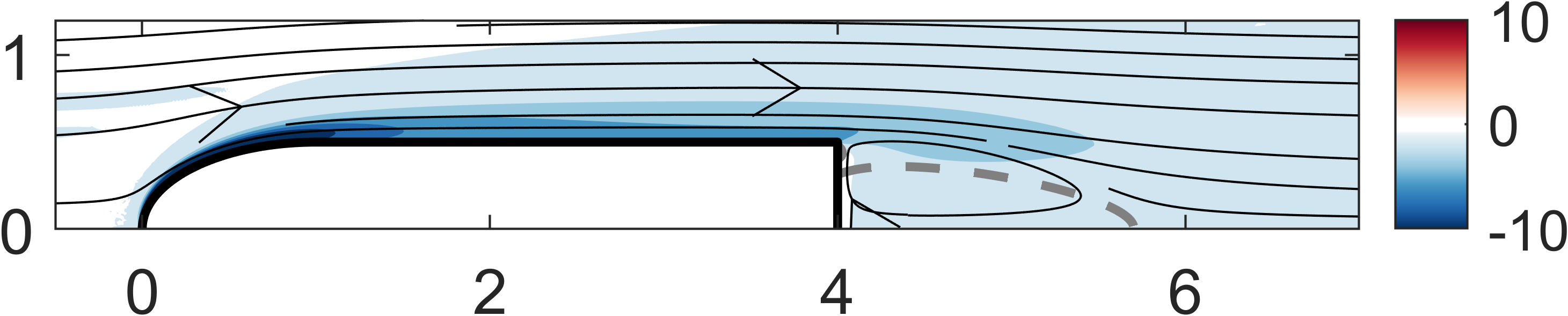}
       \put(-5,10){$r$}
       \put(47,-4){$z$}  
 	\end{overpic}
    \begin{overpic}[width=0.49\textwidth, trim=0mm 0mm 0mm 0mm, clip=false]{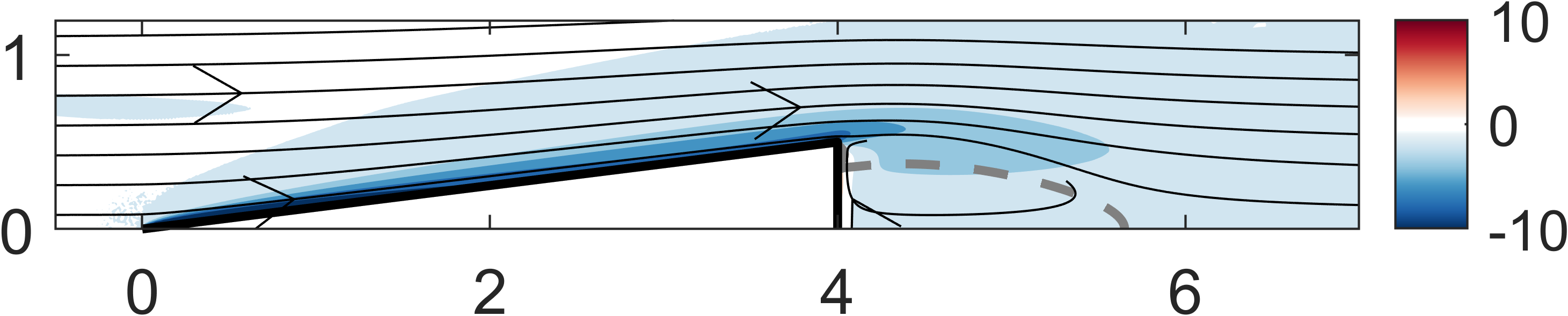} 
       \put(47,-4){$z$} 
 	\end{overpic}
}
%----------------------------------
\vspace{0.7cm}
%----------------------------------
\centerline{ 
    \begin{overpic}[width=0.49\textwidth, trim=0mm 0mm 0mm 0mm, clip=false]{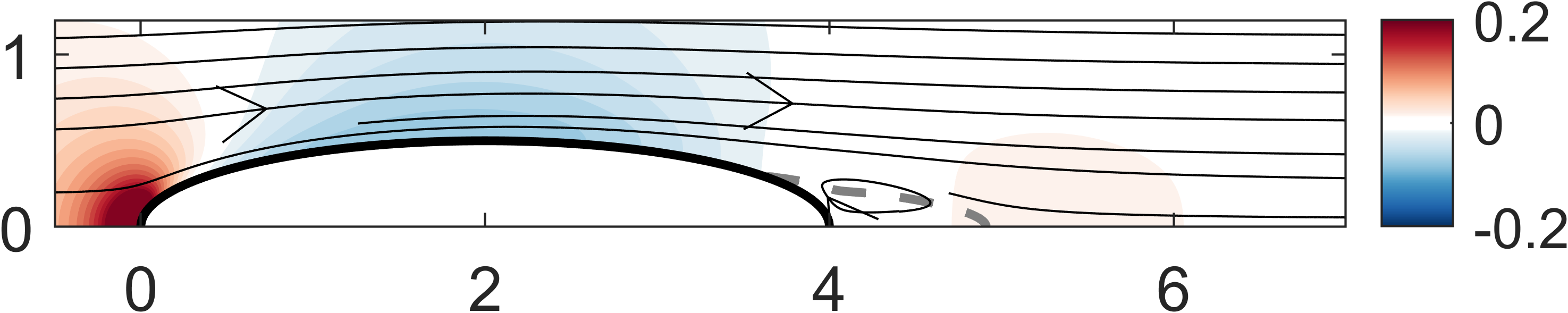} 
       \put(-8,19){$(b)$} 
       \put(-5,10){$r$}
 	\end{overpic}
    \begin{overpic}[width=0.49\textwidth, trim=0mm 0mm 0mm 0mm, clip=false]{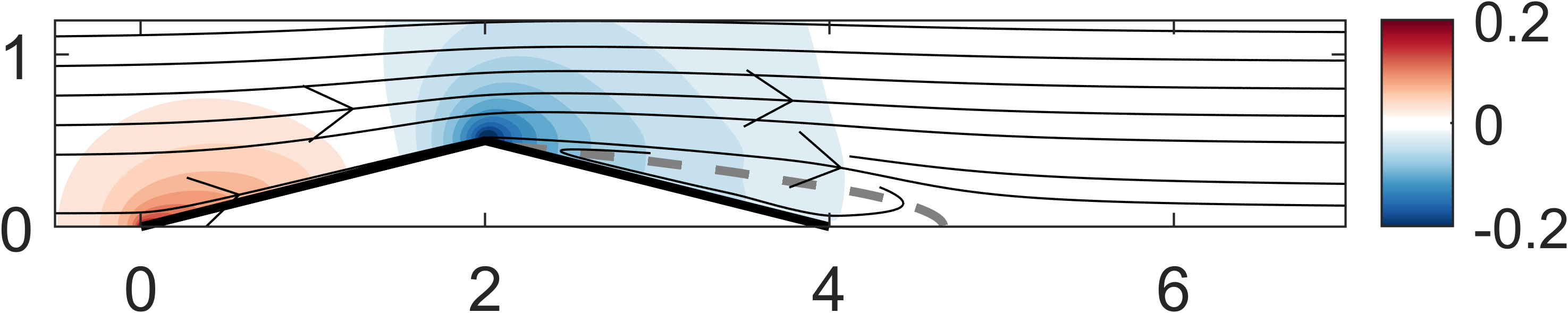}  
 	\end{overpic}
}
\centerline{ 
    \begin{overpic}[width=0.49\textwidth, trim=0mm 0mm 0mm 0mm, clip=false]{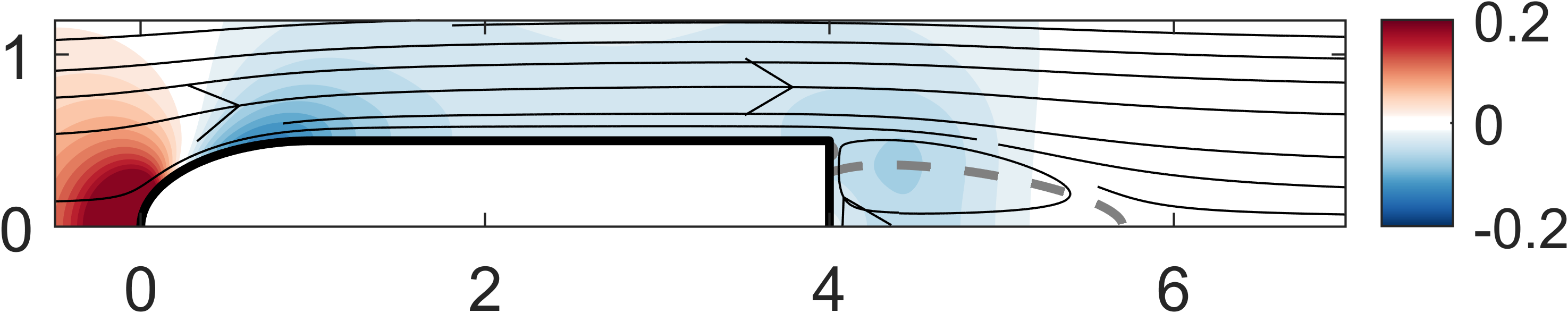} 
       \put(-5,10){$r$}
       \put(47,-4){$z$} 
 	\end{overpic}
    \begin{overpic}[width=0.49\textwidth, trim=0mm 0mm 0mm 0mm, clip=false]{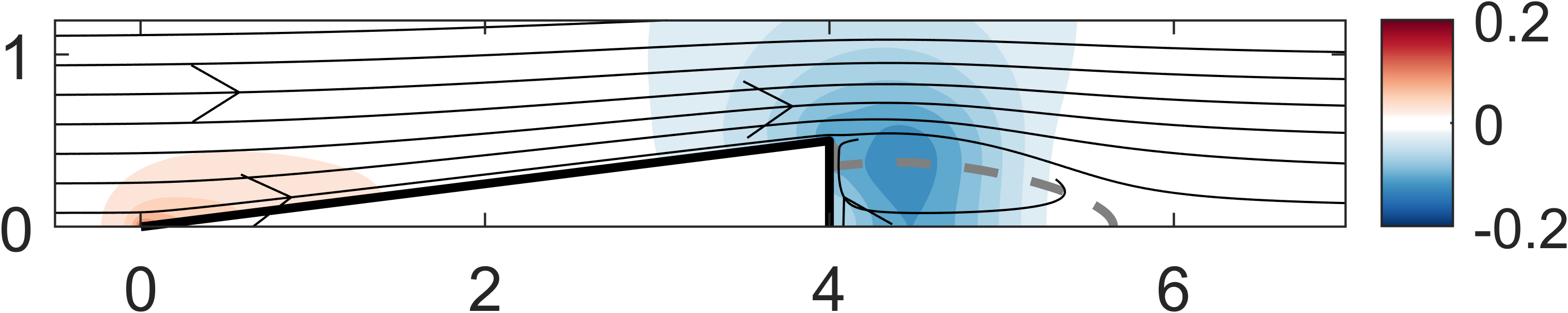} 
       \put(47,-4){$z$} 
 	\end{overpic}
}
\vspace{0.2cm}
\caption{
Same as figure~\ref{fig:BF-vort-press-AR1} for $\AR=4$.
Ellipsoid, $Re=1000$;
bicone, $Re=775$;
bullet, $Re=390$;
cone, $Re=260$. 
}
\label{fig:BF-vort-press-AR4}
\end{figure}

The axisymmetric low-$Re$ base flow for the four considered geometries is shown in figures \ref{fig:BF-vort-press-AR1} and \ref{fig:BF-vort-press-AR4} at $Re = Re_c$, corresponding to the first onset of the symmetry-breaking instability, for $\AR=1$ and $\AR=4$. Only a brief characterisation is provided here, as the effect of the different placement of the corners resembles what was found by \cite{chiarini-quadrio-auteri-2022b} for 2D symmetric bodies. 

A shear layer with negative azimuthal vorticity $\omega_{0\theta}$ separates from the rear part of the body and delimits the axisymmetric wake recirculation region. %; see figure \ref{fig:BF-vort}.
 The vorticity is maximum in the vicinity of the body surface: for cones and bicones $\omega_{0\theta}$ is maximum close to the corners, while for ellipsoids and bullets the region with intense vorticity is more spread. The pressure distribution changes accordingly (see the bottom right panels in figures \ref{fig:BF-vort-press-AR1} and \ref{fig:BF-vort-press-AR4}). For bodies with a blunt TE (cones and bullets), the minimum of the pressure is placed at the TE corners and the flow streamlines along the lateral side of the body face a favourable pressure gradient. For bodies with zero-thickness TE (ellipsoids and bicones), instead, the pressure is minimum %in correspondence of the maximum 
 where the cross-stream size of the body is maximum. 
 In this case, the flow streamlines along the lateral side of the body face first a favourable pressure gradient and then an adverse one that promotes the flow separation. Notably, the adverse pressure gradient becomes milder as $\AR$ increases. For cones and bullets the flow separation point is set by the geometry at the TE corner. The cross-stream dimension of the wake recirculation region is thus determined by the body width and its extent only slightly decreases with $\AR$. On the contrary, for ellipsoids and bicones the flow separation is driven by the pressure distribution (see figures \ref{fig:BF-vort-press-AR1} and \ref{fig:BF-vort-press-AR4}). In agreement with the milder pressure gradient,  the  separation point moves downstream for larger $\AR$, yielding a thinner and shorter wake recirculation region (see also figure~\ref{fig:BF_quantities} in Appendix \ref{sec:appendix}).

\subsection{Neutral curves}

\begin{figure}
\centerline{   
    \begin{overpic}[width=10cm, trim=30mm 126mm 30mm 95mm, clip=true]{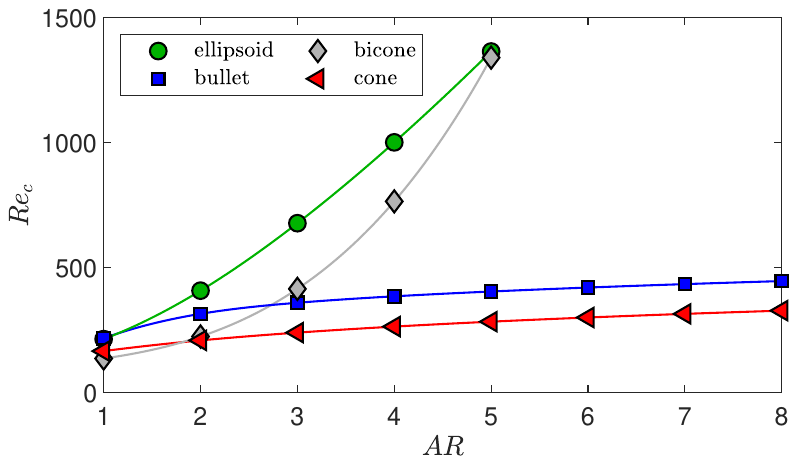}
    \put(50,-3){$\AR$}
 	\end{overpic}
}
\vspace{0.2cm}
\caption{
Critical Reynolds number as a function of the aspect ratio. 
}
\label{fig:Rec}
\end{figure}

\begin{table}
\centering
\begin{tabular}{lccccccccc}
Body                       & Ellipsoid &&   \multicolumn{6}{c}{Bullet}     \\
$\AR$                       & 1         && 1   & 2   & 3   & 4   & 5   & 6  \\ 
\\
\cite{MELIGA2009601}       & 213       && -   & -   & -   & -   & -   & -  \\
\cite{bohorquez-etal-2011} & -         && 216 & 327 & 372 & 399 & 420 & 435\\
Present study              & 213       && 216 & 314 & 358 & 384 & 403 & 419\\
\end{tabular}
\caption{Comparison of the critical Reynolds number $Re_c$ with results from the literature,  for some geometries.
}
\label{tab:Rec}
\end{table}

We now move to the results of the linear stability analysis. Figure \ref{fig:Rec} shows the neutral curves for the primary instability, that, for all cases, consists in the regular ($\lambda_i=0$) symmetry-breaking bifurcation of an eigenmode of azimuthal wavenumber $m=1$. Table \ref{tab:Rec} compares the results of our computations with those of \cite{MELIGA2009601} and \cite{bohorquez-etal-2011} for validation purposes. We find a very good agreement with the results of \cite{MELIGA2009601}, while some discrepancies are observed when comparing with \cite{bohorquez-etal-2011} (we measure a relative difference in the value of $Re_c$ of approximately $ 4\%$ for $\AR=6$). We conjecture that this discrepancy is due to the different numerical method and to the different size of the computational domain.

Looking at figure \ref{fig:Rec}, a first observation is that, although the primary bifurcation is the same for the different bodies, the value of $Re_c$ shows large variability, with $Re_c \approx 150$ for the $\AR=1$ bicone   and $Re_c \approx 1400$ for the $\AR=5$ ellipsoid. This highlights that $U_\infty$ and $H$ are not the appropriate velocity and length scales for the characterisation and prediction of this bifurcation.
Similarly to the flow past 2D symmetric cylinders \citep{chiarini-quadrio-auteri-2022b}, an increase in the aspect ratio leads to a stabilisation of the base flow regardless of the body geometry; see in figure \ref{fig:Rec} that $Re_c$ monotonously increases with $\AR$. This effect holds also for non-axisymmetric bodies, as  shown for example by \cite{zampogna-boujo-2023} and \cite{chiarini-boujo-2024} in the context of 3D rectangular prisms. Notably, the way $Re_c$ varies with $\AR$ depends on the shape of the trailing edge of the body. For bodies with a blunt TE (cones and bullets), the critical Reynolds number almost flattens as $\AR$ increases. For bodies with zero-thickness TE (ellipsoids and bicones), instead, the critical Reynolds number increases faster than linearly, and very large values of $Re_c$ are observed already at intermediate $\AR$. An explanation for the steep increase of $Re_c$ with $\AR$ for bodies with zero-thickness TE is provided in \S\ref{sec:vorticity}.

\def\field{uz}
\begin{figure}
\vspace{0.5cm}
\centerline{ 
    \begin{overpic}[width=0.445\textwidth, trim=0mm 0mm 10mm 0mm, clip=true]{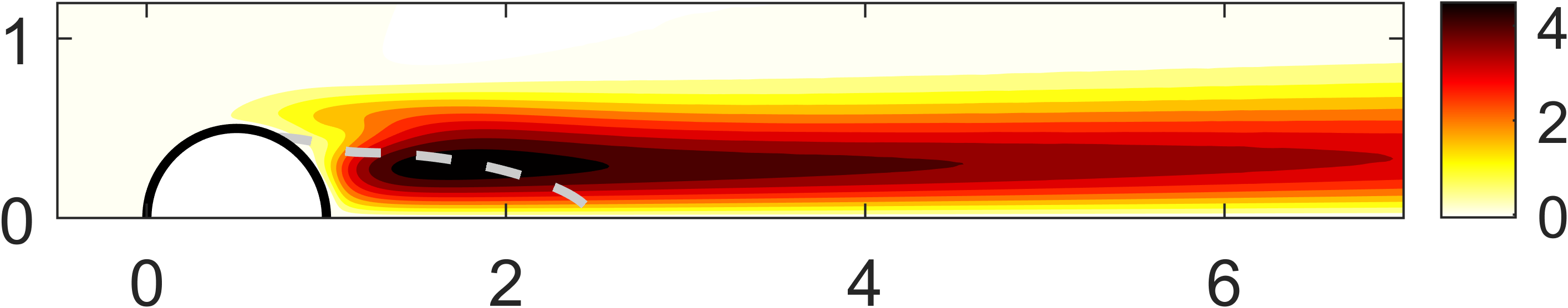}  
       \put(-5,10){$r$}
 	\end{overpic}
    \hspace{0.4cm}
    \begin{overpic}[width=0.445\textwidth, trim=0mm 0mm 10mm 0mm, clip=true]{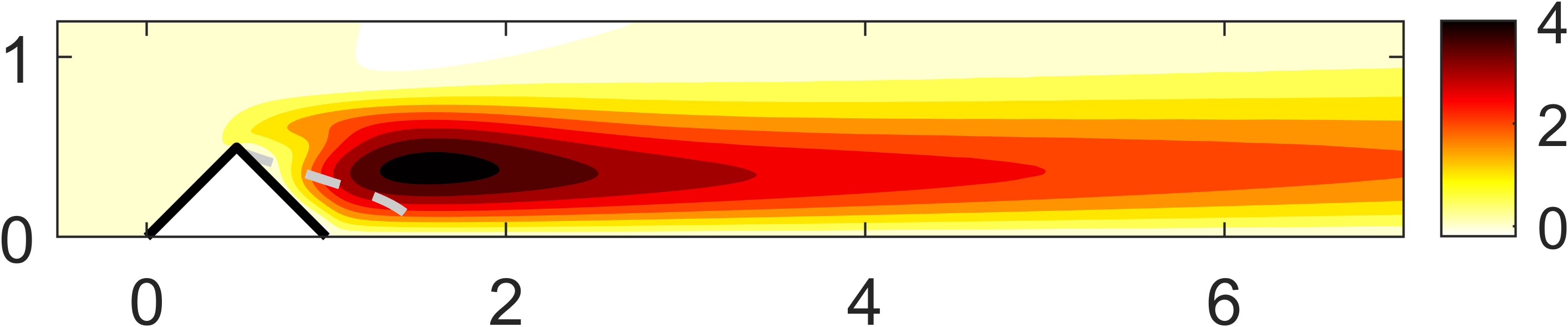}  
       \put(47,-4){$z$}
 	\end{overpic}
}
\centerline{ 
    \begin{overpic}[width=0.445\textwidth, trim=0mm 0mm 10mm 0mm, clip=true]{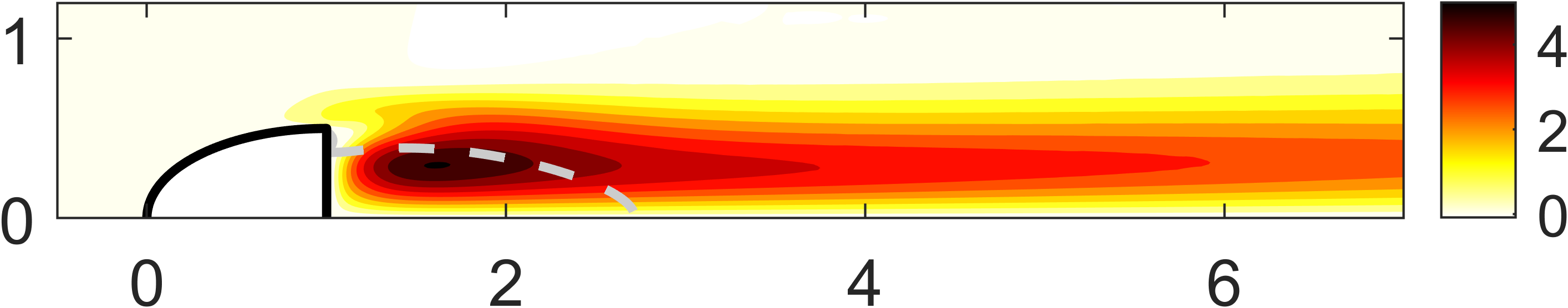}  
       \put(-5,10){$r$}
       \put(47,-4){$z$}
 	\end{overpic}
    \hspace{0.4cm}
    \begin{overpic}[width=0.445\textwidth, trim=0mm 0mm 10mm 0mm, clip=true]{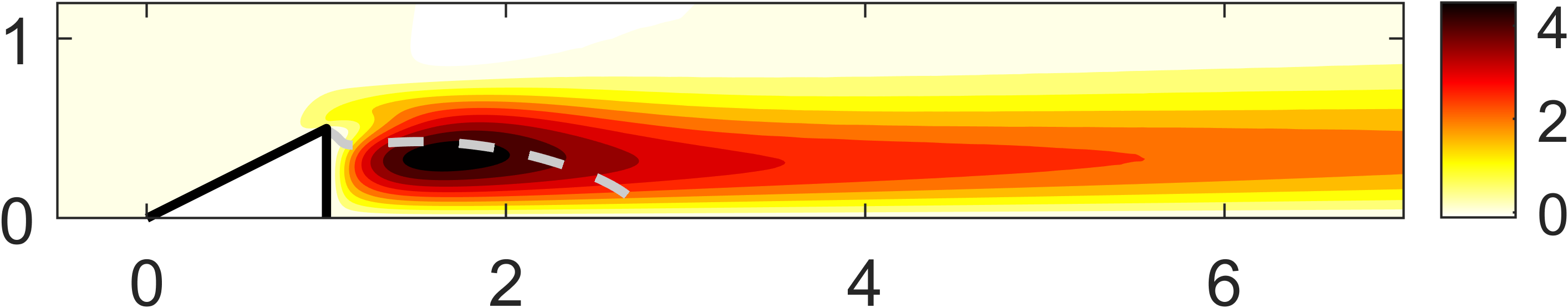}  
       \put(47,-4){$z$}
 	\end{overpic}
}
\vspace{0.2cm}
\caption{
Eigenmode for $\AR=1$ at $Re \simeq Re_c$: streamwise velocity.
Ellipsoid, $Re=210$; 
bicone, $Re=135$;
bullet, $Re=220$;
cone, $Re=160$. 
Dashed line: $u_{0z}=0$.
}
\label{fig:modes-uz-AR1}
\end{figure}
\begin{figure}
\centerline{ 
    \begin{overpic}[width=0.445\textwidth, trim=0mm 0mm 10mm 0mm, clip=true]{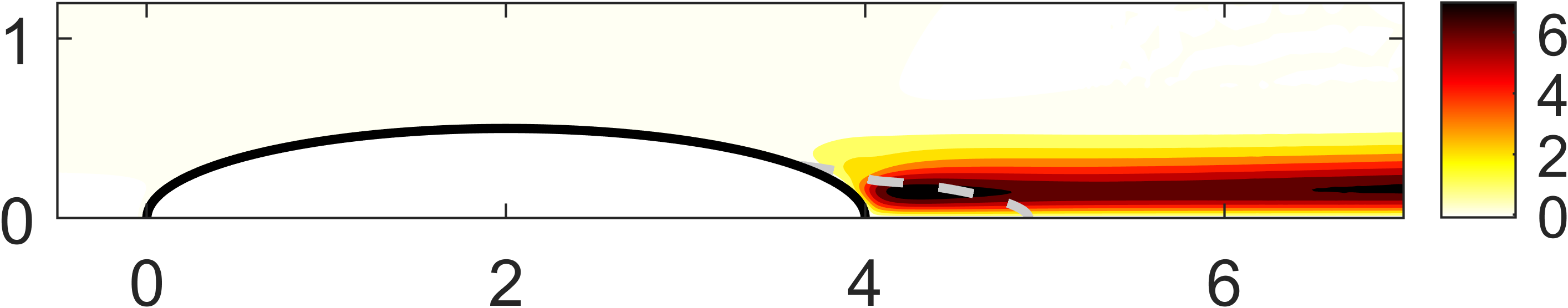}  
       \put(-5,10){$r$}
 	\end{overpic}
    \hspace{0.4cm}
    \begin{overpic}[width=0.445\textwidth, trim=0mm 0mm 10mm 0mm, clip=true]{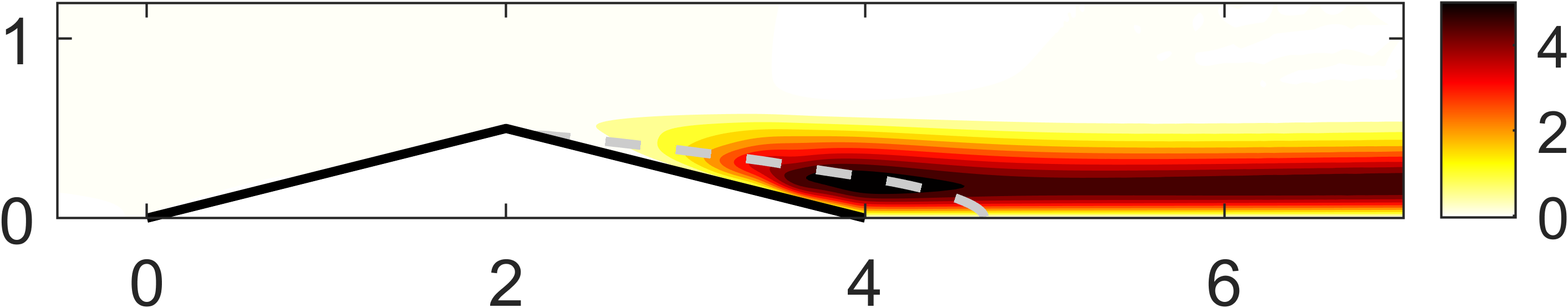}  
 	\end{overpic}
}
\centerline{ 
    \begin{overpic}[width=0.445\textwidth, trim=0mm 0mm 10mm 0mm, clip=true]{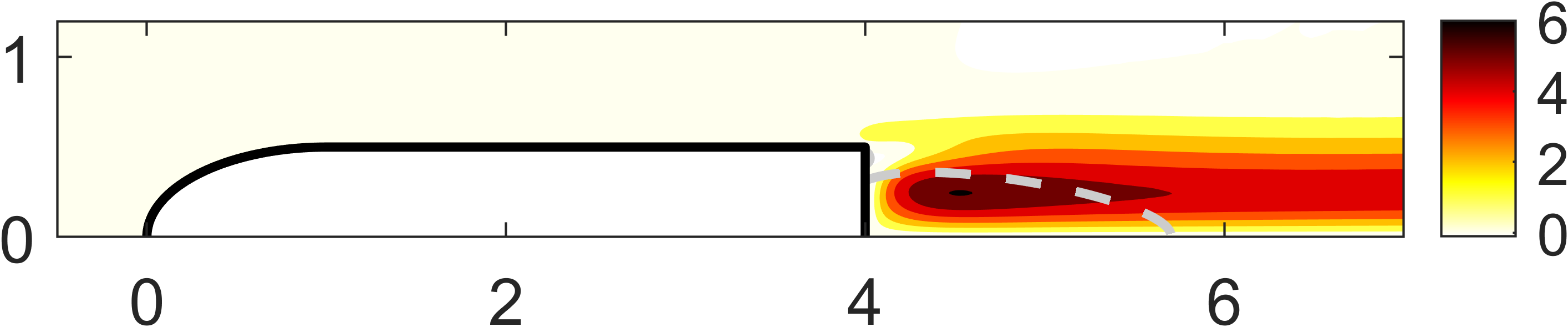}  
       \put(-5,10){$r$}
       \put(47,-4){$z$}
 	\end{overpic}
    \hspace{0.4cm}
    \begin{overpic}[width=0.445\textwidth, trim=0mm 0mm 10mm 0mm, clip=true]{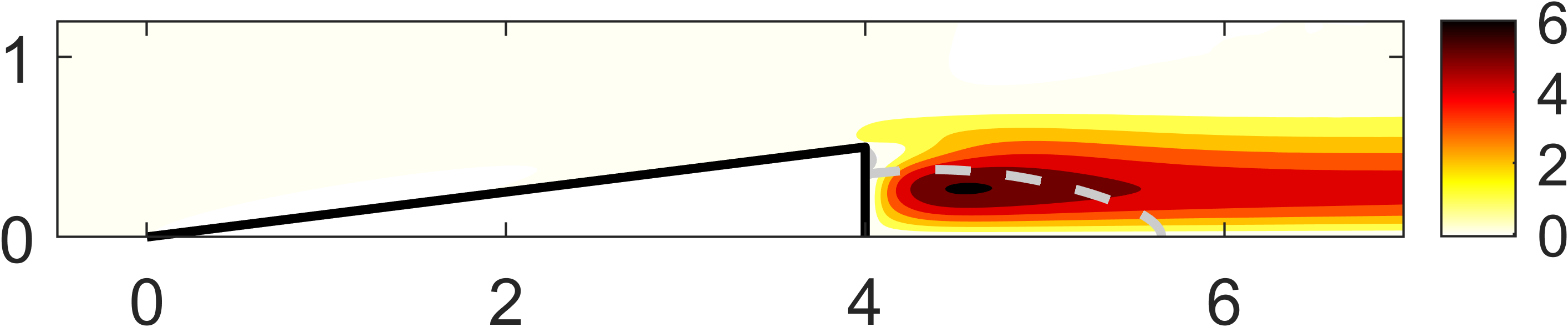}  
       \put(47,-4){$z$}
 	\end{overpic}
}
\vspace{0.2cm}
\caption{
Same as figure~\ref{fig:modes-uz-AR1} for $\AR=4$.
Ellipsoid, $Re=1000$; 
bicone, $Re=775$;
bullet, $Re=390$;
cone, $Re=260$.
}
\label{fig:modes-uz-AR4}
\end{figure}
For completeness, we show  in figures \ref{fig:modes-uz-AR1} and \ref{fig:modes-uz-AR4} the axial velocity  of the unstable mode for different geometries. Qualitatively, the structure of the mode does not change among the considered geometries, and agrees with the results of other authors; see for example figure 8 of \cite{natarajan-acrivos-1993} and figure 4 of \cite{meliga-etal-2009}. The perturbation field is confined in the wake, with the largest value found close to the TE of the body within the wake recirculation region. 
This is consistent with the appearance of a pair of streamwise counter-rotating vortices in the wake (see discussion in \S\ref{sec:introduction}), and with the instability mechanism discussed in \S\ref{sec:vorticity}. For bodies with zero-thickness TE, the region where the mode is intense extends farther downstream.

\subsection{Sensitivities}

\begin{figure}
\vspace{0.5cm}
\centerline{ 
    \begin{overpic}[width=0.49\textwidth, trim=0mm 0mm 0mm 0mm, clip=false]{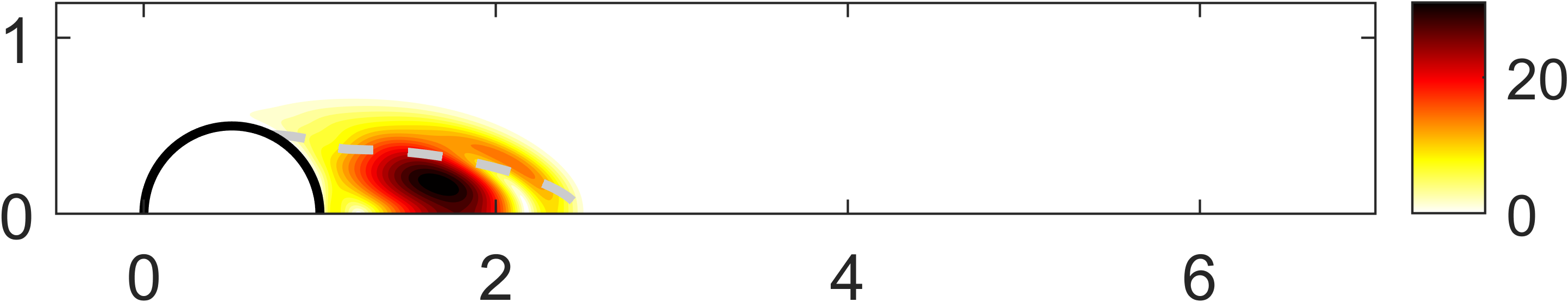}  
       \put(-5,10){$r$}
       \put(47,-4){$z$}
 	\end{overpic}
    \begin{overpic}[width=0.49\textwidth, trim=0mm 0mm 0mm 0mm, clip=false]{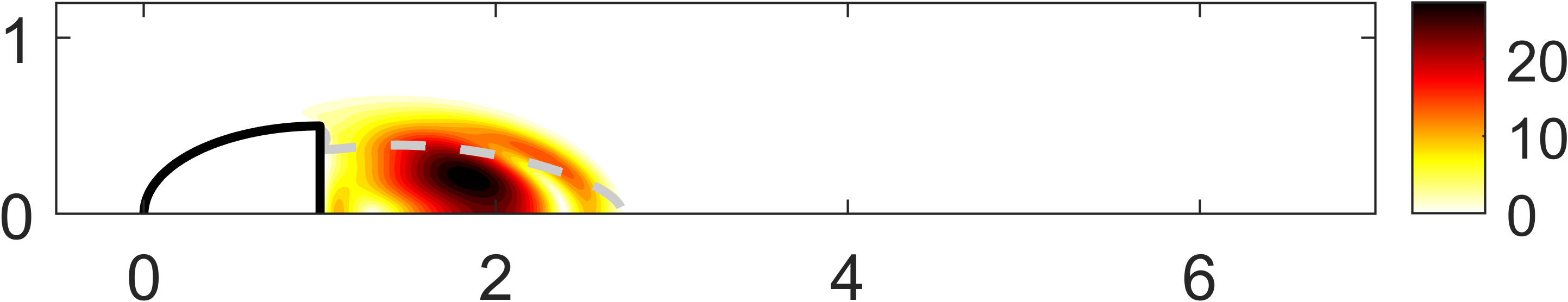}  
       \put(47,-4){$z$}
 	\end{overpic}
}
\centerline{ 
    \begin{overpic}[width=0.49\textwidth, trim=0mm 0mm 0mm 0mm, clip=false]{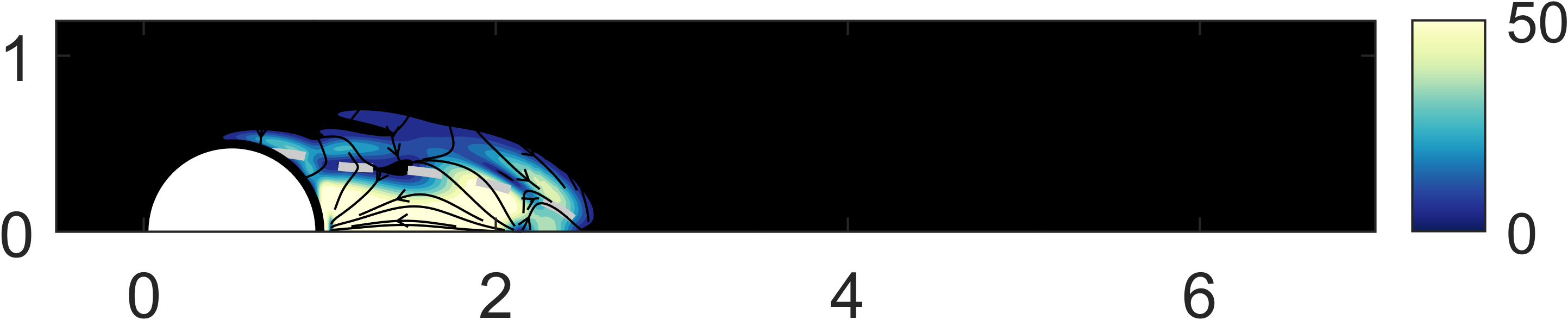}  
       \put(-5,10){$r$}
       \put(47,-4){$z$}
 	\end{overpic}
    \begin{overpic}[width=0.49\textwidth, trim=0mm 0mm 0mm 0mm, clip=false]{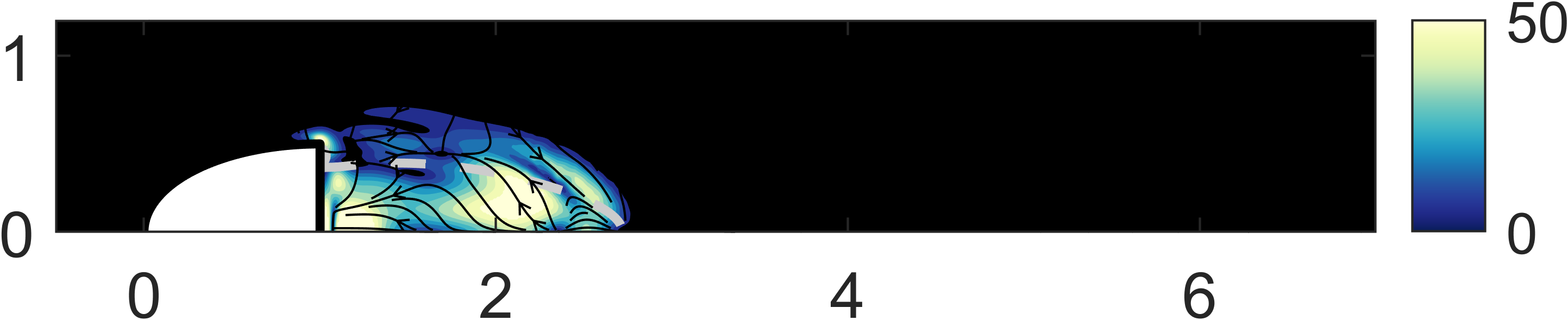}              
       \put(47,-4){$z$}
 	\end{overpic}
}
\vspace{0.2cm}
\caption{Structural sensitivity $S$ (top) and sensitivity of the growth rate to base-flow modifications $\bm{\nabla}_{\bm{u}_0} \lambda_r$ (bottom) for $\AR=1$ at $Re \simeq Re_c$. Left: Ellipsoid, $Re=210$. Right: bullet, $Re=220$. Dashed line $u_{0z} = 0$.} 
\label{fig:SS_BF_sens_AR1}
\end{figure}

\begin{figure}
%\vspace{0.5cm}
\centerline{ 
    \begin{overpic}[width=0.49\textwidth, trim=0mm 0mm 0mm 0mm, clip=false]{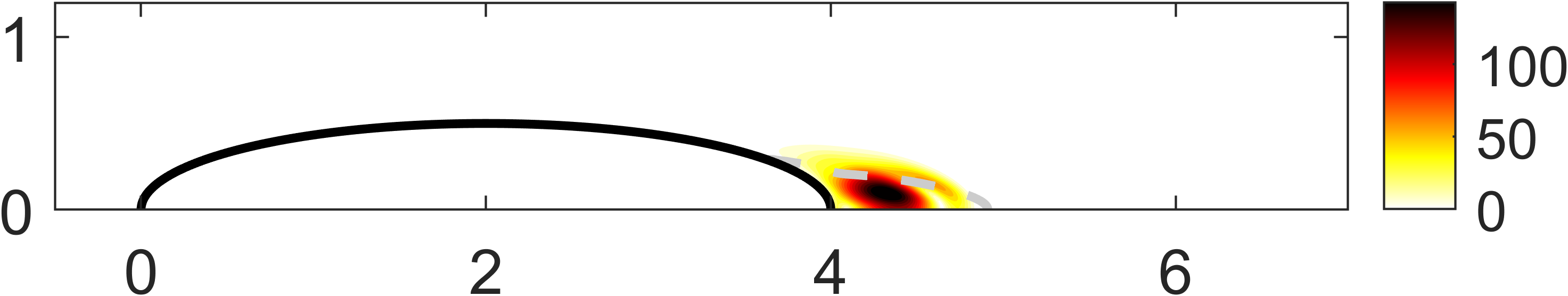}  
       \put(-5,10){$r$}
       \put(47,-4){$z$}
 	\end{overpic}
    \begin{overpic}[width=0.49\textwidth, trim=0mm 0mm 0mm 0mm, clip=false]{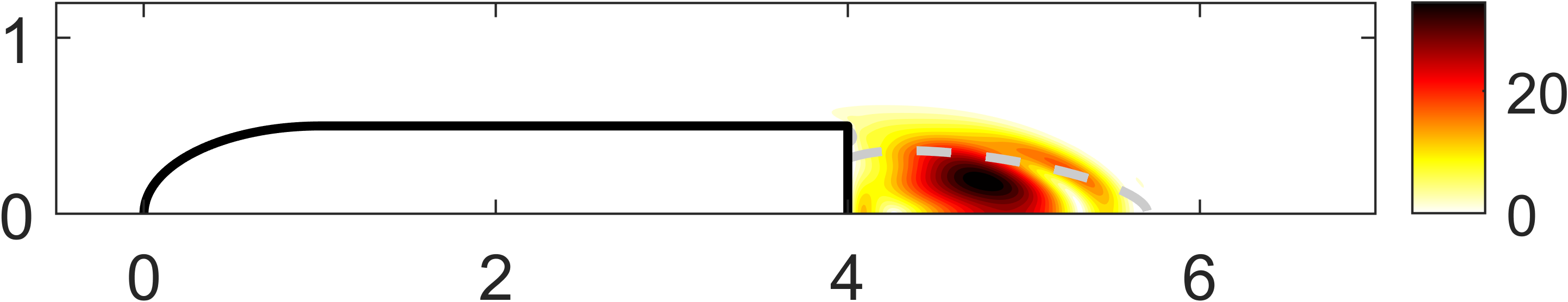}  
       \put(47,-4){$z$}
 	\end{overpic}
}
\centerline{ 
    \begin{overpic}[width=0.49\textwidth, trim=0mm 0mm 0mm 0mm, clip=false]{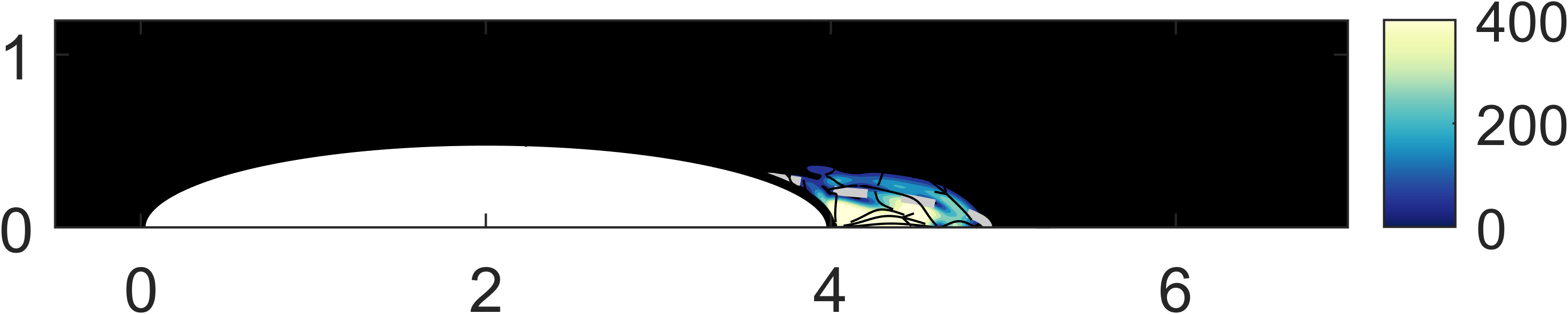}
       \put(-5,10){$r$}
       \put(47,-4){$z$}
 	\end{overpic}
    \begin{overpic}[width=0.49\textwidth, trim=0mm 0mm 0mm 0mm, clip=false]{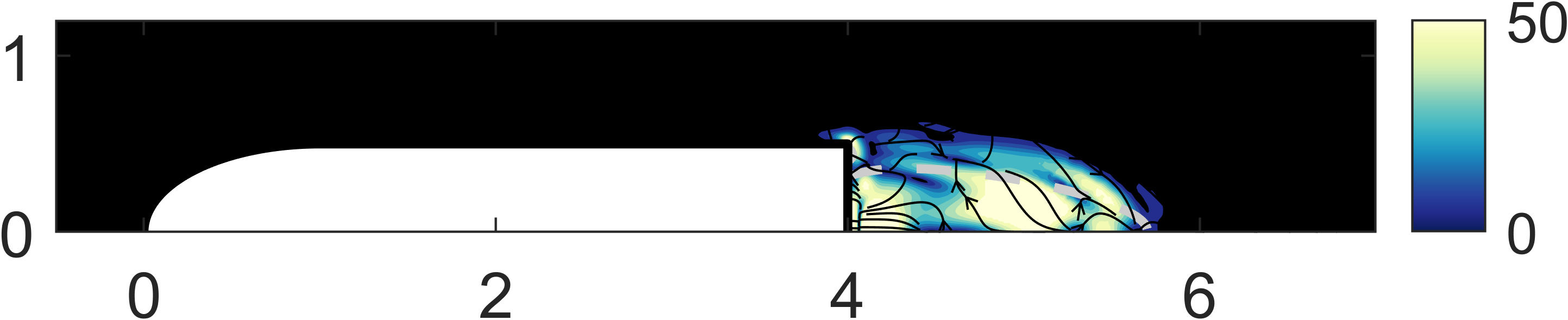}              
       \put(47,-4){$z$}
 	\end{overpic}
}
\vspace{0.2cm}
\caption{As figure \ref{fig:SS_BF_sens_AR1}, but for $\AR=4$. Ellipsoid, $Re=1000$; bullet, $Re=390$.}
\label{fig:SS_BF_sens_AR4}
\end{figure}

To further characterise the instability we consider in figures \ref{fig:SS_BF_sens_AR1} and \ref{fig:SS_BF_sens_AR4} the structural sensitivity \citep{giannetti-luchini-2007} and the sensitivity to base-flow modifications \citep{marquet.etal-2008-Sensitivityanalysis, meliga_open-loop_2010}. 
For brevity, here we only show ellipsoids and bullets, being representative of bodies with zero-thickness trailing edge and with a blunt base. 

The structural sensitivity $S(\bm{x})$ is based on the interplay between the direct and adjoint modes, and identifies the region where a generic structural modification of the stability problem provides the largest drift of the leading eigenvalue. It is an upper bound for the eigenvalue variation $|\delta \lambda|$ induced by a local body force actuation proportional to the signal of a velocity sensor located at the exact same station. The region of the flow where $S$ is large identifies the wavemaker \citep{monkewitz-huerre-chomaz-1993}. For the primary instability the structural sensitivity is defined as
\begin{equation}
  S(\bm{x}) = \frac{ || \hat{\bm{u}}_1^\dagger || || \hat{\bm{u}}_1 || }{ \langle \hat{\bm{u}}_1^\dagger, \hat{\bm{u}}_1 \rangle },
\end{equation}
where $|| \cdot ||$ represents the $R^2$ vector norm, $\hat{\bm{u}}_1^\dagger$ is the adjoint mode and $ \langle \bm{u}_A,\bm{u}_B \rangle = \int_D \bm{u}_A^* \cdot \bm{u}_B \text{d}\Omega$ is the inner product of $L^2(D)$ with the superscript $^*$ denoting the complex conjugate.
For all cases $S(\bm{x})$ is substantially close to zero everywhere except close to the body where the product between the direct and the adjoint mode is large. Large values are observed along the separation line $u_{0z}=0$ and within the recirculating region, where the maximum is observed \citep{MELIGA2009601}. The similar distribution of $S(\bm{x})$ suggests that the different geometries and $\AR$s considered share the same wavemaker and the same instability mechanism, which are spatially located within the recirculating region. Note that %, at $Re=Re_c$, 
for bodies with zero-thickness trailing edge the spatial extent of the wavemaker decreases as $\AR$ increases, in agreement with the smaller recirculating region and with the base-flow stabilisation.

Next, the sensitivity to base-flow modifications quantifies the variation of the complex eigenpair $(\lambda, \hat{\bm{u}}_1, \hat{p}_1)$ induced by a small variation of the base flow $\delta \bm{u}_0$. Specifically, the variation of the eigenvalue $\delta \lambda$ is linked with $\delta \bm{u}_0$ by the inner product $\delta \lambda = \langle \bm{\nabla}_{\bm{u}_0} \lambda, \delta \bm{u}_0 \rangle$, where
\begin{equation}
  \bm{\nabla}_{\bm{u}_0} \lambda = \frac{ - ( \bm{\nabla} \hat{\bm{u}}_1 )^H \cdot \hat{\bm{u}}_1^\dagger + \bm{\nabla} \hat{\bm{u}}_1^\dagger \cdot \hat{\bm{u}}^* }
                                        { \langle \hat{\bm{u}}_1^\dagger, \hat{\bm{u}}_1 \rangle }
\end{equation}
is indeed the sensitivity of $\lambda$ to base-flow modifications; here the superscript $^H$ indicates the transconjugate. 
The variation of the growth rate induced by $\delta \bm{u}_0$ is thus expressed as $\delta \lambda_r = ( \bm{\nabla}_{\bm{u}_0} \lambda_r, \delta \bm{u}_0)$ where the corresponding sensitivity is $\bm{\nabla}_{\bm{u}_0} \lambda_r = \Re( \bm{\nabla}_{\bm{u}_0} \lambda)$. 
Unlike the structural sensitivity  $S$, which is a scalar field, the sensitivity  $\bm{\nabla}_{\bm{u}_0} \lambda_r$ to base-flow  modification is a vector field: the field lines provide the local orientation of the sensitivity field, while the magnitude provides the intensity. As expected, far from the bodies $\bm{\nabla}_{\bm{u}_0} \lambda_r$ decays to zero due to the spatial separation of the direct and adjoint modes. Large values are instead observed close to the $u_{0z} = 0$ line and within the recirculating region. For bodies with zero-thickness trailing edge, $\bm{\nabla}_{\bm{u}_0} \lambda_r$ is large within the entire recirculating region, while for bodies with a blunt base the sensitivity is maximum at the downstream end of the recirculating region, close to the base and in correspondence of the corners \citep[this resembles what observed for 2D rectangular cylinders; see][]{chiarini-quadrio-auteri-2021a}. For all cases, an increase of the backflow within the recirculating region ($ \delta u_{0z}<0$) largely destabilises the flow. Again, for bodies with zero-thickness trailing edge the spatial extent where the sensitivity is large decreases with $\AR$.

\subsection{Effect of the Reynolds number}
\label{sec:Reeffect}

\begin{figure}
\vspace{0.5cm}
\centerline{ 
    \begin{overpic}[width=0.49\textwidth, trim=0mm 0mm 0mm 0mm, clip=false]{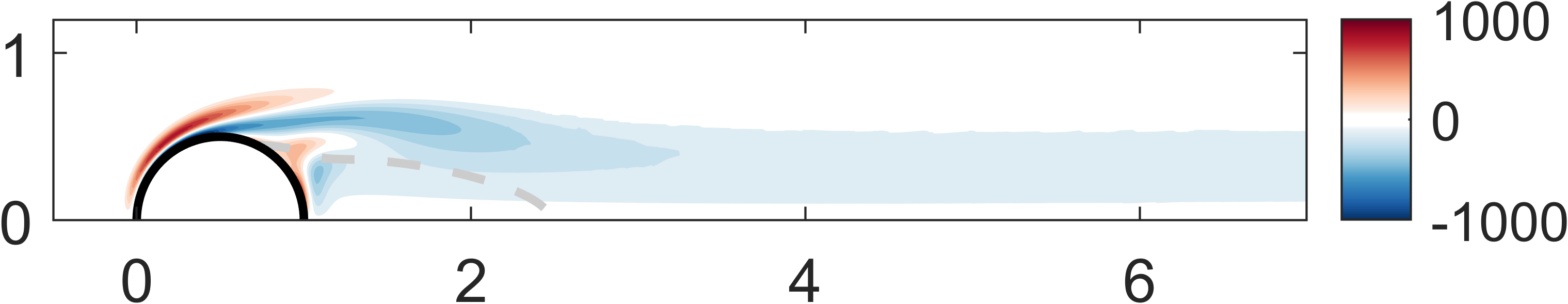}  
       \put(-5,10){$r$}
       \put(47,-4){$z$}
 	\end{overpic}
    \begin{overpic}[width=0.49\textwidth, trim=0mm 0mm 0mm 0mm, clip=false]{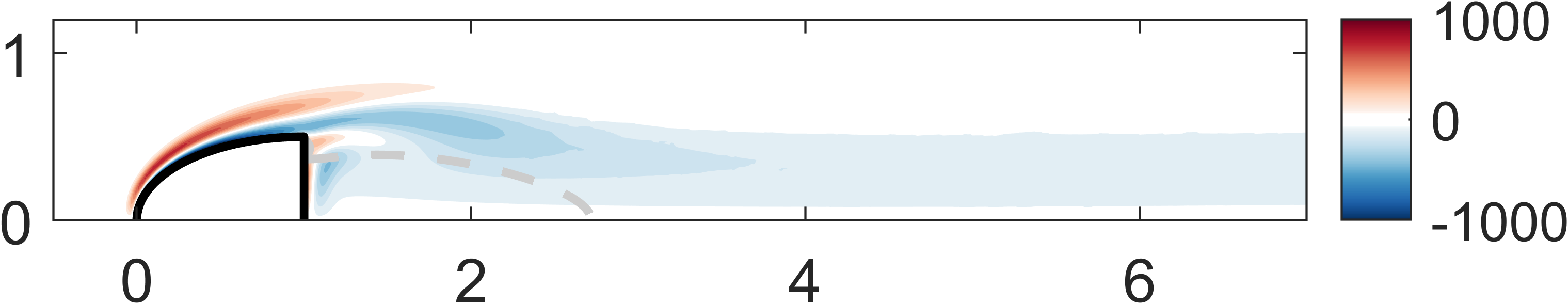}              
       \put(47,-4){$z$}
 	\end{overpic}
}
\centerline{ 
    \begin{overpic}[width=0.49\textwidth, trim=0mm 0mm 0mm 0mm, clip=false]{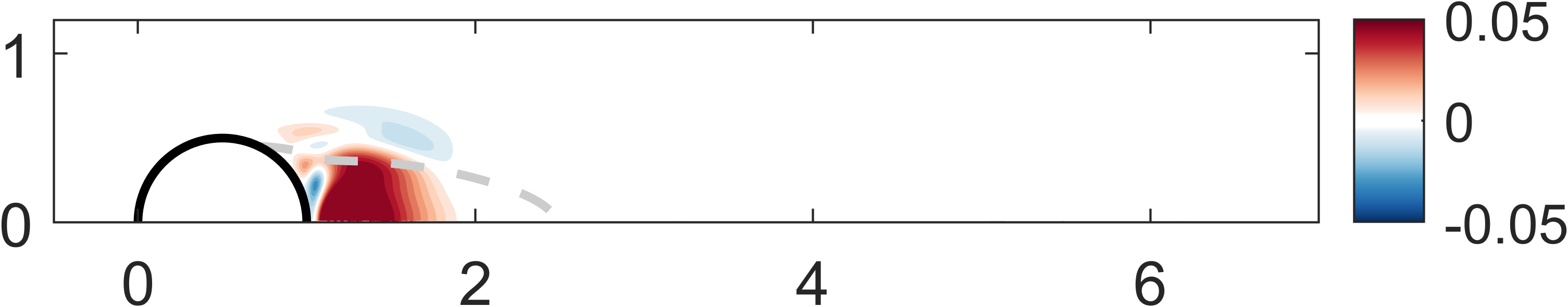}  
       \put(-5,10){$r$}
       \put(47,-4){$z$}
 	\end{overpic}
    \begin{overpic}[width=0.49\textwidth, trim=0mm 0mm 0mm 0mm, clip=false]{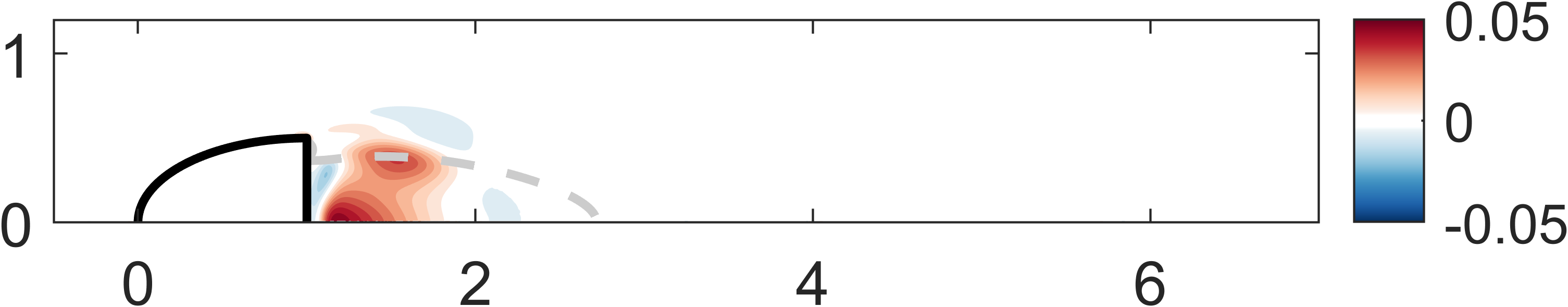}              
       \put(47,-4){$z$}
 	\end{overpic}
}
\vspace{0.2cm}
\caption{
Effect of a small increase in $Re$ on the base flow and on the growth rate for $\AR=1$ at $Re \simeq Re_c$. Top: spatial distribution of $\omega_{\theta,0}^{\varepsilon}$. Bottom: spatial distribution of the integrand of $\delta \lambda_r$ (see text and equation \ref{eq:delta_lambdar}). Left: Ellipsoid, $Re=210$. Right: bullet: $Re=220$. Dashed line $u_{0z} = 0$.}
\label{fig:Re_effect}
\end{figure}

We now investigate the effect of a small variation of $Re$ on the growth rate $\lambda_r$. We start by looking for an expression for $\partial \lambda_r/\partial Re$. We consider a small departure from criticality such that
\begin{equation}
  \frac{1}{Re_c} - \frac{1}{Re} = \varepsilon
\end{equation}
where $|\varepsilon| \ll 1$ ($\varepsilon<0$ for $Re<Re_c$ and $\varepsilon>0$ for $Re>Re_c$). A small-amplitude change in $Re$ induces small changes in the base flow, in the eigenmode and in the eigenvalues, i.e.
\begin{equation}
  \{     \bm{u}_0,      p_0 \} \rightarrow \{     \bm{u}_0,      p_0 \} + \{\delta      \bm{u}_0,  \delta p_0       \}, \
  \{\hat{\bm{u}}_1,\hat{p}_1\} \rightarrow \{\hat{\bm{u}}_1,\hat{p}_1\} + \{\delta \hat{\bm{u}}_1, \delta \hat{p}_1 \}, \
  \lambda                      \rightarrow \lambda + \delta \lambda.
\end{equation}
We now set $Re=Re_c$ as reference, and inject these changes into the steady nonlinear NS equations and into the LNSE. By keeping the first-order terms only we then obtain the equations for the base-flow modification $\{\delta \bm{u}_0,\delta p_0\}$ and for the eigenmode modification $\{\delta \hat{\bm{u}}_1, \delta \hat{p}_1\}$, i.e.
\begin{equation}
  \begin{pmatrix}
    \mathcal{L}_0\{\bm{u}_0,Re\}(\cdot) & - \bm{\nabla}_0(\cdot) \\
    \bm{\nabla}_0 \cdot (\cdot)         &            0                         
  \end{pmatrix}
  \begin{pmatrix}
  \delta \bm{u}_0 \\
  \delta     p_0
  \end{pmatrix} = 
  \begin{pmatrix}
  - \varepsilon \bm{\nabla}_0^2 \bm{u}_0 \\
   0 
  \end{pmatrix}
  \label{eq:BF_mod}
\end{equation} 
and
\begin{equation}
  \lambda
  \begin{pmatrix}
  \mathcal{I} & \bm{0} \\
  \bm{0}^T    &  0
  \end{pmatrix}
  \begin{pmatrix}
    \delta \hat{\bm{u}}_1 \\
    \delta \hat{p}_1
  \end{pmatrix} +
  \begin{pmatrix}
    \mathcal{L}_m\{\bm{u}_0,Re\}(\cdot) & - \bm{\nabla}_m(\cdot) \\
    \bm{\nabla}_m \cdot ( \cdot )       &      0
  \end{pmatrix}
  \begin{pmatrix}
  \delta \hat{\bm{u}}_1 \\
  \delta \hat{p}_1
  \end{pmatrix} 
  = 
  \begin{pmatrix}
  \mathcal{F}_m(\delta \bm{u}_0, \hat{\bm{u}}_1) \\
  0
  \end{pmatrix}
  \label{eq:EM_mod}
\end{equation}
accompanied by 
homogeneous Dirichlet boundary conditions at the inlet and at the surface of the body and stress-free conditions at the outlet and at the farfield; here 
\begin{equation*}
\mathcal{F}_m(\delta \bm{u}_0,\hat{\bm{u}}_1) = - \mathcal{C}_m( \hat{\bm{u}}_1, \delta \bm{u}_0 ) -  \varepsilon \bm{\nabla}^2_m \hat{\bm{u}}_1 - \delta \lambda \hat{\bm{u}}_1.
%-  \delta \bm{u}_0 \cdot \bm{\nabla}_m \hat{\bm{u}}_1 - \hat{\bm{u}}_1 \cdot \bm{\nabla}_m \delta \bm{u}_0 - \varepsilon \bm{\nabla}^2_m \hat{\bm{u}}_1 - \delta \lambda \hat{\bm{u}}_1.
\end{equation*}
Equation \eqref{eq:BF_mod} is divided by $\varepsilon$ and solved for $\bm{u}_0^\varepsilon = \delta \bm{u}_0/\varepsilon$. Equation \eqref{eq:EM_mod} is projected on the adjoint mode $\{\hat{\bm{u}}_1^\dagger,\hat{p}_1^\dagger\}$ to eliminate $\delta \hat{\bm{u}}_1$ and obtain an expression for $\delta \lambda$, i.e.
\begin{equation}
  \delta \lambda = -\varepsilon \frac{ \langle \hat{\bm{u}}_1^\dagger, \mathcal{C}_m( \hat{\bm{u}}_1, \bm{u}_0^\varepsilon ) + \bm{\nabla}^2_m \hat{\bm{u}}_1 \rangle }
                                     { \langle \hat{\bm{u}}_1^\dagger, \hat{\bm{u}}_1 \rangle }.
\end{equation}
Interestingly, this expression matches exactly the one for the linear coefficient of the amplitude equation obtained with a standard weakly nonlinear stability analysis \citep{sipp-lebedev-2007,zampogna-boujo-2023}. At this point we can write the $\partial \lambda_r/\partial Re$ derivative as
\begin{equation}
  \frac{\partial \lambda_r}{\partial Re} =
  \frac{\partial \lambda_r}{\partial \varepsilon}
  \frac{\partial \varepsilon}{\partial Re} = 
  \frac{1}{Re^2}
  \frac{\partial \lambda_r}{\partial \varepsilon} =
  - \frac{1}{Re^2} \Re \left(  \frac{ \langle \hat{\bm{u}}_1^\dagger,       \mathcal{C}_m( \hat{\bm{u}}_1, \bm{u}_0^\varepsilon ) + \bm{\nabla}^2_m \hat{\bm{u}}_1 \rangle }
                                     { \langle \hat{\bm{u}}_1^\dagger, \hat{\bm{u}}_1 \rangle } \right).
\label{eq:delta_lambdar}
\end{equation}
A small change in the Reynolds number thus leads to a modification of the growth rate that depends on the effect of the base-flow modification on the eigenmode, i.e. $\langle \hat{\bm{u}}_1^\dagger, \mathcal{C}_m( \hat{\bm{u}}_1, \bm{u}_0^\varepsilon ) \rangle$, and on the direct effect of $Re$ on the viscous term of the eigenmode, i.e. $ \langle \hat{\bm{u}}_1^\dagger, \bm{\nabla}^2_m \hat{\bm{u}}_1 \rangle $. The spatial distribution of the integrand in the inner product of equation \eqref{eq:delta_lambdar} provides information regarding the region that contributes most to the increase of the growth rate when $Re$ increases (recall that we set $Re=Re_c$ as reference such that here $\partial \lambda_r/\partial Re>0$).

Figure \ref{fig:Re_effect} shows the results for the ellipsoid (left) and the bullet (right) with $\AR=1$. The top panels show the modification of the base-flow azimuthal vorticity $\omega_{0\theta}^\varepsilon$ (to be compared with figure \ref{fig:BF-vort-press-AR1}). For all considered geometries, the $\varepsilon$ Reynolds increase leads to a more negative vorticity in the layer along the separating streamline and to a more positive vorticity in the base region and in a tiny elongated region along the $u_{0z} = 0$ line. Interestingly, the latter region closely matches the region where $\partial\omega_{0\theta}/\partial r$ first changes sign (see the discussion in \S\ref{sec:vorticity}). The bottom panels show the integrand of the inner product in equation \eqref{eq:delta_lambdar} and reveal that the destabilisation due to the increase of $Re$ mostly comes from the upstream region of the recirculating region, which closely matches the wavemaker identified by the structural sensitivity. Moreover, a close inspection of the terms in equation \eqref{eq:delta_lambdar} shows that almost the complete destabilisation is due to the convective-like term resulting from the base-flow modification rather than due to the direct effect of $Re$ on the eigenmode, as $|\hat{\bm{u}}_1^{\dagger,*} \cdot \mathcal{C}_m( \hat{\bm{u}}_1, \bm{u}_0^\varepsilon ) | \gg |\hat{\bm{u}}_1^{\dagger,*} \cdot \bm{\nabla}^2_m \hat{\bm{u}}_1|$ at all positions (not shown).

\section{Azimuthal vorticity and mechanism of the primary bifurcation}
\label{sec:vorticity}

\begin{figure}
%\centerline{   
%    %\begin{overpic}[width=7cm, trim=0mm 0mm 0mm 0mm, clip=true]{figure/curves/tmp_BF_omega_theta_max_at_Rec_vs_Re}
% 	%\end{overpic}
%    %\begin{overpic}[width=7cm, trim=30mm 90mm 30mm 90mm, clip=true]{figure/curves/max_surface_vorticity.pdf}
% 	%\end{overpic}  
%    \begin{overpic}[width=7cm, trim=30mm 90mm 30mm 90mm, clip=true]{figure/curves/max_surface_vorticity_new.pdf}
%\end{overpic} 
%}  
\centerline{    
   \begin{overpic}[width=10cm, trim=30mm 120mm 30mm 95mm, clip=true]{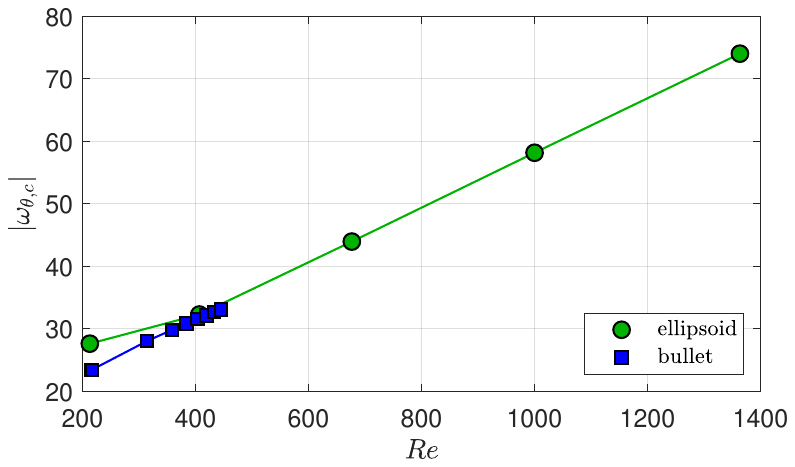}
 	\end{overpic} 
}
\caption{
Maximum surface azimuthal vorticity at $Re=Re_c$. 
}
\label{fig:omega_max_vs_Re}
\end{figure}
In this section we show that the instability mechanism proposed by \cite{magnaudet-mougin-2007} for oblate spheroidal objects with a free-slip surface extends to axisymmetric bodies with a no-slip surface.

\subsection{Maximum surface vorticity}
\label{sec:max_vort}

As discussed in \S\ref{sec:introduction}, MM's arguments are based on the idea that the bifurcation is driven by the vorticity generated at the body surface and transported into the wake. Therefore, we start by  assessing the relation between the base-flow surface vorticity and the onset of the bifurcation. To simplify the notation, in this and in the following section (\S\ref{sec:grad}) we drop the ``$0$" subscript.
Introducing $\omega_{\theta,max}(Re)$ the maximum vorticity measured on the surface of a body of given geometry and aspect ratio, figure \ref{fig:omega_max_vs_Re} shows $\omega_{\theta,c}(Re)$, i.e. the maximum surface vorticity $\omega_{\theta,max}$ at the critical Reynolds number $Re=Re_c$.% = \omega_{\theta,max}(Re)$, the maximum surface vorticity at the critical Reynolds number $Re = Re_c$.
%We translate the neutral curves $Re_c(\AR)$ in $\omega_c(Re)$, where $\omega_c(Re) = \omega_{\theta,\max}(Re,\AR)$ is the maximum surface vorticity at $Re = Re_c$; see figure \ref{fig:omega_max_vs_Re}. 
%
We report values for the ellipsoids and bullets, which have a smooth geometry. For cones and bicones, we note that the maximum surface vorticity
diverges at the sharp edges, which leads to $\omega_{\theta,max}$ growing unbounded when refining the numerical mesh, while the critical Reynolds
number is well converged. It is therefore unlikely that a simple relationship between $\omega_{\theta,max}$ and $Re_c$ holds for bodies with a sharp
geometry.
See Appendix~\ref{sec:appendix_sharp_LE} for more details.

% When trying to get rid of the geometrical singularity, we have tried to introduce a small rounding at the corners. However, $\omega_{\theta,\max}$ shows a strong dependence on the curvature radius, making again the measure unreliable. 
%
Similarly to what was observed by MM for free-slip surfaces, figure \ref{fig:omega_max_vs_Re} shows that $\omega_{\theta,c}$ has a linear dependence on $Re$, and that the results for the two considered geometries collapse rather well onto the same curve 
\begin{equation}
\omega_{\theta,c}(Re) \approx  a + b Re,
\label{eq:lin-fit}
\end{equation}
where $a$ and $b$ are two constants. %The maximum surface vorticity at criticality linearly increases with $Re$. In other words, t
The amount of vorticity that has to be produced at the body surface to promote the flow instability  increases linearly with the Reynolds number, in a way that does not depend on the geometry of the body. This provides a simple criterion for predicting the onset of the instability, which only requires the knowledge of the maximum azimuthal vorticity at the surface (which is however not always easily available, as later discussed in \S\ref{sec:scaling}): at a given Reynolds number the flow is stable when $\omega_{\theta,\max}(Re) < \omega_{\theta,c}(Re)$ and unstable when $\omega_{\theta,\max}(Re) > \omega_{\theta,c}(Re)$. 
The linear fit of our data with equation (\ref{eq:lin-fit}) gives $a = 14$ and $b = 4.4 \times 10^{-2}$. These values differ from the constants $a = 12.5$ and $b=4.3 \times 10^{-3}$ found by MM for ellipsoids with free-slip surface. This is expected, as the maximum surface vorticity at criticality depends on the slip at the body surface; see for example figure 4 of \cite{LEGENDRE_LAUGA_MAGNAUDET_2009} for the flow past a circular cylinder at different Knudsen numbers. 
Notably, the different values 
%$Re$-prefactor in equation (\ref{eq:lin-fit}) (the $b$ coefficient) 
of $b$ show that the critical %maximum 
vorticity at the body surface %that promotes the flow bifurcation 
exhibits a faster variation with $Re$ for axisymmetric bodies with a no-slip surface than for those with a free-slip surface.

\subsection{Vorticity gradient}

\label{sec:grad}

\begin{figure}
\centerline{  
    \begin{overpic}[width=0.49\textwidth, trim=0mm 15mm 9mm 15mm, clip=true]{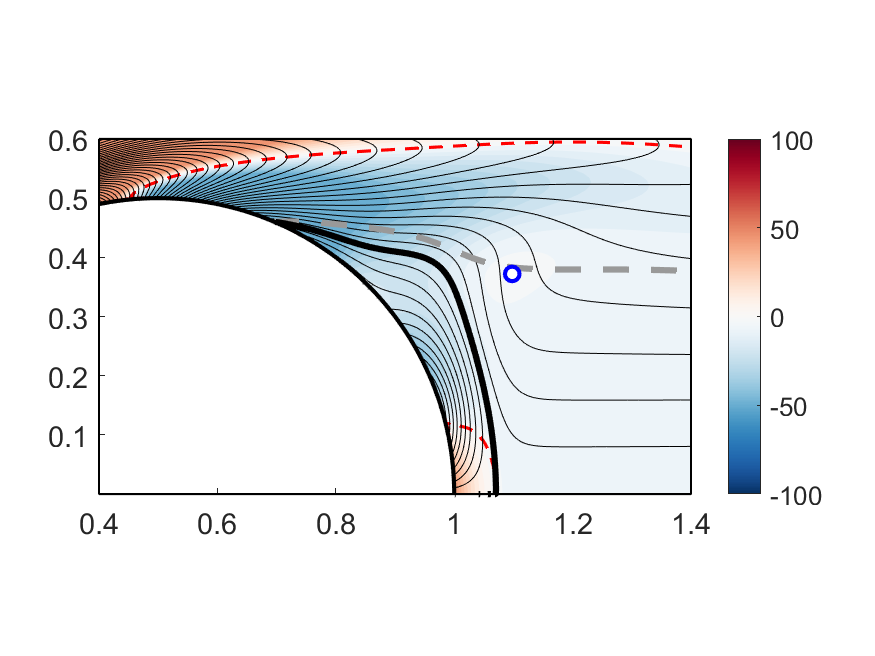}
        \put(-1,53){$(a)$}    
        \put(1,30){$r$} 
        \put(48,0){$z$}    
        \put(99,23){\rotatebox{90}{$\partial \omega_\theta/\partial r$}}
 	\end{overpic} 
  \hspace{0.1cm}
    \begin{overpic}[width=0.49\textwidth, trim=0mm 15mm 9mm 15mm, clip=true]{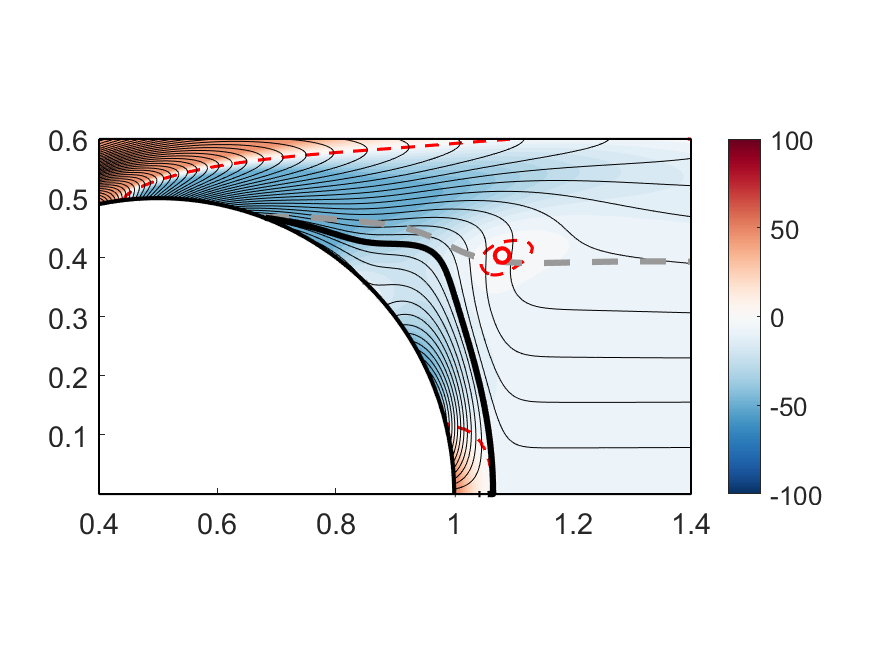}
        \put(-1,53){$(b)$}
        \put(48,0){$z$}    
        \put(99,23){\rotatebox{90}{$\partial \omega_\theta/\partial r$}}
 	\end{overpic}
}
\caption{
%As figure \ref{fig:MM} for an
Near-wake distribution of the azimuthal vorticity around an  ellipsoid with no-slip surface and $\AR=1$. 
Panel $(a)$ is for $Re=250 < \widetilde{Re}$, and panel $(b)$  for $Re=290 > \widetilde{Re}$, where $\widetilde{Re}$ is the Reynolds number corresponding to the first appearance of a region with $\partial \omega_\theta/\partial r \geq 0$ where $\omega_\theta<0$. 
Thin black lines are isocontours of the azimuthal vorticity $\omega_\theta$.
The thick black line delimits the ``boundary layer'' where $\omega_\theta>0$.
Coloured contours show $\partial \omega_\theta/\partial r$.
The grey dashed line shows $u_z = 0$, and the red dashed line  $\partial \omega_\theta/\partial r = 0$.  
Blue/red circles indicate the negative/positive maximum of $\partial \omega_\theta/\partial r$ in the near-wake region.
}
\label{fig:contours-zoom-before-after}
\end{figure}

We now focus on the near-wake region and investigate the correlation between the flow bifurcation and the appearance of points where $\partial \omega_\theta/\partial r = 0$. Figure \ref{fig:contours-zoom-before-after} shows the vorticity distribution in the near wake for the $\AR=1$ ellipsoid (with the same representation as in figure \ref{fig:MM}). 
Due to the no-slip boundary condition, a thin boundary layer arises on the rear side of the body, where the azimuthal vorticity is positive, $\omega_\theta>0$, i.e. has the opposite sign as in the rest of the flow (the black thick line in figure \ref{fig:contours-zoom-before-after} denotes $\omega_\theta=0$). 
Outside this boundary layer, the vorticity distribution closely resembles that of free-slip bodies: isocontours of $\omega_\theta$  are essentially parallel to the solid surface in the vicinity of the body, 
%(see the region immediately outside the grey line), 
and farther downstream they align with the symmetry axis. 
%Figure \ref{fig:contours-zoom-before-after} also shows that 
In the transition region in-between, the turning of the $\omega_\theta$ isocontours becomes sharper and sharper as $Re$ gets larger. 
Two values of the Reynolds numbers are considered in figure \ref{fig:contours-zoom-before-after}, one below (figure \ref{fig:contours-zoom-before-after}$a$) and one above (figure \ref{fig:contours-zoom-before-after}$b$)  the critical Reynolds number $Re_c$. One can immediately notice that $\partial \omega_\theta/\partial r < 0$ (blue) everywhere in the transition region when $Re < Re_c$, while a region where $\partial \omega_\theta/\partial r  \ge 0$ (red) appears when $Re > Re_c$, in agreement with  MM's arguments.
\begin{figure}
\centerline{   
    \begin{overpic}[width=0.49\textwidth, trim=0mm 15mm 9mm 15mm, clip=true]{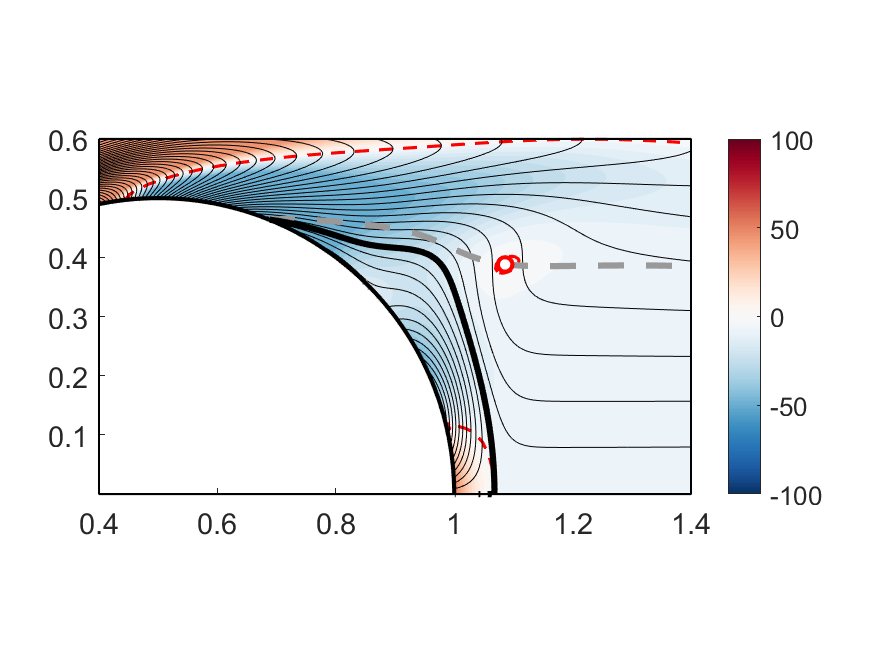}
        %\put(-1,53){$(a)$}    
        \put(1,30){$r$} 
        %\put(48,0){$z$}       
        \put(99,23){\rotatebox{90}{$\partial \omega_\theta/\partial r$}}
 	\end{overpic} 
  \hspace{0.1cm}
    \begin{overpic}[width=0.49\textwidth, trim=0mm 15mm 9mm 15mm, clip=true]{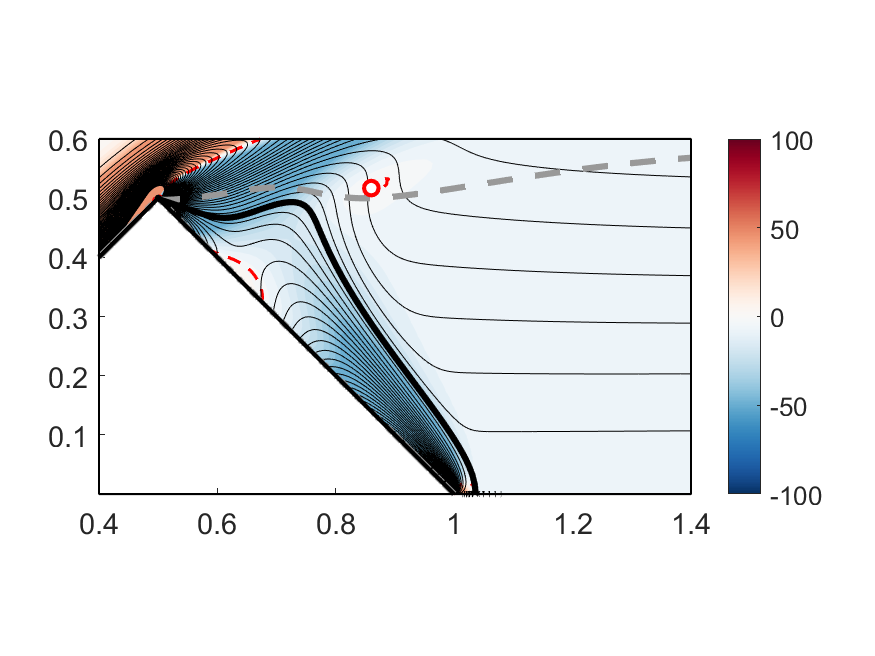}
        %\put(-1,53){$(b)$}
        %\put(48,0){$z$}       
        \put(99,23){\rotatebox{90}{$\partial \omega_\theta/\partial r$}}
 	\end{overpic}
}
\centerline{  
    \begin{overpic}[width=0.49\textwidth, trim=0mm 15mm 9mm 15mm, clip=true]{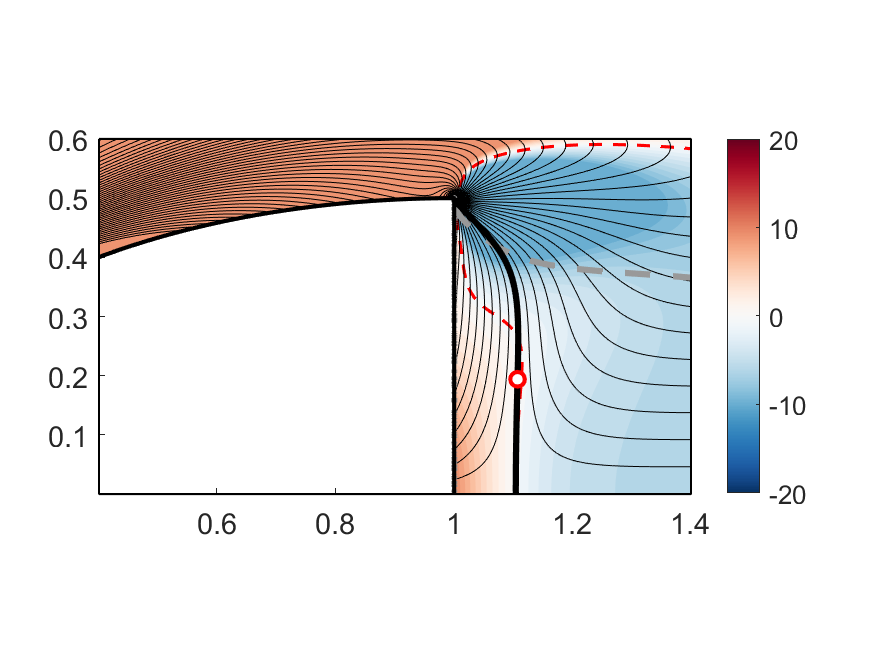}
        %\put(-1,53){$(c)$}    
        \put(1,30){$r$} 
        \put(48,1){$z$}    
        \put(99,23){\rotatebox{90}{$\partial \omega_\theta/\partial r$}}
 	\end{overpic} 
  \hspace{0.1cm}
    \begin{overpic}[width=0.49\textwidth, trim=0mm 15mm 9mm 15mm, clip=true]{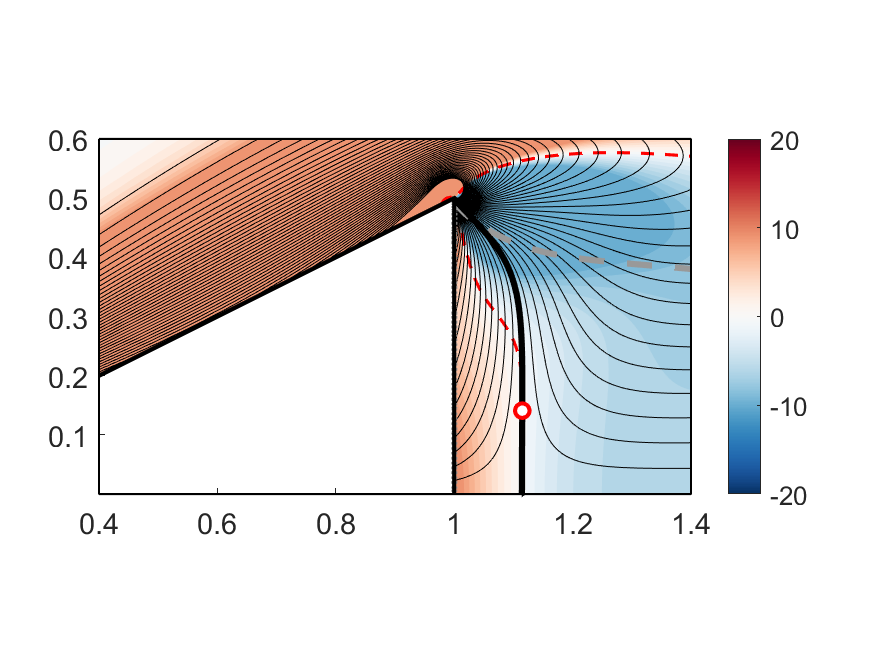}
        %\put(-1,53){$(d)$}   
        \put(48,1){$z$}    
        \put(99,23){\rotatebox{90}{$\partial \omega_\theta/\partial r$}}
 	\end{overpic}
}
\caption{As figure \ref{fig:contours-zoom-before-after} for the four considered geometries with $\AR=1$. The Reynolds number is slightly above $\widetilde{Re}$, i.e. after the first appearance of a region with $\partial \omega_\theta/\partial r \geq 0$ where $\omega_\theta<0$. Top left: ellipsoid at $Re=270$. Top right: bicone at $Re=295$. Bottom left: bullet at $Re=120$. Bottom right: cone at $Re=80$.}
%
%Dashed line: $u_z=0$.
%Black contours: azimuthal vorticity $\omega_\theta$ (the thick contour delimits the ``boundary layer'' where $\omega_\theta>0$).
%Filled contours: $\partial \omega_\theta/\partial r$.
%Circle symbol: maximum of $\partial \omega_\theta/\partial r$ in the region $\omega_\theta<0$, i.e. outside the ``boundary layer''.
%}
\label{fig:contours-zoom}
\end{figure}
\begin{figure}
\centerline{   
    \begin{overpic}[width=0.49\textwidth, trim=0mm 15mm 9mm 15mm, clip=true]{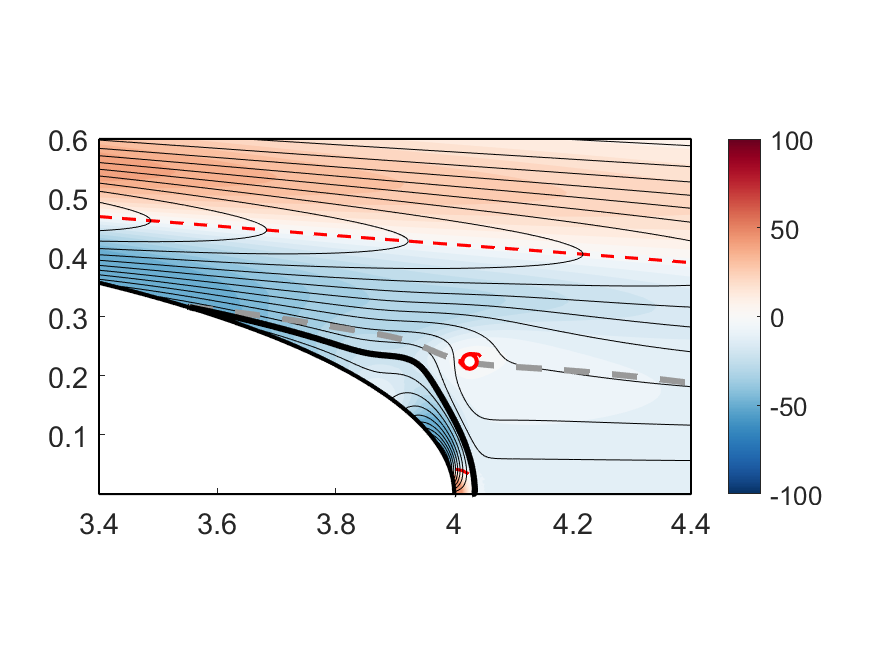}
        %\put(-1,53){$(a)$}    
        \put(1,30){$r$}  
        \put(99,23){\rotatebox{90}{$\partial \omega_\theta/\partial r$}}
 	\end{overpic} 
  \hspace{0.1cm}
    \begin{overpic}[width=0.49\textwidth, trim=0mm 15mm 9mm 15mm, clip=true]{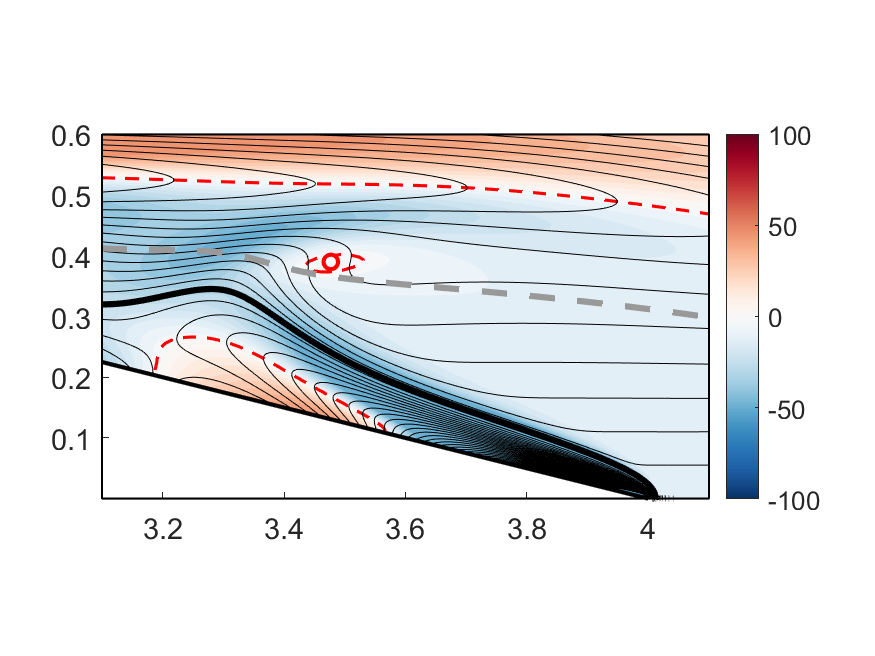}
        %\put(-1,53){$(b)$} 
        \put(99,23){\rotatebox{90}{$\partial \omega_\theta/\partial r$}}
 	\end{overpic}
}
\centerline{  
    \begin{overpic}[width=0.49\textwidth, trim=0mm 15mm 9mm 15mm, clip=true]{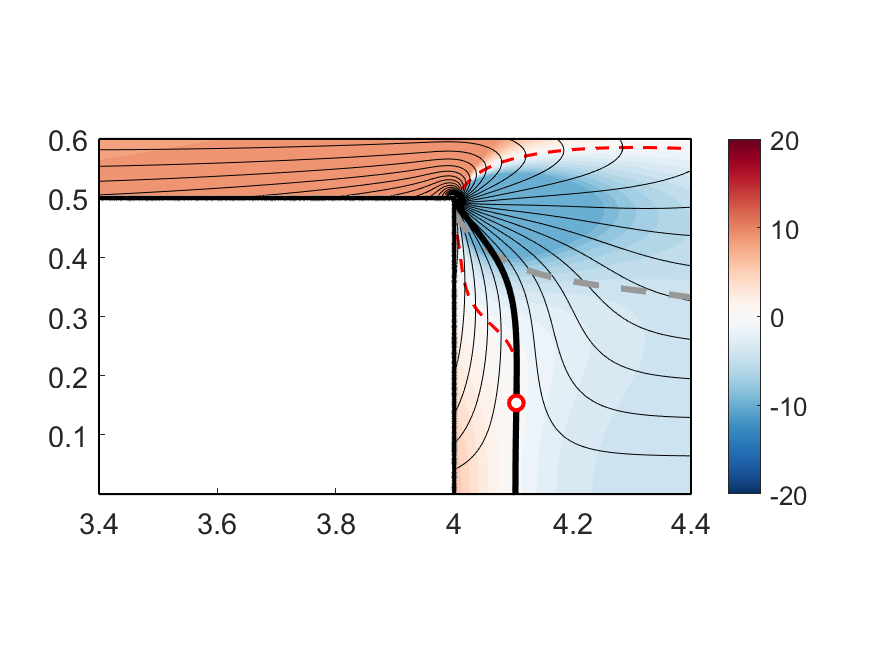}
        %\put(-1,53){$(c)$}    
        \put(1,30){$r$} 
        \put(48,1){$z$} 
        \put(99,23){\rotatebox{90}{$\partial \omega_\theta/\partial r$}}
 	\end{overpic} 
  \hspace{0.1cm}
    \begin{overpic}[width=0.49\textwidth, trim=0mm 15mm 9mm 15mm, clip=true]{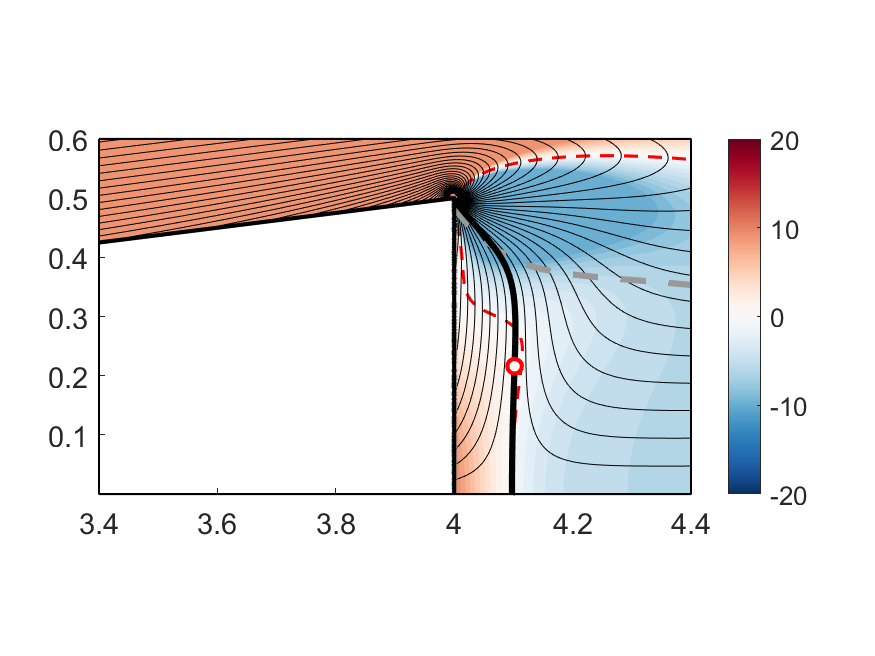}
        %\put(-1,53){$(d)$}   
        \put(48,1){$z$} 
        \put(99,23){\rotatebox{90}{$\partial \omega_\theta/\partial r$}}
 	\end{overpic}
}
\caption{
As figure~\ref{fig:contours-zoom} for $\AR=4$. Top left: ellipsoid at $Re=1250$. Top right: bicone at $Re=950$. Bottom left: bullet at $Re=175$. Bottom right: cone at $Re=150$.}
\label{fig:contours-zoom-AR4}
\end{figure}
Figures \ref{fig:contours-zoom} and \ref{fig:contours-zoom-AR4} are as figure \ref{fig:contours-zoom-before-after}$(b)$, for different geometries and aspect ratios. %, for $Re$ slightly above $Re_c$. 
Notably, the location where the $\partial \omega_\theta/\partial r \ge 0$ region appears depends on the shape of the TE. For blunt TE, %the $\partial \omega_\theta/\partial r  \ge 0$ region 
it is placed along the $\omega_\theta = 0$ isocontour delimiting the base boundary layer  due to the no-slip boundary condition.
For bodies with zero-thickness TE, instead, %the $\partial \omega_\theta/\partial r \ge 0$ region arises 
it is located farther downstream within the wake recirculation region and close to the contour of zero streamwise velocity $u_z=0$.

It is worth noting that the vorticity distribution for bodies with zero-thickness TE explains the steep increase of $Re_c$ with $\AR$ observed in figure~\ref{fig:Rec}. Due to the specific shape of these bodies,  the $\omega_\theta$ isocontours close to the body surface become more aligned with the symmetry axis $r=0$ as $\AR$ increases. 
For these bodies, therefore, the condition $\partial \omega_\theta / \partial r = 0$ ($\omega_\theta$ isocontours being locally perpendicular to the symmetry axis) for the instability to occur requires more vorticity to be produced at the body surface, i.e. a larger Reynolds number.

\begin{figure}
\centerline{   
    \begin{overpic}[width=10cm, trim=30mm 120mm 30mm 95mm, clip=true]{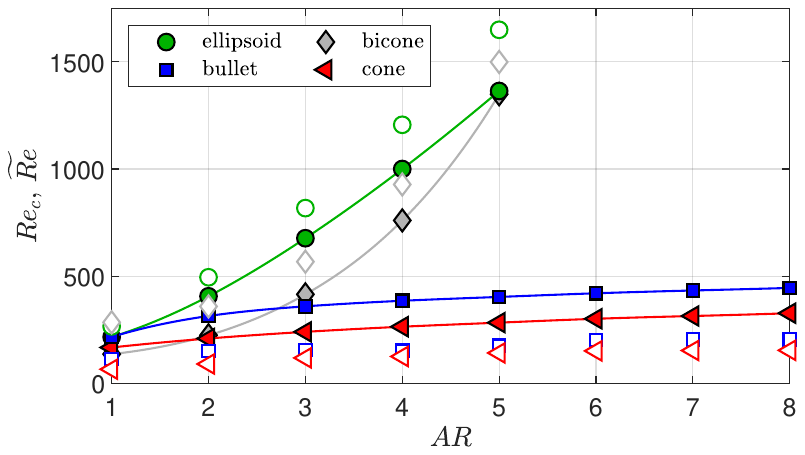}
 	\end{overpic} 
}
\caption{
Critical Reynolds number $Re_c$ (lines) and lowest Reynolds number $\widetilde{Re}$ (open symbols) such that $\partial \omega_\theta/\partial r \geq 0$ where $\omega_\theta<0$ in the near wake. 
}
\label{fig:Rec_and_Revort}
\end{figure}
Figure \ref{fig:Rec_and_Revort} quantifies the correlation between the onset of the instability and the first appearance of points where $\partial \omega_\theta/\partial r = 0$. We compare the critical Reynolds number $Re_c$ obtained from our stability analysis with the Reynolds number $\widetilde{Re}_c$ corresponding to the first appearance of a point where $\partial \omega_\theta/\partial r = 0$ in the near wake (we only consider %$\partial \omega_\theta/\partial r = 0$ points in 
the near wake outside the boundary layer that develops along the TE). 
Although the collapse with $Re_c$ is not perfect, $\widetilde{Re}_c$ captures the trend of the critical Reynolds number rather nicely for all the considered geometries,  increasing sharply  with $\AR$ for bodies with a zero-thickness TE, %while it tends to an asymptote 
and much more slowly for bodies with a blunt TE. 

To summarise, our results show that the instability mechanism proposed by \cite{magnaudet-mougin-2007} in the context of ellipsoids with free-slip surfaces also extends to no-slip surfaces, and describes fairly well the primary symmetry-breaking bifurcation of the flow past axisymmetric bodies with different geometries and aspect ratios. 
This clearly hints to the fact that the vorticity generated at the body surface is at the root of the instability mechanism, while the way it is produced, i.e. the nature of the %boundary condition at the 
surface, does not play a major role.

\subsection{Free-slip vs. no-slip}

It is worth spending few words on the fact that, as shown in the previous sections, axisymmetric wakes past free-slip and no-slip  bodies share the same instability mechanism. It is well known that the mechanism of vorticity generation on a surface changes with the boundary condition; see for example \cite{truesdell-1954, lighthill-1963, morton-1983, leal-1989, wu-wu-1993, wu-1995, lundgren-koumoutsakos-1998, brons-etal-2014, terrington-etal-2020}. 
This is clear when looking at the ``boundary vorticity flux'' (normal diffusion flux of vorticity) at the surface, $\bm{\sigma} = \nu \bm{n} \cdot \bm{\nabla}\bm{\omega}$, which is a measure of the vorticity creation %process 
at the wall. 
For stationary rigid walls, it reads \citep[see][]{wu-wu-1993}
\begin{equation}
  \bm{\sigma} =
  % \nu \bm{n} \cdot \bm{\nabla}\bm{\omega} = 
  \underbrace{ \ \bm{n} \times  \left( \frac{\bm{\nabla}p}{\rho} \right)  }_{\bm{\sigma}_p} +
  \underbrace{ - \left( \bm{n} \times \bm{\nabla} \right) \cdot \left( \bm{\tau} \bm{n} \right) }_{\bm{\sigma}_{\tau} },
\end{equation}
where $\bm{\tau} = \nu \bm{\omega} \times \bm{n}$ is the tangential shear stress.
Here $\bm{\sigma}$ depends on the non-uniform distribution of the normal ($\bm{\sigma}_p$)  and shear ($\bm{\sigma}_\tau$)  stresses. 
In the limit case of a two-dimensional flow  $\bm{u}(x,y)=(u_x,u_y,0)$ near a stationary, no-slip flat wall of normal $\bm{n}=\bm{e}_y$,  
the boundary vorticity flux reduces to $\bm{\sigma} = \bm{\sigma}_p = -(1/ \rho) (\partial p /\partial x) \bm{e}_z$ while $\bm{\sigma}_\tau=\bm{0}$ \citep{lighthill-1963}. 
By contrast, on a free-slip surface the vorticity appears as a consequence of the continuity of the tangent stresses and is non null only in the case of curved surfaces; in the limit case of a  two-dimensional stationary surface, $\omega = 2 U \kappa$, where $U$ is the tangential velocity and $\kappa$ is the local curvature; 
accordingly, the boundary vorticity flux
%normal diffusion flux of the vorticity 
is non null only for curved surfaces \citep[see][]{wu-1995}. %, and reads \citep{wu-1995}
%
\iffalse
\begin{equation}
\begin{bmatrix}
\nu \bm{n} \cdot \bm{\nabla} \bm{\omega}_S \\
\nu \bm{n} \cdot \bm{\nabla} \omega_n
\end{bmatrix} = 
\begin{bmatrix}
\bm{n} \left( \frac{D \bm{V}}{D t} + \bm{\nabla}  \frac{p}{\rho} \right) +
\nu \left( \bm{\nabla}_S \omega_n - \bm{\omega}_S \cdot \bm{\nabla}_S \bm{n} \right) \\
- \nu \left( \omega_n \bm{\nabla}_S \cdot \bm{n} + \bm{\nabla}_S \cdot \bm{\omega}_S \right)
\end{bmatrix}.
\end{equation}
%
\tcb{$\bm{V}$ is not defined. 
Actually, it would be easier to stay in the stationary case, just like (6.1). 
More generally, I don't understand very well the purpose of (6.2). Do you mean to emphasize that $\bm{\sigma}$ is different for no-slip and free-slip surfaces? Currently, it's not very clear (partly because it's not easy to compare (6.1) and (6.2): we introduce the decomposition, and the curvature is hidden in the differential operator, etc).
Perhaps we should add a sentence explaining why $\bm{\sigma}$ is different for no-slip and free-slip surfaces, in the light of (6.2); otherwise I would vote for removing this equation. }
%
Here the vorticity has been decomposed in the form $\bm{\omega} = \bm{\omega}_S + \omega_n$, where  $\bm{\nabla}_S = \bm{\nabla} - \bm{n} \cdot \bm{\nabla}$ is the surface gradient operator and $\omega_n = \bm{\omega} \cdot \bm{n}$. 
\fi

The fact that axisymmetric bodies with no-slip and free-slip surfaces exhibit the same instability mechanism means that, although the vorticity is at the root of the instability, the way it is produced at the wall does not play a dominant role in the overall triggering mechanism \citep[see also the discussion in][]{magnaudet-mougin-2007}. In other words, the flow becomes unstable once the amount of vorticity brought into the near wake is large enough, %notwithstanding 
irrespective of the production mechanism. 
Clearly, the different vorticity creation mechanism has an impact on the critical Reynolds number and on the actual onset of the instability, as it influences the size of the wake recirculation region and the amount of vorticity  brought into the wake. 
The different vorticity creation mechanism, indeed, leads to some substantial differences between no-slip and free-slip surfaces in terms of boundary vorticity dynamics \citep{moore-1963,wu-1995}. 
For large Reynolds numbers,  dimensional analysis shows that
(i)~in the no-slip case, a boundary vorticity flux $\sigma \sim O(1)$  is generated, which implies that  the surface vorticity increases as $\omega \sim Re^{1/2}$;
(ii)~in the free-slip case, the  boundary vorticity flux  decreases as $\sigma \sim Re^{-1/2}$ and the amount of surface vorticity is independent of the Reynolds number, $\omega \sim O(1)$. 

\subsection{On the amplification mechanism}

We now investigate the contribution of various physical mechanisms on the exponential growth of the mode in the linear regime ($Re \approx Re_c$). We start by looking at the energy equation for the perturbation $\bm{u}_1$ \citep[see][]{lanzerstorfer-kuhlmann-2012}. We sum equation \eqref{eq:LNSE} multiplied by $\hat{\bm{u}}_1^*$ with the complex conjugate of the same equation multiplied by $\hat{\bm{u}}_1$ and, after some manipulation and using the incompressibility constraint, we obtain the equation for $\hat{\bm{u}}_1 \cdot \hat{\bm{u}}_1^*$, i.e.
\begin{equation}
  \begin{gathered}
  2 \lambda_r \hat{\bm{u}}_1 \cdot \hat{\bm{u}}_1^* =
  \underbrace{ - ( \hat{\bm{u}}_1 \otimes \hat{\bm{u}}_1^*  + \hat{\bm{u}}_1^* \otimes \hat{\bm{u}}_1 ) : \bm{\nabla}_0 \bm{u}_0 }_{ \mathcal{P} }
  \underbrace{ - \frac{1}{Re} \bm{\nabla}_0^2 ( \hat{\bm{u}}_1 \cdot \hat{\bm{u}}_1^* ) }_{ \mathcal{D} } \\
  \underbrace{ - \bm{\nabla}_0 \cdot ( \hat{p}_1 \hat{\bm{u}}_1^* + \hat{p}_1^* \hat{\bm{u}}_1 ) }_{ \mathcal{T}_p } 
  \underbrace{ - \bm{\nabla}_0 \cdot( \bm{u}_0 \hat{\bm{u}}_1 \cdot \hat{\bm{u}}_1^* ) }_{ \mathcal{A} } -
  \underbrace{ \frac{2}{Re} \left( \bm{\nabla}_m \hat{\bm{u}}_1 : \bm{\nabla}_m^* \hat{\bm{u}}_1^* \right) }_{ \mathcal{E} }.
  \end{gathered}
\end{equation}
On the right-hand side, in order we find the production term $\mathcal{P}$ which accounts for the exchange of energy between the base flow and the perturbation, the viscous diffusion term $\mathcal{D}$, the pressure transport term $\mathcal{T}_p$, the advection associated with the base flow $\mathcal{A}$, and the viscous dissipation $\mathcal{E}$. All the terms contribute to the total budget, but differently depending on the flow region.

\begin{figure}
\vspace{0.5cm}
\centerline{ 
    \begin{overpic}[width=0.49\textwidth, trim=0mm 0mm 0mm 0mm, clip=false]{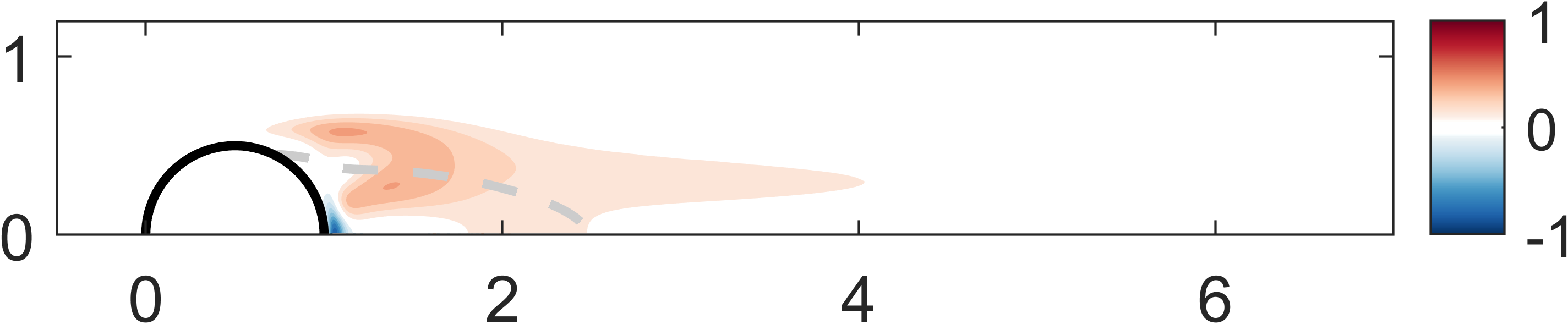}  
       \put(-5,10){$r$}
       \put(47,-4){$z$}
 	\end{overpic}
    \begin{overpic}[width=0.49\textwidth, trim=0mm 0mm 0mm 0mm, clip=false]{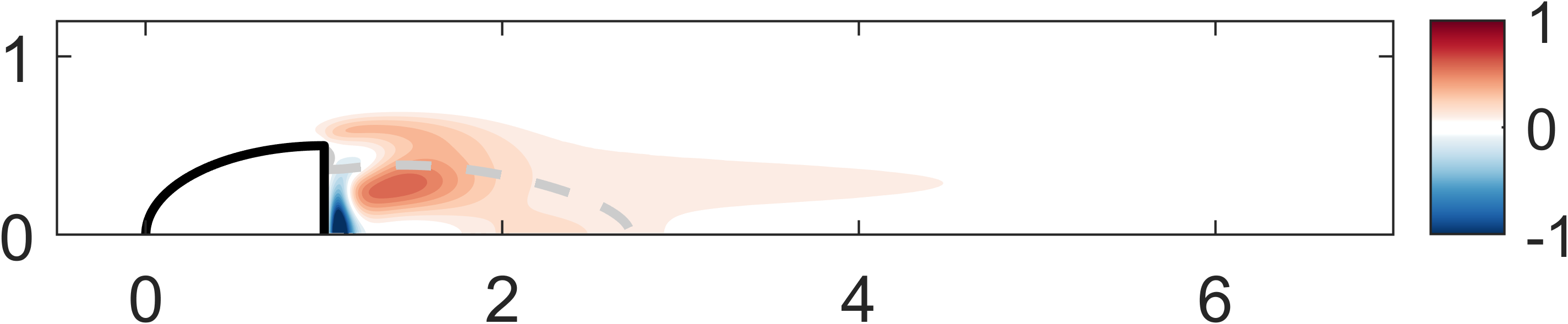}              
       \put(47,-4){$z$}
 	\end{overpic}
}
\centerline{ 
    \begin{overpic}[width=0.49\textwidth, trim=0mm 0mm 0mm 0mm, clip=false]{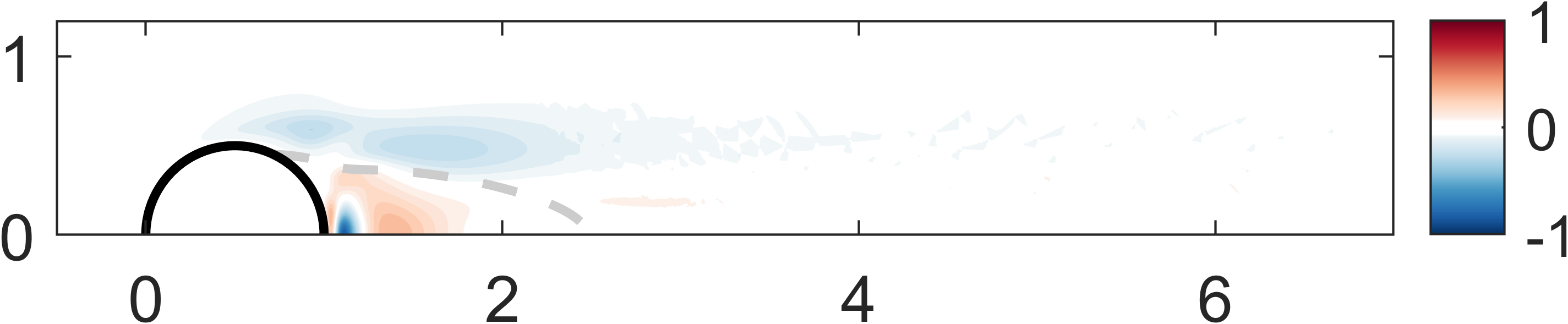}  
       \put(-5,10){$r$}
       \put(47,-4){$z$}
 	\end{overpic}
    \begin{overpic}[width=0.49\textwidth, trim=0mm 0mm 0mm 0mm, clip=false]{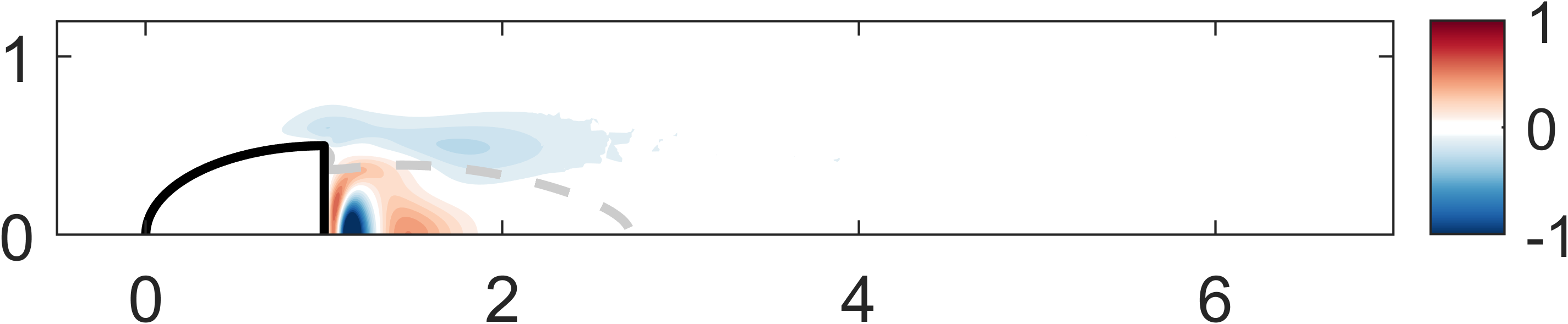}              
       \put(47,-4){$z$}
 	\end{overpic}
}
\vspace{0.2cm}
\caption{Energy budget for $\AR=1$ at $Re \simeq Re_c$. Top: Production term $\mathcal{P}$. Bottom: Advection term $\mathcal{A}$. Left: ellipsoid, $Re=210$. Right: bullet, $Re=220$. The dashed line is for $u_{0z} = 0$.}
\label{fig:en_bud}
\end{figure}
Figure \ref{fig:en_bud} considers the energy equation for the ellipsoid and the bullet with $\AR=1$ at $Re=Re_c$. In this case $\lambda_r = 0$ and the Reynolds-Orr equation reduces to $\int_D \mathcal{P} \text{d}\Omega = \int_D \epsilon \text{d}\Omega$ with $\int_D \mathcal{A} + \mathcal{T}_p + \mathcal{D} = 0$. We focus on the spatial distribution of $\mathcal{P}$ and $\mathcal{A}$ to locate the flow regions where the mode amplification is driven by the production and/or the advection. We observe that $\mathcal{P}>0$ downstream the bodies, with the maximum placed along the separating streamline and within the recirculating region, where the base-flow velocity gradient is indeed maximum. At the base, the small $\mathcal{P}<0$ region indicates that energy is moved from the perturbation to the base flow there. For bodies with a blunt base we observe a more intense production activity, in agreement with the stronger base-flow velocity gradients. The advection term $\mathcal{A}$, instead, behaves differently depending on the direction of $\bm{u}_0$. It is a sink ($\mathcal{A}<0$) and tends to stabilise the flow along the separating shear layer where $u_{0z}>0$, while it is a source ($\mathcal{A}>0$) and tends to destabilise the flow within the recirculating region where $u_{0z}<0$. In the recirculating region the base flow transports the perturbations backwards in the near-wake so that they can grow more compared to the situation where they are advected downstream. Note that this agrees with \S\ref{sec:Reeffect} where the effect of a small increase in $Re$ on $\lambda_r$ is discussed (see in figure \ref{fig:Re_effect} the large positive $\partial \lambda_r/\partial Re>0$ region in the core of the recirculating region).

Similar conclusions can be drawn by using the concept of endogeneity introduced by \cite{marquet-lesshafft-2015}. The endogeneity $E(\bm{x})$ characterises the contributions of localised flow regions to the global dynamics of the eigenmode, and allows to separate the contributions from individual mechanisms such as production, base-flow advection, pressure forces and viscous diffusion. The endogeneity is defined as $E(\bm{x}) = \hat{\bm{u}}_1^{\dagger*} \cdot ( - \mathcal{L}_m \hat{\bm{u}}_1 - \bm{\nabla}_m \hat{p}_1 )$ and has the property that its integral equals the eigenvalue, i.e $\int_D E(\bm{x}) \text{d}\Omega = \lambda$. We now replace the definition of $\mathcal{L}_m$ in the above relation to obtain
\begin{equation}
  E(\bm{x}) = - \hat{\bm{u}}_1^{\dagger*} \cdot ( (\bm{u}_0 \cdot \bm{\nabla}_m ) \hat{\bm{u}}_1 )
              - \hat{\bm{u}}_1^{\dagger*} \cdot ( (\hat{\bm{u}}_1 \cdot \bm{\nabla}_m ) \bm{u}_0 ) 
              - \hat{\bm{u}}_1^{\dagger*} \cdot \bm{\nabla} \hat{p}_1
              + \frac{1}{Re} \hat{\bm{u}}_1^{\dagger*} \cdot \bm{\nabla}_m^2 \hat{\bm{u}}_1
  \label{eq:end}
\end{equation}
where in order the different terms account for the contribution of the base-flow advection, the production due to the base-flow velocity gradients, pressure forces and viscous diffusion. The real part of the endogeneity and of equation \eqref{eq:end} isolates the contributions to the growth rate $\lambda_r$, while the imaginary part the contributions to the eigenfrequency $\lambda_i$.
\begin{figure}
\vspace{0.5cm}
\centerline{ 
    \begin{overpic}[width=0.49\textwidth, trim=0mm 0mm 0mm 0mm, clip=false]{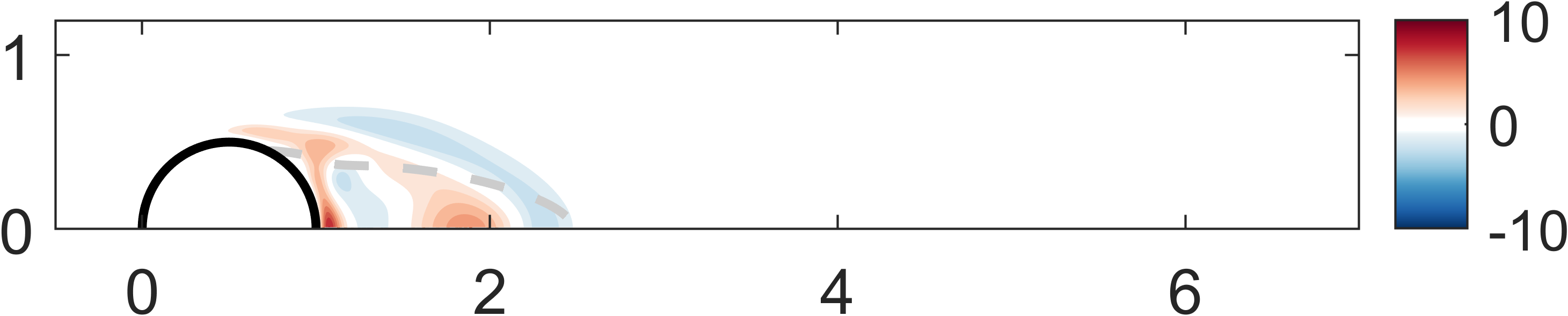}  
       \put(-5,10){$r$}
       \put(47,-4){$z$}
 	\end{overpic}
    \begin{overpic}[width=0.49\textwidth, trim=0mm 0mm 0mm 0mm, clip=false]{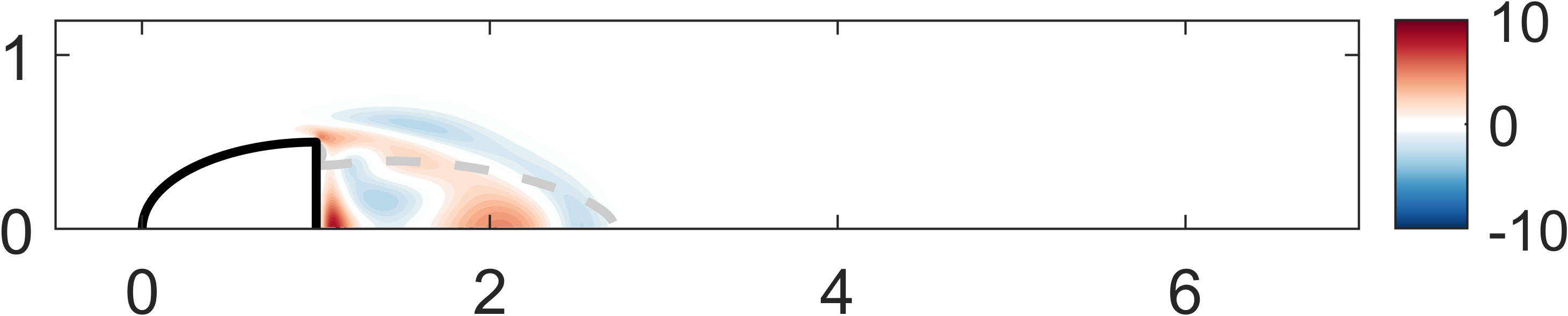}              
       \put(47,-4){$z$}
 	\end{overpic}
}
\centerline{ 
    \begin{overpic}[width=0.49\textwidth, trim=0mm 0mm 0mm 0mm, clip=false]{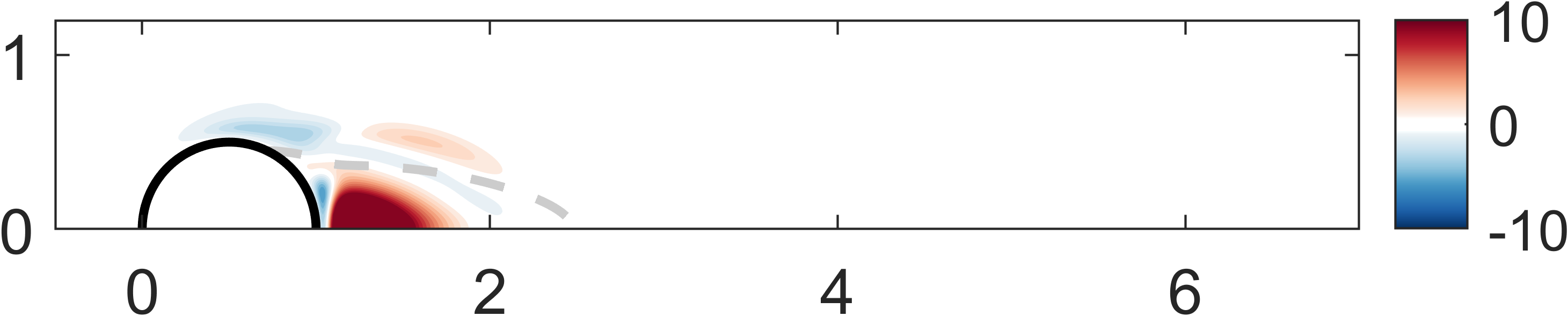}  
       \put(-5,10){$r$}
       \put(47,-4){$z$}
 	\end{overpic}
    \begin{overpic}[width=0.49\textwidth, trim=0mm 0mm 0mm 0mm, clip=false]{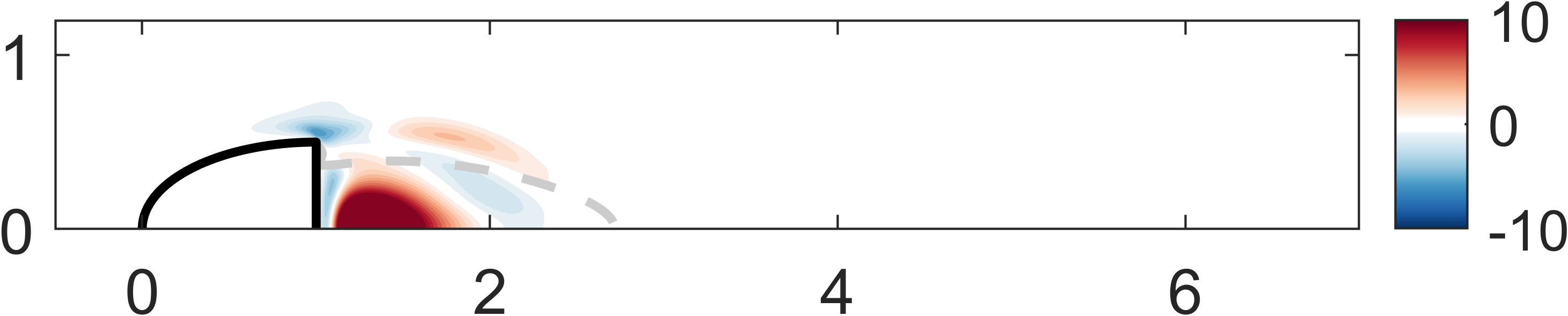}              
       \put(47,-4){$z$}
 	\end{overpic}
}
\vspace{0.2cm}
\caption{Spatial distribution of the production (top) and advection (bottom) contribution to the endogeneity for $\AR=1$ at $Re \simeq Re_c$. Left: ellipsoid, $Re=210$. Right: bullet, $Re=220$. The dashed line is for $u_{0z} = 0$.}
\label{fig:end}
\end{figure}
Figure \ref{fig:end} shows the production and advection contribution to $\Re(E(\bm{x}))$ for the ellipsoid and the bullet with $\AR=1$ at $Re=Re_c$. The endogeneity is concentrated close to the body around the shear layers and within the recirculating region. The production has a stabilising/destabilising effect depending on the region, with the positive contribution to $\Re(E(\bm{x}))$ being maximum at the centre of the recirculating region. The advection, instead, has a strongly destabilising effect within the recirculating region where $u_{0z}<0$, while a stabilising effect in the first portion of the shear layers separating from the bodies where $u_{0z}>0$. This partially agrees with the energy budget analysis.

\section{A new scaling}
\label{sec:scaling}

In the previous section we have shown that the base-flow azimuthal vorticity drives the primary instability of the flow past 3D axisymmetric bodies. We now use this information to define a new Reynolds number which is more suitable for predicting the onset of the instability than the standard Reynolds number based on $U_\infty$ and $H$. 
This is done in the same way as \cite{chiarini-quadrio-auteri-2022b}, who proposed a new scaling for the description of the primary instability of the flow past 2D symmetric bluff bodies. They used a measure of the spatial extent of the separation bubble as a length scale and the largest reverse flow speed within it as velocity scale, and found that the ensuing Reynolds number evaluated at criticality only marginally changed with the geometry and aspect ratio of the body. 
This choice of scales %descends from 
was inspired by their link to the local amplification of the unstable wave packets and to the extension of the absolute instability region \citep{hammond-redekopp-1997,chomaz-2005}. 
Here, we introduce a new scaling that can be used to estimate whether the low-$Re$ steady flow past axisymmetric bodies is unstable, without the need for a computationally expensive stability analysis. This new scaling is based on the inspection of the base flow only (here we again drop the ``$0$" subscript), and relies on quantities that are related to the physics of the problem and are easily accessible in both experiments and numerical simulations.

\begin{figure}
\centerline{   
    \begin{overpic}[width=7cm, trim=30mm 90mm 30mm 90mm, clip=true]{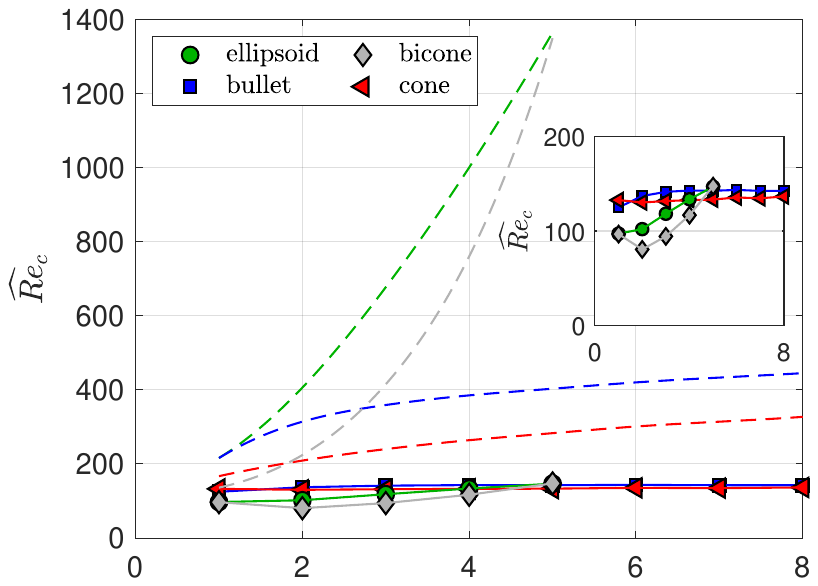}
       \put(-1,70){$(a)$}
       \put(49,2){\scriptsize{$\AR$}}
       \put(75,31.5){\tiny{$\AR$}}
 	\end{overpic}
    \begin{overpic}[width=7cm, trim=30mm 90mm 30mm 90mm, clip=true]{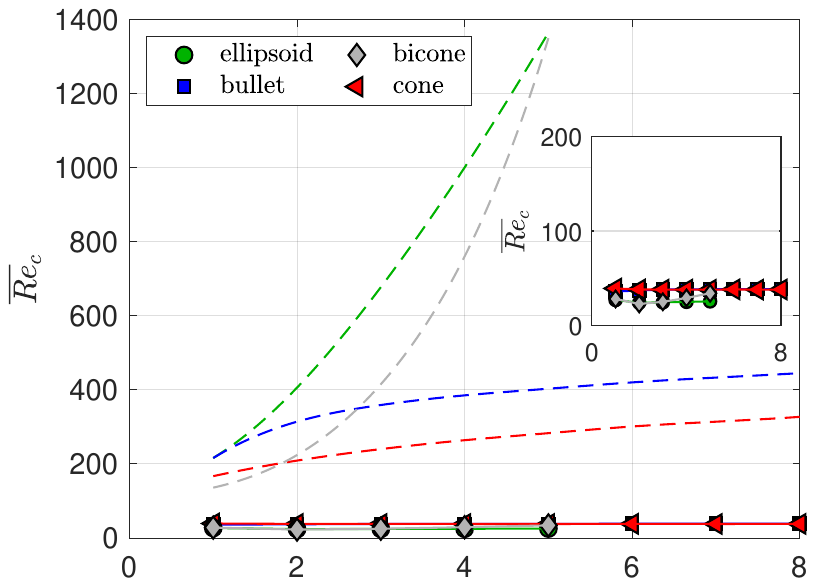}
       \put(-1,70){$(b)$}
       \put(49,2){\scriptsize{$\AR$}}
       \put(75,31.5){\tiny{$\AR$}}
 	\end{overpic}
}
\caption{Critical Reynolds number $Re_c$ (dashed lines) and newly defined Reynolds numbers  
(solid lines):
$(a)$~$\widehat{Re}_c = \widetilde{\omega}_\theta r_0^2/ \nu$,
$(b)$~$\overline{Re}_c = U_{rev} r_0/\nu$.
}
\label{fig:scaling}
\end{figure}
The onset of the primary symmetry-breaking bifurcation depends on the thickness of the transition region where the instability mechanism takes place (see \S\ref{sec:vorticity}). 
A thinner transition region, indeed, means that the $\omega_\theta$ isocontours exhibit sharper turnings, and points with $\partial \omega_\theta/\partial r = 0$ are more likely to arise. The thickness of this region depends on (i)~the maximum vorticity $\omega_{\theta,\max}$ generated at the body surface (see figure \ref{fig:omega_max_vs_Re}) and (ii)~the radial extent of the wake recirculation region \citep{magnaudet-mougin-2007}. 
One may be thus tempted to use $\omega_{\theta,\max}$ and a measure of $r_0$ to define a new Reynolds number. 
However the measure of $\omega_{\theta,\max}$ at the body surface is not always available. 
First, vorticity diverges at a geometric singularity such as a sharp leading edge, as observed for cones and bicones. 
Second, in experiments, the flow region close to the body surface is not easily accessible. 
A different estimation of the surface vorticity is thus needed. 
Here we propose $\widetilde{\omega}_\theta = \max_r \{\omega_\theta(r,z_{max})\}$ where $z_{max}$ is the streamwise location of maximum reverse flow $U_{rev}$ within the wake recirculation region. 
We use $\widetilde{\omega}_\theta$ as a vorticity-related scale and the radius of the $u_z=0$ isocontour at the same $z_{max}$ streamwise location as a measure of $r_0$, and define the new Reynolds number as $\widehat{Re} = \widetilde{\omega}_\theta r_0^2/\nu$. 
Figure \ref{fig:scaling}$(a)$ shows that at criticality this new Reynolds number $\widehat{Re}$ collapses %quite nicely 
approximately to the same value for all the considered bodies. 
The collapse is not perfect, yet the relative variation is one order of magnitude smaller than that observed with the standard Reynolds number based on $U_\infty$ and $H$. 
To be quantitative, we measure for example $\widehat{Re}_c \approx 133$ for the $\AR = 1$ cone   and $\widehat{Re}_c \approx 147$ for the $\AR=5$ ellipsoid, to be compared with $Re_c \approx 167$ and $Re_c \approx 1360$, respectively. 
More generally,  the new proposed Reynolds number provides a simple and effective global criterion for the prediction of the instability onset: 
considering a variation in $\widehat{Re}_c$ of one standard deviation around its mean value (both  computed with all the cases considered in this study and with weights such that each of the four geometries contributes equally), we obtain $125 \pm 20$ and can therefore conclude that the flow is likely to be stable when $\widehat{Re} \lessapprox 105$ and likely to be unstable when $\widehat{Re} \gtrapprox 145$.
For comparison, the original critical Reynolds number has a much larger relative spread, $Re_c = 480 \pm 340$.

It is worth noting that the new scaling works well also when other vorticity-related quantities are used instead of $\widetilde{\omega}_\theta$, provided that they are a measure of the vorticity generated at the body surface. 
As an example, in figure \ref{fig:scaling}$(b)$ we keep $r_0$ as the length scale and use $\widehat{\omega}_\theta = U_{rev}/r_0$ as a vorticity scale, that may indeed be seen as a rough estimate of $\widetilde{\omega}_\theta$. 
Again, at criticality the new proposed Reynolds number $\overline{Re}_c = \widehat{\omega}_\theta r_0^2/\nu = U_{rev} r_0/\nu$ collapses rather well for all the considered aspect ratios and geometries to $ 34
\pm 6 $. The relevance of $U_{rev}$ on the physics of the instability agrees with the map of the sensitivity to base-flow modifications shown in figures \ref{fig:SS_BF_sens_AR1} and \ref{fig:SS_BF_sens_AR4}.

\section{Conclusion}
\label{sec:conclusions}

In this study we have investigated the primary symmetry-breaking bifurcation of the flow past axisymmetric bodies with a no-slip surface. 
We have computed the neutral curves $Re_c(\AR)$ and investigated the instability mechanism using various quantities: structural sensitivity, sensitivity to a base-flow modification or to an increase in $Re$, dynamics of the perturbation kinetic energy, and endogeneity.
We have also proposed a new scaling that is suitable for predicting the onset of the bifurcation. The generality of our conclusions has been assessed by considering bodies with different geometries (ellipsoids, bullets, cones and bicones) and different aspect ratios ($1 \le \AR \le 8$).

The instability is driven by the azimuthal vorticity generated at the body surface. We have shown that the mechanism proposed by \cite{magnaudet-mougin-2007} in the context of free-slip spheroidal bubbles well describes the onset of the instability for bodies with a no-slip surface too. In both cases, indeed, the instability arises in a thin region in the near wake, and its onset is strongly related to the occurrence of points where $\partial \omega_\theta/\partial r = 0$, i.e. points where isocontours of $\omega_\theta$  align with the radial direction and are perpendicular to the symmetry axis. 
At these points, the streamwise vorticity gradient $\partial \omega_\theta/\partial z$ largely increases with the Reynolds number, favouring %the onset of 
the instability.

Having characterised the physical mechanism, we have then proposed a new scaling for the prediction of the instability, in the spirit of the study of \cite{chiarini-quadrio-auteri-2022b} in the context of 2D bluff bodies. The new scaling is based on measures of the near-wake azimuthal vorticity, $\widetilde{\omega}_\theta$, and the radial extent of the wake recirculation region, $r_0$. 
When computed at criticality, the ensuing Reynolds number $\widehat{Re} = \widetilde{\omega}_\theta r_0^2/\nu$ has been shown to collapse 
approximately to the same value across all geometries and aspect ratios. 
This observation can be used to readily estimate whether the steady base flow past axisymmetric bodies is unstable, avoiding a computationally expensive stability analysis and relying only on an inspection of the base flow.

While the present study focuses on the first instability, it is worth concluding with  a word about turbulent wakes.
In the turbulent regime, while the mean (time-averaged) flow past an axisymmetric bluff body is itself axisymmetric, the most probable instantaneous flow state is a static $m=1$ symmetry-breaking deflection \citep{grandemange-etal-2014, rigas2014}. 
The wake visits different azimuthal orientations $\theta$ through sudden switches that are triggered by turbulent fluctuations, and that are both random and rare events: the residence time $\Delta t$ between successive switches follows an exponentially decreasing distribution with no preferred frequency, and the mean residence time  $\langle \Delta t\rangle$ is orders of magnitude larger than the convective time $L/U_\infty$.
Qualitatively similar dynamics are observed in non-axisymmetric bodies with a planar symmetry, like the Ahmed body \citep{grandemange-etal-2013, barros-etal-2017}.
For both types of bluff bodies, static symmetry-breaking  deflections of the wake are reminiscent of the primary, laminar pitchfork bifurcation. 
The understanding of the mechanism of the low-$Re$ instabilities may thus play an important role for the understanding of the flow physics at larger $Re$ and, for example, to locate the most sensitive regions in order to develop new flow control strategies. 
While the notion of critical Reynolds number may not be relevant in the turbulent regime, it would be interesting to analyse mean turbulent wakes through the lens of the present study.
For example, one could evaluate the newly defined Reynolds numbers $\widehat{Re}$ and $\overline{Re}$ and investigate how they depend on the geometry, aspect ratio and $Re$. 
It may also be worth determining whether regions with $\partial \omega_\theta/\partial r<0$  can be sustained or if the competition between the  production and advection of vorticity is such that  $\partial \omega_\theta/\partial r >0$ everywhere in the near wake.

\section*{Funding} 
This research received no specific grant from any funding agency, commercial or not-for-profit sectors.

\section*{Declaration of Interests} 
The authors report no conflict of interest.

\appendix
\section{Characteristic base-flow quantities at $Re_c$}
\label{sec:appendix}

\begin{figure}
\iffalse
\centerline{   
    \begin{overpic}[width=7cm, trim=30mm 90mm 30mm 90mm, clip=true]{figure/curves/umin.pdf}
       \put(0,70){$(a)$}
 	\end{overpic}  
    \begin{overpic}[width=7cm, trim=30mm 90mm 30mm 90mm, clip=true]{figure/curves/xmin.pdf}
       \put(0,70){$(b)$}
 	\end{overpic} 
}
\fi 
\centerline{   
    \begin{overpic}[width=7cm, trim=30mm 90mm 30mm 90mm, clip=true]{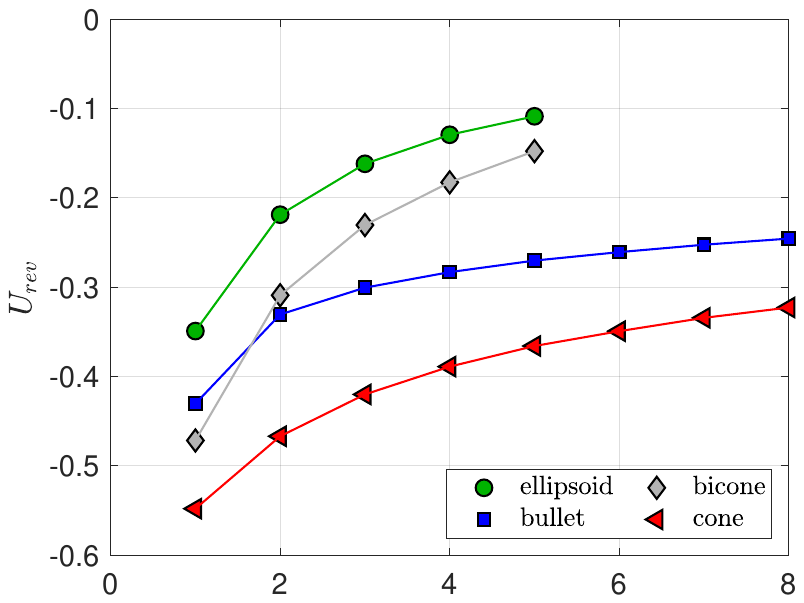}
       \put(0,70){$(a)$}
       \put(49,1){\scriptsize{$\AR$}}
 	\end{overpic}  
    \begin{overpic}[width=7cm, trim=30mm 90mm 30mm 90mm, clip=true]{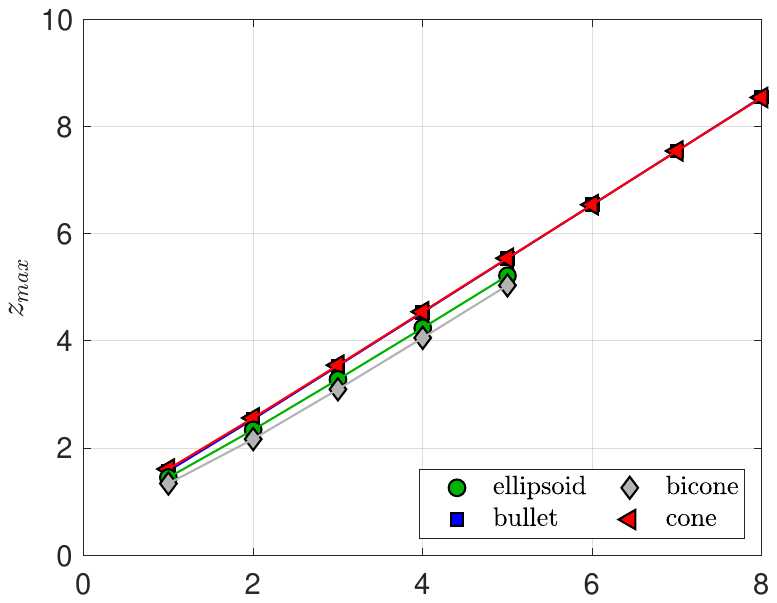}
       \put(0,70){$(b)$}
       \put(49,1){\scriptsize{$\AR$}}
 	\end{overpic} 
}
\centerline{   
    \begin{overpic}[width=7cm, trim=30mm 90mm 30mm 90mm, clip=true]{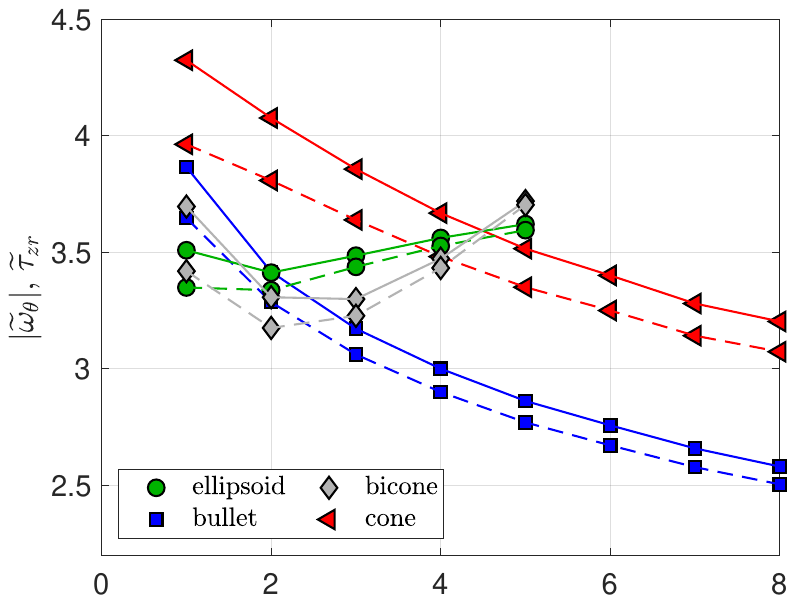}
       \put(0,70){$(c)$}
       \put(49,1){\scriptsize{$\AR$}}
 	\end{overpic}   
    \begin{overpic}[width=7cm, trim=30mm 90mm 30mm 90mm, clip=true]{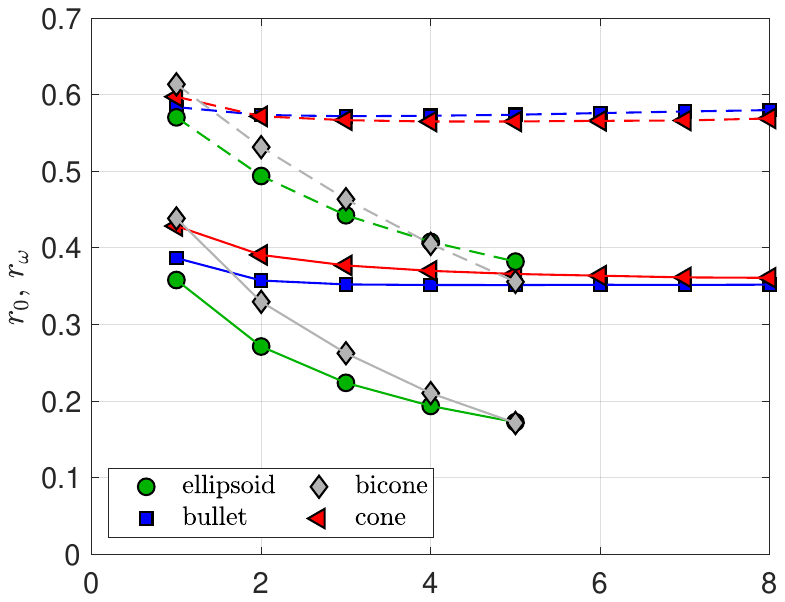}
       \put(0,70){$(d)$}
       \put(49,1){\scriptsize{$\AR$}}
 	\end{overpic}  
}
\iffalse
\centerline{   
    \begin{overpic}[width=7cm, trim=30mm 90mm 30mm 90mm, clip=true]{figure/curves/max_vort_max_tau_at_xmin.pdf}
       \put(0,70){$(c)$}
 	\end{overpic}   
    \begin{overpic}[width=7cm, trim=30mm 90mm 30mm 90mm, clip=true]{figure/curves/rayon_max_vort_max_tau_at_xmin.pdf}
       \put(0,70){$(d)$}
 	\end{overpic}  
} 
\fi
\caption{
Base-flow quantities at the critical Reynolds number $Re_c$.
$(a)$~Maximum reverse flow $U_{rev}$.
$(b)$~Streamwise location $z_{max}$ of maximum reverse flow.
$(c)$~Maximum azimuthal vorticity $\widetilde{\omega}_\theta$ (solid lines) and maximum shear $\widetilde{\tau}_{zr}$ (dashed lines) at $z=z_{max}$.
$(d)$~Radius $r_0$ of zero streamwise velocity (solid lines)  and radius $r_\omega$ of maximum azimuthal vorticity (dashed lines) at $z=z_{max}$.
}
\label{fig:BF_quantities}
\end{figure}

In \S\ref{sec:scaling} we have introduced two new  Reynolds numbers, i.e. 
%$\widehat{Re} = \tilde{\Omega} r_0^2/\nu$ 
$\widehat{Re} = \widetilde{\omega}_\theta r_0^2/\nu$ and $\overline{Re} = \widehat{\omega}_\theta r_0^2/\nu = U_{rev} r_0/\nu$.
Figure~\ref{fig:BF_quantities} shows the base-flow quantities involved in the definitions of $\widehat{Re}$ and $\overline{Re}$ and measured at $Re=Re_c$. They are
the maximum reverse flow $U_{rev}$ (figure~\ref{fig:BF_quantities}$a$),
the streamwise location $z_{max}$ of maximum reverse flow (figure~\ref{fig:BF_quantities}$b$),
the maximum azimuthal vorticity $\widetilde{\omega}_\theta$ (and, for reference, the maximum shear $\widetilde{\tau}_{zr}$) at $z=z_{max}$  (figure~\ref{fig:BF_quantities}$c$),
and the radius $r_0$ of zero streamwise velocity    (and, for reference, the radius $r_\omega$ of maximum azimuthal vorticity) at $z=z_{max}$ (figure~\ref{fig:BF_quantities}$d$).
We note in particular that, for bodies with a zero-thickness TE, $\widetilde{\omega}_\theta$ increases with $\AR \geq 2$ and $r_0$ decreases, while for bodies with a blunt TE, $\widetilde{\omega}_\theta$ decreases  and $r_0$ remains approximately constant.

%---------------------------------------------------
\section{Mesh convergence}
\label{sec:appendix_convergence}

Table~\ref{tab:mesh_convergence} reports the critical Reynolds numbers obtained on three different numerical meshes, for all the geometries considered in the present study and for $\AR=1$ and $\AR=4$.
From one mesh to the next, all mesh sizes are divided by $\sqrt{2}$, such that the number of elements $N_{elmts}$ increases by a factor of approximately~2. 
Additionally, table~\ref{tab:mesh_convergence_2} reports the base-flow quantities involved in the definition of the newly introduced Reynolds numbers $\widehat{Re}$ and $\overline{Re}$. For illustration purposes, we arbitrarily chose the flow past the $\AR=1$ cone at $Re=160$.
In the present study we used mesh M2, for which all results are very well converged.

\begin{table}
\centering
\begin{tabular}{l cc ccccccccc} 
          &&    & & \multicolumn{2}{c}{$\AR=1$} && \multicolumn{2}{c}{$\AR=4$} \\ 
Geometry  &&  Mesh && $N_{elmts}$ & $Re_c$ && $N_{elmts}$ & $Re_c$\\ \\
%--------------------------------------------
          &&   M1 & &  42141  & 212.67 &&  60643 & 999.64 \\
Ellipsoid &&   M2 & &  79198  & 212.67 && 113796 & 999.66 \\
          &&   M3 & & 150697  & 212.66 && 217854 & 999.66 \\ \\
%--------------------------------------------
          &&   M1 & &  44697  & 135.13 &&  70250 & 758.94 \\
Bicone    &&   M2 & &  82757  & 135.13 && 129564 & 759.02 \\
          &&   M3 & & 156943  & 135.13 && 244945 & 758.90 \\ \\
%--------------------------------------------
          &&   M1 & &  40207  & 216.13 &&  51755 & 383.87 \\
Bullet    &&   M2 & &  73940  & 216.13 &&  93773 & 383.87 \\
          &&   M3 & & 139788  & 216.14 && 174449 & 383.88 \\ \\  
%--------------------------------------------
          &&   M1 & &  41976  & 164.69 &&  58107 & 262.74 \\
Cone      &&   M2 & &  77044  & 164.69 && 104849 & 262.75 \\
          &&   M3 & & 145009  & 164.69 && 194017 & 262.74 \\  
%--------------------------------------------
\end{tabular} 
\caption{Values of the critical Reynolds numbers obtained on three different meshes M1-M3 for the four different geometries, and two aspect ratios each.}
\label{tab:mesh_convergence}
\end{table}

\begin{table}
\centering
\begin{tabular}{c ccccccccc}   
Mesh && $N_{elmts}$ & $U_{rev}$ & $z_{max}$ & $\widetilde\omega_\theta$ & $\widetilde\tau_{zr}$ & $r_0$ & $r_\omega$ \\ \\
%--------------------------------------------
M1 & &  41976  & -0.5737 & 1.5985 & -4.2795 & 3.9234 & 0.4255 & 0.5936 \\
M2 & &  77044  & -0.5737 & 1.5985 & -4.2796 & 3.9235 & 0.4255 & 0.5937 \\
M3 & & 145009  & -0.5737 & 1.5985 & -4.2795 & 3.9235 & 0.4255 & 0.5937 \\  
%--------------------------------------------
\end{tabular} 
\caption{Values of the base-flow quantities of figure~\ref{fig:BF_quantities} obtained on three different meshes M1-M3 for the $\AR=1$ cone at $Re=160$.}
\label{tab:mesh_convergence_2}
\end{table}

%---------------------------------------------------
\section{Rounding sharp leading edges}
\label{sec:appendix_sharp_LE}

In section~\ref{sec:max_vort}, we reported the maximum vorticity at the wall for ellipsoids and bullets, i.e. bodies with a smooth leading edge. Here, we give more details about bodies with a sharp leading edge.

\begin{figure}
\centerline{    
    \begin{overpic}[width=7cm, trim=30mm 90mm 30mm 90mm, clip=true]{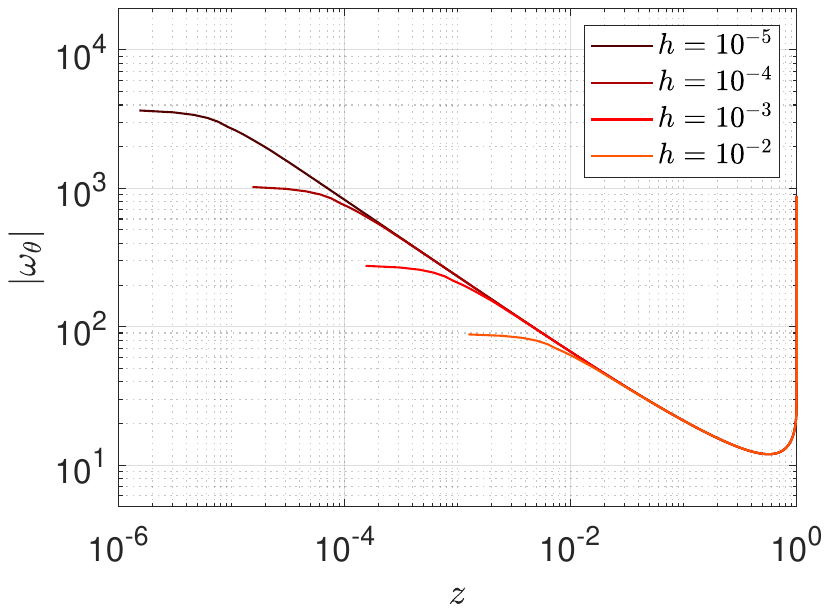}
       \put(0,70){$(a)$}
 	\end{overpic}  
    \begin{overpic}[width=7cm, trim=30mm 90mm 30mm 90mm, clip=true]{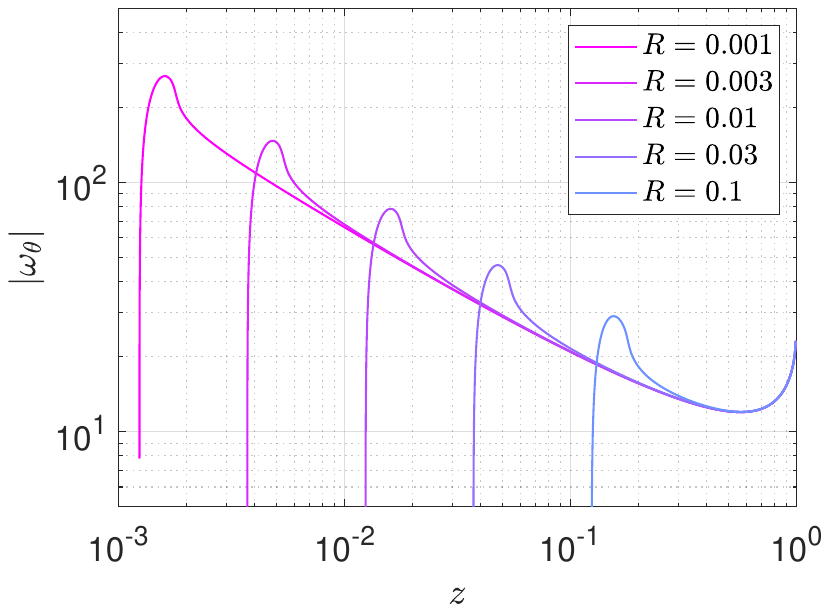}
       \put(0,70){$(b)$}
 	\end{overpic}  
}
\caption{
Azimuthal vorticity $\omega_\theta(z)$ on the lateral surface of a cone for $\AR=1$ at $Re=160$.
$(a)$~Sharp leading edge, varying local mesh size  $h$.
$(b)$~Rounded leading edge, varying fillet radius $R$.
}
\label{fig:vorticity_rounded_LE}
\end{figure}

Figure~\ref{fig:vorticity_rounded_LE}$(a)$ shows the azimuthal vorticity $\omega_\theta(z)$ obtained numerically on the surface of an $\AR=1$  cone with a sharp leading edge, at $Re=160$. 
Different curves correspond to different values of the local mesh size $h$ at the leading edge.
The vorticity computed at the sharp leading edge keeps increasing when the numerical mesh is refined locally.

Next, we investigate the effect of smoothing the leading edge with a fillet of radius $R$. 
We use a local mesh size $h=R/10$, and verify convergence with $h=R/30$.
Figure~\ref{fig:vorticity_rounded_LE}$(b)$ shows that the maximum vorticity is obtained slightly downstream of the fillet and increases as $R$ decreases.  
However, the critical Reynolds number remains essentially constant (not shown), which is consistent with all the curves $\omega_\theta(z)$ collapsing downstream enough (for $z \gtrsim 0.4$ in the present case). 
Therefore,  $Re_c$ and the maximum vorticity are not related.

\bibliographystyle{jfm}
\bibliography{Stability}

\end{document}